\documentclass[aps,prc,showpacs,nofootinbib]{revtex4}

\usepackage{amsmath,amssymb}
\usepackage{mathrsfs}
\usepackage{graphicx}
\usepackage{amsthm}
\usepackage{amscd}
\usepackage{bm}
\usepackage{dsfont}

\newcommand{\beq}{\begin{eqnarray}}
\newcommand{\eeq}{\end{eqnarray}}
\newcommand{\be}{\begin{equation}}
\newcommand{\ee}{\end{equation}}

\newcommand{\la}{\lesssim}

\newcommand{{\SD}}{\rm SD}

\newcommand{\vex}{\mbox{\boldmath${\rm x}$}}
\newcommand{\vey}{\mbox{\boldmath${\rm y}$}}
\newcommand{\ver}{\mbox{\boldmath${\rm r}$}}

\newcommand{\vep}{\mbox{\boldmath${\rm p}$}}

\newcommand{\veK}{\mbox{\boldmath${\rm K}$}}

\newcommand{\vem}{\mbox{\boldmath${\rm m}$}}
\newcommand{\vel}{\mbox{\boldmath${\rm l}$}}
\newcommand{\veR}{\mbox{\boldmath${\rm R}$}}

\newcommand{\vek}{\mbox{\boldmath${\rm k}$}}
\newcommand{\ven}{\mbox{\boldmath${\rm n}$}}

\newcommand{\lan}{\langle}
\newcommand{\ran}{\rangle}
\newcommand{\WID}[2]{{#1}_{#2}}
\newcommand{\WIT}[3]{{#1}_{#2}^{#3}}
\newcommand{\xmn}[2]{{#1} \times 10^{#2}}

\begin{document}

\title{Nuclear matter at high density: Phase transitions, \\multiquark states, and supernova outbursts}
\author{ M.~I.~Krivoruchenko, D.~K.~Nadyozhin, T.~L.~Rasinkova,
 Yu.~A.~Simonov, M.~A.~Trusov, A.~V.~Yudin}
\affiliation{Institute for Theoretical and Experimental Physics,
Moscow, Russia}

\begin{abstract}
Phase transition from hadronic matter to quark-gluon matter is discussed for various
regimes of temperature and baryon number density.
For small and medium densities, the phase transition is accurately described in the framework of the Field Correlation Method,
whereas at high density predictions are less certain and leave room for the phenomenological models.
We study formation of multiquark states (MQS) at zero temperature and high density.
Relevant MQS components of the nuclear matter can be described using a previously
developed formalism of the quark compound bags (QCB).

Partial-wave analysis of nucleon-nucleon scattering indicates the existence of 6QS which
manifest themselves as poles of $P$-matrix. In the framework of the QCB model,
we formulate a self-consistent system of coupled equations for the nucleon and 6QS propagators in nuclear matter and the G-matrix. The approach provides
a link between high-density nuclear matter with the MQS components and the cumulative effect observed in reactions on the nuclei, which requires the admixture
of MQS in the wave functions of nuclei kinematically.

6QS determine the natural scale of the density for a possible phase transition into the MQS phase of nuclear matter.
Such a phase transition can lead to dynamic instability of newly born protoneutron stars and dramatically
affect the dynamics of supernovae. Numerical simulations show that the phase transition may be a good remedy for
the triggering supernova explosions in the spherically symmetric supernova models. A specific signature of the
phase transition is an additional neutrino peak in the neutrino light curve.
For a Galactic core-collapse supernova, such a peak could be resolved by the present neutrino detectors.
The possibility of extracting the parameters of the phase of transition from observation of the neutrino signal is discussed also.
\end{abstract}

\pacs{12.38.Mh, 12.39.Mk, 26.30.-k, 25.75.Nq}

\maketitle

\newpage
\tableofcontents
\newpage

\renewcommand{\baselinestretch}{1.0}
\renewcommand{\theequation}{\thesection.\arabic{equation}}

\section{Introduction}
\setcounter{equation}{0}

The problem of transition from the confined phase of hadronic matter to the
deconfined quark-gluon phase is now widely discussed both on theoretical
\cite{1} and experimental levels \cite{2} and also using lattice simulations
\cite{3}. The transition process is nonperturbative in nature and requires for
a realistic description the use of nonperturbative models.
Nambu-Jona-Lasinio (NJL) model and its generalizations describe successfully the spontaneous breaking
of chiral symmetry but, unfortunately, do not lead to confinement and can not tell much about
the deconfinement phase transition.

Recently the Field Correlator Method (FCM) \cite{4}, successfully applied
earlier to the description of confinement and  hadrons in the vacuum, has been
extended for the description of hadron matter at nonzero temperature and
density including phase transition into the deconfined phase. In this method,
the quark-hadron phase transition occurs due to in-medium modifications of the
QCD vacuum. The energy density of the QCD vacuum is determined by the gluonic
condensate,  and the colorelectric component of it is responsible for the
confinement. In course of the quark-hadron phase transition, the colorelectric
component evaporates, whereas the colormagnetic component stays intact (for a
review see \cite{5}).

In this way one obtains the critical transition temperature as a function of
the quark flavors in agrement with the lattice data \cite{6}. The phase diagram
extrapolated to small temperatures points toward the critical quark chemical
potentials of order $\mu_{cr} \cong 0.6$ GeV. This value is large enough to
allow the ordinary nuclei to exist as collection of nucleons rather than
quarks. The low-temperature phase transition gives rise to strong
nonperturbative attraction in colorless channels \cite{7} that provides
dominance of the $q \bar q$ correlations over the diquark $qq$ ones, making
thereby conjecture on color superconductivity not quite realistic, in line with
earlier findings \cite{8}. In two-color QCD, where $qq$ and $q\bar q$ are
equivalent, superconducting phase of quark matter with colorless quark Cooper
pairs \cite{KGK91,GKK91,KKR92} is confirmed by lattice simulations \cite{HAN06}.

The low-temperature region of the phase diagram is, however, less
certain because no interaction in the hadron medium (nuclear
matter) was taken into account yet. The standard models of nuclear
matter (see e.g. \cite{RIFF}) based on the realistic $NN$
potentials predict energy per nucleon $E/A$ depending rather
sensitively on details of $NN$ interaction.

An additional uncertainty comes from the $3N$ forces and nucleon-hyperon
transitions, which make computation of equation of state (EoS) of the nuclear matter
a rather complicated task, which calls for experimental verification. At this point an important
piece of information can be obtained from astrophysics.

During the last decade, great progress is made in observational
astrophysics towards the study of properties of neutron stars (see e.g.
\cite{BARR05,OZEL06} and, for a review, \cite{LATT}). EoS of nuclear matter determined
by the density dependence of $E/A$ is important
for calculation of structure and cooling rates of neutron stars and possible
appearance of quark matter in interiors of neutron stars as was suggested for the first time by Ivanenko
and Kurdgelaidze \cite{IVKU69}. The current status of the quark stars
hypothesis is discussed in Refs. \cite{LATT,WEBE05}.

The recent interest to the nuclear matter EoS is also connected to new constraints
obtained from the collective flow data and subthreshold
kaon production in heavy-ion collisions \cite{DANI02,FUCH07}.

One of the main purposes of the present paper is to consider a generalized
picture of $NN$ interaction and of nuclear matter, which includes from the very
beginning the quark degrees of freedom in the form of MQS. This implies a
revision of the standard picture, based on the Yukawa mechanism of meson
exchanges.

In 1935 Yukawa \cite{Yuk35} proposed a hypothesis that the interaction between the
nucleons may be due to the exchange of finite-mass meson. The experimental searches
have led to the discovery of pions, heavier mesons, and eventually to the emergence
of One Boson Exchange (OBE) model of nucleon-nucleon interaction. In this model,
pions are responsible for long-range part of the nucleon-nucleon potential, while
heavy mesons are responsible for the interaction at intermediate and short distances
\cite{BREI60}.

OBE models provide an accurate quantitative description of nucleon-nucleon interaction.
The physical meaning of the meson exchange at short distances is, however, not entirely
clear because of the finite size of nucleons and mesons. The values of the proton and
pion charge radii
$<r_{p}^{2}>^{1/2}=0.875\pm 0.007$ fm and $<r_{\pi}^{2}>^{1/2}= 0.659 \pm 0.025$ fm
\cite{PDG08} indicate that the mechanism of pion exchange is justified at distances
 $r \gtrsim  1$ fm while at smaller distances there is overlap between hadrons, where
the quark-gluon degrees freedom come into play and affect the dynamics.

Restrictions of this kind should be taken into account when calculating EoS of nuclear matter. The maximum density of nuclear matter below which the OBE models
can be applied is determined by comparing the proton charge radius with the average distance
between the nucleon with its nearest neighbor. The nearest neighbor is localized between
$r$ and $r + dr $, while the sphere of radius $r$ with the probing nucleon in the center is empty.
The probability distribution of the nearest neighbor is the probability
of not finding a nucleon inside the sphere multiplied by the probability of finding a nucleon
in the volume element $dV=4\pi r^{2}dr$. The probability do not find a nucleon inside the
sphere is given by the Poisson law $P­_{0} = \exp(-\rho V)$, where $\rho$ is nuclear matter density.
For the saturation density $\rho_{0} = 0.16$ fm$^{-3}$, a simple calculation gives $<r> = 1.02 \pm 0.37$ fm,
where $ <r> $ is the mean distance, the second number is the standard error.
When the density is close to saturation density, OBE mechanism, obviously, can not be applied.

The problem of nuclear matter EoS is studied in the realistic OBE models of nucleon-nucleon interaction
since the mid-1970's. One can distinguish three main approaches: A popular class of models based
on the mean-field approximation \cite{Walecka:1974qa,Chin:1977iz}. The field-theoretic Dirac-Brueckner-Hartree-Fock
method goes beyond mean-field approximation (see \cite{RIFF} and references therein).
Variational method is described in Ref. \cite{PAND}.

One should note that in all approaches of OBE-type additional fine tuning is required
to obtain the realistic values of energies and densities, e.g., in the variational approach,
$3N$ forces have to be added to the $NN$ potentials.

OBE models are in reasonable agreement with laboratory data \cite{DANI02,FUCH07}, but predict a surprisingly
low value of the maximum mass of neutron stars in the $\beta$-equilibrium,
if one includes hyperons \cite{GLEN91,BURG03,ISHI08,SCHA08,DAPO08}.
$\beta$-equilibrium leads to the occurrence of the hyperons when density increases to about $(2\div 3)\rho_{0}$
and, consequently, leads to a softening of the nuclear matter EoS.

It has earlier been noted \cite{Krivoruchenko:2009kx} that observational data on the rotation
speed of X-ray transient XTE J1739-285, which point to a soft EoS, and mass of the pulsar PSR J1748-2021B,
which points to a very stiff EoS, are almost mutually exclusive. In a recent study \cite{Ozel:2010fw}
data on compact sources 4U 1608-52, 1820-30 and 4U EXO 1745-248 were re-analyzed. The authors came to
conclusion that only a very soft EoS is consistent with values of the mass and radius of the stars.
A similar technique was used previously in the analysis of data from a compact source EXO 0748-676.
In conjunction with the analysis \cite{OZEL06}, one can assume the existence of two classes
of compact stars, e.g., neutron stars metastable against conversion to exotic stars,
such as quark stars, strange stars, or perhaps dibaryon stars. The results of Refs. \cite{OZEL06,Ozel:2010fw}
still require confirmation.

The difficulty in describing the massive neutron (hyperon) stars
as well as the possibility of existence of various classes of compact objects
stimulate the search for new concepts and models in which quark-gluon dynamics
at short distances plays a more prominent role.
One can expect that the microscopic models are more adequate at high density
and can provide a physically satisfactory picture of the short-distance dynamics and high-density EoS.

The standard realistic $NN$ forces (and also $3N$ forces) used for nuclear matter
calculations exploit only baryonic and mesonic degrees of freedom. However,
multiquark states (MQS) can appear in nuclear matter also, yielding degrees
of freedom of their own.

This idea is discussed, e.g., in Refs. \cite{BALD84,Nazmitdinov:1985qd,12}. In most papers,
it is associated with multiquark bags similar to the MIT bags \cite{MIT}, where only perturbative interquark forces
act inside bags. This invokes immediately the idea of deconfined pieces of matter
inside of hadrons and the quark-hadron phase transition caused by the increased number
of multiquark bags and their final overlap.

However, the MIT bag is only a crude model of hadrons. Lattice and other
analytical models show that the correlation length in the QCD vacuum is very
small $\sim 0.1$ fm. This fact is due to large mass of nonperturbative
structures called gluelumps (for a review see \cite{Simonov:2009nf}). The
string (confinement) between quarks acts therefore already at distances $\sim
0.1$ fm or less and instead of multiquark bags one gets for MQS
strongly bound systems with inner density growing with the number of quarks
and a radius twice smaller than in the MIT bag model \cite{13}.
Moreover, a detailed study of the dependence of the confinement potential on the density  \cite{14}
showed that the medium effects lead to an additional attraction, as a result of which MQS are becoming smaller and lighter.
A growing admixture of MQS can produce another minimum in the
curve of $E/A$ vs density providing a phase transition into the heterophase
nuclear matter with substantial MQS component or the quark matter.

MQS can play an important role in the so-called cumulative and subthreshold processes
i.e., reactions on nuclei that cannot proceed on single nucleons \cite{15}. The
corresponding cumulative number $N_{cum}$ defines the minimal number of
nucleons needed kinematically for the reaction. It is clear,
that MQS are appropriate objects to provide the cumulative effect.
The MQS density enters directly cross-sections of the cumulative and subthreshold
reactions. There is a close connection between the MQS density in nuclei and neutron star interiors.

Multiquark configurations that appear first when overlap of nucleons becomes significant
can be considered as a kind of dibaryons. The experimental searches of dibaryons in the past did
not give conclusive results. Recently, resonance behavior of the double pionic fusion reaction
$pn \to d \pi^0\pi^0$ measured at CELSIUS-WASA has been interpreted as
evidence for a $\Delta\Delta$ dibaryon \cite{Bashkanov:2008ih}. Such dibaryon has,
however, more features in common with deuteron rather than a compact 6QS.

The possibility for occurrence of a Bose condensate of dibaryons in nuclear
matter is discussed in Refs.
\cite{BALD84,Nazmitdinov:1985qd,Krivoruchenko:1987gv,Faessler:1996ta,Faessler:1997jg,Buchmann:1996pd,Faessler:1997kz,Faessler:1997mb}.
The ground state of the heterophase nucleon-dibaryon matter is stable for a wide
range of parameters of the models discussed and nuclear matter densities. Using the
mean-field approximation of the OBE models, constraints for the $\omega $- and
$\sigma $-meson coupling constants with dibaryons were extracted from properties
of nuclear matter at saturation, stability condition of the binary mixture
of nucleons and dibaryons, and from the existence of massive neutron stars.

To give a quantitative method for study of the nucleon and dibaryon fields,
we are using the so-called Quark Compound Bag (QCB) model \cite{SIMO81},
appeared in the development of the $P$-matrix formalism of Jaffe and Low \cite{JALO}.
Jaffe and Low proposed to identified MQS with the so-called "primitives"
which appear as poles of $P$ matrix rather than $S$ matrix.
The $P$-matrix formalism was used to yield an accurate
description of the nucleon-nucleon systems in Ref. \cite{NARO94}
(see also \cite{SIMO81,SIMO84,BHGU,FALE,MULD}).

The dynamical character of the QCB model allows to apply it for the description of
nucleon-nucleon interaction at finite density and temperature and for the study
of nuclear matter EoS.

It should be noted also that the QCB model is very economical since it uses only a few parameters,
such as mass and radius of the MQS with a fixed orbital and total angular momentum of quarks.

The average orbital momentum of a nucleon pair in the nuclear matter can be estimated as
$L \lesssim r p_{F} = 1.4 \pm 0.5$. This estimate does not depend on the density.
The $s$-channel dibaryon exchange as the mechanism of nucleon-nucleon interaction
is therefore restricted in nucleon matter by a few lowest partial waves.
An additional suppression of the high partial waves appears in the low-density limit
because of the threshold behavior $\sim a (p_{F})^L$ of the partial wave amplitudes,
where $a$ is the scattering length. The $S$-wave interaction is therefore dominant
in the low-density limit.


The comparison of the observational data with the calculations of static properties of neutron
stars and the dynamic simulations of core-collapse supernovae rise questions which have not found
a definite answer yet. One such issue is the quantitative properties of the phase transition. The inclusion of MQS in
the $NN$ interaction dynamics, e.g., in the QCB model is an alternative approach to describe nuclear forces
and nuclear matter EoS. In this approach, the modeling of heterophase states of high-density baryonic matter is possible.

The physical conditions that appear at the final stages of the core collapse in interiors
of newly born hot protoneutron stars are unique to constrain parameters of the quark-hadron
phase transition, which cannot be constrained otherwise from laboratory data (see, e.g., \cite{KRMA91}).
Observation of neutrino bursts from supernovae, which may genetically be related to a phase transition,
will play an important role in the identification of physical nature and properties of the exotic states of nuclear matter at high density.

Massive stars end their life with gravitational collapse of iron central cores that according
to astronomical observations must result in the explosion of so-called core-collapse
supernovae (SN). However, an extensive hydrodynamic modeling during already more than thirty years
has demonstrated that in case of spherical symmetry it is very hard to simulate the explosion:
stellar envelope would not separate and finally fell back on the collapsed core.
Therefore, the SN theorists began to concentrate special attention on basically nonspherical effects
such as rotation, magnetic fields, large scale convection, jet streams, etc
(e. g., see \cite{Nadyozhin 2005,Nadyozhin 2008} and references therein).

Here, we discuss the calculations of gravitational collapse that was undertaken to estimate
the effect of the nuclear phase transition on the core collapse dynamics.
Although the property of phase transition to destabilize
the hydrostatic equilibrium of stars and giant planets is well known for a long time
\cite{Ramsey 1950,Lighthill 1950, Seidov 1967,Bisnovatyi-Kogan et. al. 1975,
Bisnovatyi-Kogan 1989} the hydrodynamic consequences of such a destabilization
were yet rarely addressed in the investigation of SN mechanism.
For a historical review and as the starting point for understanding
the instability induced by phase transition, remarkable essays by Seidov
\cite{Seidov 1999,Seidov 1999a} can be recommended.
It was understood that the phase transition onset in stellar center can generate
a shock wave \cite{Takahara 1988,Gentile 1993}.
However the hydrodynamic effects on the SN mechanism expected from
such a shock required further detailed study.
Recently there appeared a detailed research of possible phase transition
influence on the dynamics of the collapse and SN mechanism
\cite{Sagert 2009,Sagert 2009a}. The research is based on
sophisticated hydrodynamic code including neutrino transport
and quark-hadron phase transition simulated with the aid of
the MIT bag model. It was shown that an additional neutrino peak appears in
the neutrino light curve as a specific signature of the phase transition.
For the core-collapse supernovae of our Galaxy, this peak could be resolved by the existing neutrino detectors.

The QCD phase transitions may be a good remedy for triggering
the SN explosion in spherically symmetric SN models, while the
detected neutrinos can provide important information on physical
nature and parameters of the QCD phase transition.

This work reviews phenomenological and microscopic approaches to $NN$
interactions and nuclear matter properties with inclusion of quark-gluon
degrees of freedom, quark-hadron phase transitions and influence of phase
transitions on physics of neutron stars, supernova outbursts and neutrino
signals from supernovae.

In Sect. II we discuss the $\mu -T$ QCD phase diagram, give estimates of
critical temperature of the deconfinement phase transition at low baryon
chemical potentials, and discuss the role of diquark correlations in cold quark
matter. In Sect. III we discuss modification of the confinement force at nonzero
chemical potential and estimate critical density above which formation of MQS
becomes energetically preferable. In Sect. IV the QCB model is introduced in its
original nonrelativistic form, while Sects. V-VII are devoted to the
relativistic formulation of QCB and the accurate description of $NN$ data in
the QCB framework. In Sect. VIII the EoS of nuclear matter are given in the
standard OBE form, in Sect. IX  in a simplified nonrelativistic QCB form,
while section X is devoted to the relativistic mean field treatment based on
the QCB model of Sect. V. In Sect. X  we provide equations to calculate the $G$-matrix
and describe qualitative features of the phase transition to the new MQS phase.
In Sect. XI astrophysical aspects of (proto)
neutron star dynamics are given and the evolution of the supernova outbursts is
presented based on detailed calculations. Section XII is devoted to conclusions
and perspectives of the present approach. Appendices contain technical
details necessary to derive some of the equations in the text.


\begin{figure} [htb!]
\includegraphics[width = 0.618\textwidth]{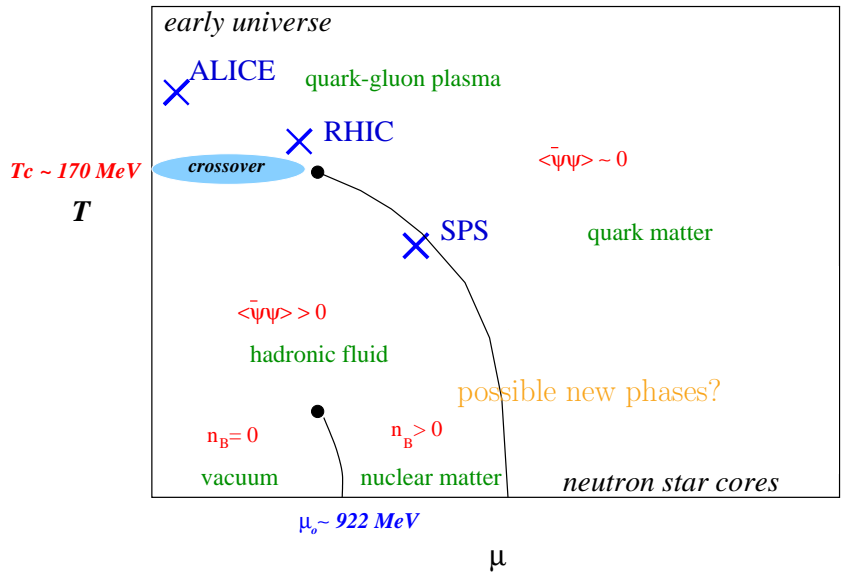}
\caption{QCD phase diagram in the temperature-chemical
potential plane. First-order phase transitions are shown by solid curves.
Filled circles are second-order phase transitions.
Crossover at $\mu \approx 0$ shows transition from hadronic phase to quark-gluon plasma.
Crosses depict heavy-ion collision experiments.
}
\label{fig:qcd}
\end{figure}

\section{Quark-hadron phase transition in QCD}
\setcounter{equation}{0}

In Fig. \ref{fig:qcd} one can see the generic picture of the  phase transition
expected in QCD. Only the low-$\mu$ region is accessible to lattice QCD, while
mostly NJL-type models are used to predict the phase curve and the
superconducting phases. We shall show below that the phase curve can be
obtained from the Field Correlator method (FCM) \cite{4,5} and indeed looks
like in Fig. \ref{fig:qcd}, whereas the superconducting phases are unlikely. We
shall also find the critical values $T_c$ and $\mu_c$ for symmetric nuclear
matter ($\mu_p = \mu_n \equiv 3 \mu_q$) with switched off weak interactions in
terms of the gluon condensate, following \cite{6}.

\subsection{Gibbs' criterion}

According to the Gibbs' criterion, two phases in thermodynamic equilibrium
have the same temperatures, pressures, and balanced chemical potentials.
Phase with highest pressure is thermodynamically preferred provided other conditions are equal.
Chemical equilibrium leads to a set of relations for the chemical potentials
of constituents. If weak interactions are switched off, chemical equilibrium
for the substance with $n_{f}$ flavors reduces to
\begin{equation}
\mu_A = C_{A}^{a}\mu_{a},
\label{MATCSTRO}
\end{equation}
where $\mu_{A}$ and $\mu_{a}$ are the chemical potentials
in the hadron phase with $A= p, n, \Lambda \ldots $, $\pi^{\pm}$- and $\pi^{0}$-mesons, etc.
and in the quark phase with $a = u$-, $d$-quarks for $n_{f}=2$ and $u$-, $d$-, $s$-quarks
for $n_{f}=3$.
The matrix $C_{A}^{a}$ determines the composition of particles in the hadron phase
and, in particular, allows to express electric charge of particles in the hadron phase $e_{A}$
in terms of electric charge of the constituents $e_{a}$:
$e_{A} = C^{a}_{A}e_{a}$.
If leptons appear in the substance then $A$ and $a$ run over the leptons too.
For a representative set of hadrons and electrons and muons
$C_{A}^{a}$ has the form
\[
C_{A}^{a}=
\begin{array}{ll}
\begin{array}{ll}
\;\;\; & a\rightarrow
\end{array}
& \;
\begin{array}{lllll}
\;u & \;\;d & s & e & \mu
\end{array}
\\
\begin{array}{l}
A \\
\downarrow  \\
\; \\
\; \\
\; \\
\; \\
\; \\
\end{array}
\begin{array}{l}
p \\
n \\
\Sigma ^{-} \\
\pi ^{-} \\
e \\
\mu  \\
\vdots
\end{array}
& \left\|
\begin{array}{lllll}
2 & 1 & 0 & 0 & 0 \\
1 & 2 & 0 & 0 & 0 \\
0 & 2 & 1 & 0 & 0 \\
-1 & 1 & 0 & 0 & 0 \\
0 & 0 & 0 & 1 & 0 \\
0 & 0 & 0 & 0 & 1 \\
&  & \vdots  &  &
\end{array}
\right\|
\end{array}.
\]
The critical temperature without the lepton component
depends on the $n_{f}$ chemical potentials of the quarks, that are independent parameters.

If the relaxation time is longer than the typical time scale of the weak processes,
the weak interactions come into play. This is the case of phase transitions in neutron
stars whose lifetimes are comparable with the age of our Galaxy. Even in supernovae the
free fall time is much longer than time needed to keep substance under the chemical
equilibrium with respect to the weak interactions.

The weak interactions modify the matching conditions (\ref{MATCSTRO}). The lepton component
appears in the substance and the electric neutrality is to be imposed. The modified
conditions look like (see e.g. \cite{GIBB})
\begin{equation}
\mu_A = C_{A}^{a} \mu_{a} + e_{A}V,
\label{MATCWEAK}
\end{equation}
where $e_{A}$ is electric charge of particle $A$, the indices $A$
and $a$ run over leptons also. The chemical equilibrium generates
at the boundary of two phases a jump of the electrostatic
potential $V$ formed by a double layer of leptons to match chemical
potentials under the condition of the bulk electroneutrality. The
similar jump exists e.g. at boundaries of two different metals in
contact. The chemical equilibrium with respect to the weak
interactions and the electroneutrality condition leave each phase
with one free parameter only that can be chosen to be, e.g., the
neutron chemical potential in the hadron phase and the $d$-quark
chemical potential in the quark phase. The matching conditions
(\ref{MATCWEAK}) are reduced to one equation for $\mu_{n}$ and
$\mu_{d}$ at the phase transition. The critical temperature is
thereby a function of one parameter only.

Below we start with the system of quarks and gluons without  weak interactions
and where interaction in the hadron phase is also switched off, as it happens
in the large $N_c$ limit. One can as a first step neglect also interaction
between quarks and gluons as compared to their interaction with vacuum fields.
This approximation can be called the Vacuum Dominance Model (VDM).

\subsection{QCD phase diagram in FCM}

The basic notion for the phase transition in QCD is the nonzero vacuum energy
density $\varepsilon_{vac}$ introduced in QCD sum rules \cite{5*}.
The conformal anomaly  gives \be \varepsilon_{vac} = 1/4 \theta_{\mu\mu} =
\frac{\beta(\alpha_s)}{16\alpha_s} \lan (F^a_{\mu\nu})^2\ran\cong
-\frac{(11-\frac23 n_f)}{32} G_2^{(n_f)} \label{1s} \ee where \cite{5*,8*} \be
G_2=(0.01 \pm 0.002)~ {\rm GeV}^4. \label{3s} \ee

The quark-hadron phase transition occurs due to reconstruction of the vacuum in
the course of which the colorelectric part of $\varepsilon_{vac}$, ensuring
confinement, is thermodynamically not advantageous and therefore vanishes in
the temperature-affected QCD vacuum \cite{6*}.

In the confining phase the pressure looks like
\begin{eqnarray}
P_1(T, \mu_1) &=& \varepsilon_{vac} + T^4\chi_1(T, \mu_1 ),
\label{4s}
\end{eqnarray}
where $\mu_1$ denotes the set of the chemical potentials, the second term
stands for pressure of the hadron gas. In the deconfined phase, the
colorelectric confining correlator $D^E(x) =0$ \cite{6*}, as confirmed by
lattice data \cite{7*}, so that the vacuum energy density
$\varepsilon^{dec}_{vac}$ is decreased by about a factor of two, $\Delta G_2
\approx \frac12 G_2$ within 10\%.

The pressure of the deconfined phase equals
\be
P_2(T, \mu_2 ) = \varepsilon^{dec}_{vac} + T^4 \chi_2(T, \mu_2 )
\label{6s}
\ee
where $\mu_2$ is the set of the quark chemical potentials,
$\chi_2(T, \mu_2 ) = p_{gl} + p_q$ is a function weakly depending on $T$
for low chemical potentials, $p_i={P_i}/{T^4}$ are reduced gluon and quark pressures.

The Gibbs' criterion,
\be
P_1 (T_c, \mu_1 ) =P_{2}(T_c, \mu_2 ),
\label{7s}
\ee
together with the matching conditions (\ref{MATCSTRO}) allow to find $T_c$ in terms
of $\Delta \varepsilon_{vac} \sim \Delta G_2$.
By neglecting the hadron pressure $\chi_1(T, \mu_2 )$, which is $O(10\%)$ of $\Delta \varepsilon_{vac}$
for low chemical potentials, one gets
\be
T_c =\left( \frac{\Delta \varepsilon_{vac}}{\chi_2(T_c, \mu_2 )}\right)^{1/4}.
\label{8s}
\ee
$\chi_1(T, \mu_2 )$ is, however, important for the order of quark-hadron phase transition.

Next step is the evaluation of $p_q$ and $p_{gl}$. It is shown in \cite{9*},
that  in VDM the main dynamical contribution  to quark and gluon pressure can
be written in the form  of the Polyakov line (\ref{9s}), creating effective
selfenergies  $V_1(\infty,T) $ for quarks and $\frac94 V_1(\infty, T)$ for
gluons, and otherwise quarks and gluons are free. The resulting expressions for
$p_{gl}, p_q$ are in (\ref{10s}), (\ref{11s}). Neglecting quark mass  one
obtains the form (\ref{12s}), where $\Phi_{+,-}^{(k)}$ are given in
(\ref{13s}).

Equations (\ref{10s}), (\ref{12s}) contain all necessary dynamical
information for quark-hadron phase transition  and dynamics of
quark-gluon plasma (QGP). Basic quantity is the potential
$V_1(\infty, T)$, entering Polyakov line (\ref{9s}), which is
calculated on lattice and analytically see \cite{10*}. \be
|L_{fund}|=\exp\left(-\frac{V_1}{2T}\right).\label{9s}\ee \be
p_{gl}=\frac{16}{\pi^2} L_{adj} (T)=\frac{16}{\pi^2 }\exp
\left(-\frac{9V_1(\infty, T)}{8T}\right)\label{10s}\ee

\be p_q=\frac{12 n_f}{\pi^2} \sum^\infty_{n=1}
\frac{(-)^{n+1}}{n^4} (L_f)^n \varphi^{(n)}_q (T) \cosh \frac{\mu
n}{T}\label{11s}\ee

$$p_q(m_q=0) = \frac{12 n_f}{\pi^2} \sum^\infty_{n=1}\frac{(-)^{n+1}}{n^4} L_f^n  \cosh \frac{\mu
n}{T}=$$ \be
 = \frac{n_f}{\pi^2} \left[ \Phi^{(3)}_- \left(
 \frac{\mu-\frac{V_1}{2}}{T}\right) + \Phi^{(3)}_+ \left(
 \frac{\mu+\frac{V_1}{2}}{T}\right)\right]\label{12s}\ee

 \be
 \Phi^{(k)}_-(z) =\int^\infty_0 \frac{x^kdx}{e^{x-z}+1};~~\Phi^{(k)}_+(z) =\int^\infty_0
 \frac{x^kdx}{e^{x+z}+1}.\label{13s} \ee

Now one can compute $T_c(\mu)$ for different $n_f$ from Eq. (\ref{8s}), where
$\frac{11}{32}\Delta G_2\to \frac{(11-\frac23 n_f)}{32} \frac12 G_2$ and
$p_q\sim n_f$ in (\ref{11s}).


In this way one obtains a simple formula for the transition temperature \be
T_c\simeq\left(\frac{(11-\frac23 n_f) G_2}{64(p_{gl}+ p_q)}
\right)^{1/4}\label{1} \ee where $p_{gl} (p_q)$ are gluon (quark)  pressure
divided by $T^4$ in  the  deconfined phase, calculated in the  same method (see
\cite{5} for details). This simple formula yields for different number of
flavors $n_f$ correct values $T_c(n_f)$, in good agrement with lattice data
\cite{6}.

In doing so one  assumes that temperature and moderate density do not affect
strongly the vacuum fields in $G_2$  and for the latter one can use the  same
phenomenological values as for  zero temperature, i.e. $G_2 \cong 0.01$
GeV$^4$.

\begin{figure}[htb]
\includegraphics[height=5cm]{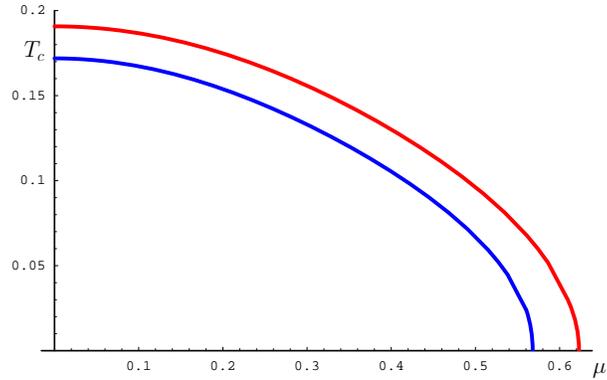}
\caption{Phase transition curves in $(T,\mu)$ plane. $T$ and $\mu$ are in GeV. Upper curve: $n_f=2$; lower curve: $n_f=3$.}
\label{fig:2}
\end{figure}

Proceeding  in this  way one obtains the phase diagram in  the $\mu_q-T$  plane
\cite{6},  see  Fig. \ref{fig:2}, where the critical value $\mu_{cr} = \mu_q (T=0) \cong
0.6$ GeV. This value is large enough
 and it will be shown below that vacuum structure is affected by the chemical potential $\mu_q$ and
 confinement is modified. As it is, one has for
small $T$ the first order density phase transition to the deconfined
quark-gluon matter with strong np attraction in the white systems \cite{7}
preferring $q\bar q$ system over diquarks and thus making the proposal of quark
superconductivity not very realistic, in line with earlier criticism of
\cite{8}.


Then one obtains the numbers in Table \ref{table1}, where the
asterisk corresponds roughly to 1/2 of standard condensate. The
values $T_c(n_f=0) =0.27$ GeV and $T_c(n_f=2)=0.19$ GeV are in
good agreement with numerous lattice data, while $\mu_c\approx
0.6$ GeV cannot be obtained on the lattice. The resulting curves
$T_c(\mu)$ are shown in Fig. \ref{fig:2}.

The critical density of the phase transition
into the strange quark matter ($n_f = 3$) at zero temperature gets positive. This means
that the strange quark matter is not absolutely stable. The condition of absolute stability
of the strange quark matter restricts the critical temperature for vanishing chemical potentials by $T_c(n_f = 3) < 122 \pm 7$ MeV \cite{KOMK}.
According to the lattice data, this restriction is not satisfied with a large certainty.

\begin{table}
\caption{Critical values of $T$ and $\mu$ for various $n_f$ and $\Delta G_2$ } \label{table1}
 \begin{center}
\begin{tabular}{|l|l|l|l|l|}
 \multicolumn{5}{c}{~~~~~~~~~~~~~~~~~~~$\bigstar$}\\\hline \hline
   &&&&\\
$\frac{\Delta G_2}{0.01~{\rm  GeV}^4}$& 0.191&0.341&0.57&~~~1\\
&&&&\\
\hline \hline
&&&&\\
 $T_c({\rm ~ GeV})$~~ $n_f=0$ &0.246&0.273&0.298&0.328\\ &&&&\\
\hline
&&&&\\
 $T_c({\rm ~ GeV})$~~ $n_f=2$ &0.168&0.19&0.21&0.236\\ &&&&\\
\hline
&&&&\\
$T_c({\rm~ GeV})$~~  $n_f=3$ &0.154&0.172&0.191&0.214\\&&&&\\
\hline
&&&&\\
$\mu_c( {\rm ~ GeV})$ ~~ $n_f=2$ & 0.576& 0.626&0.68&0.742\\&&&&\\
\hline
&&&&\\
$\mu_c( {\rm ~ GeV})$ ~~ $n_f=3$ & 0.539& 0.581&0.629&0.686\\&&&&\\
\hline \hline

 \end{tabular}

\end{center}
\end{table}

In the derivation of the phase curve in Fig.~\ref{fig:2} it was assumed, that the
pressure of the hadronic phase can be given with good accuracy by the vacuum
fields only (the Vacuum Dominance Model (VDM)), while the contribution of the
hadronic gas is neglected. This approximation is reasonable for $\mu=0 $, where
the effect of pionic gas is around 10\%, but maybe crude for large $\mu$, where
hadronic matter is dense nuclear matter. Therefore the region of large $\mu$ and
small $T$ can be modified, when interaction in nuclear matter is properly
taken into account. Since the main emphasis of the present paper is the
treatment of the possible new phase of MQS at smaller $\mu$, it will be
reasonable to reconsider the hadron-quark phase transition with account of
possible MQS contribution.

\subsection{Suppression of color superconductivity}

As was shown in \cite{10*,11*} the diquark nonperturbative interaction is
expressed via $V_1(r,T)$ as follows (for  a large size superconducting diquark)
\be V_{QQ} (r) =\frac12 V_1 (\infty, T) + \frac12 V_1 (r, T)\to  V_1 (\infty,
T_c) \approx 0.5 {\rm GeV} \label{14s}\ee For each quark it gives the factor

\be L_q(T) =\exp \left( -\frac{V_1(\infty, T)}{2T}\right).\label{15s}\ee Hence
each quark carries Boltzmann factor \be L_q \approx \exp \left( -\frac{0.25
GeV}{T}\right).\label{16s}\ee

Assuming that the dependence on $\mu$ is weak one expects that   approximately
the same factor holds for $\mu\to \mu_{crit}$.

For the white $Q\bar Q$ system this factor is missing \cite{11*}, hence one
expects \be \frac{\lan V_{Q\bar Q}\ran}{\lan V_{QQ}\ran} \approx L^2_q\approx
\exp \left( - \frac{0.5 GeV}{T} \right) \ll  1\label{17s}\ee

Therefore one expects $Q\bar Q$ pairing,  and not   $QQ$ pairing, in $qgp$.

\subsection{Critical baryon chemical potential in cold nuclear matter}

One can expand (\ref{13s}) for $T\to 0, ~a\to \infty$

\be \Phi_-^{(3)} (a\to \infty) =\frac{a^4}{4} +\frac{\pi^2}{2} a^2
+\frac{7\pi^4}{60}+...,~~
a=\frac{\mu\mp\frac{V_1}{2}}{T};\label{18s}\ee For the pressure
one obtains \be P= \frac{n_f}{4\pi^2} \left(\left(
\mu-\frac{V_1}{2}\right)^4 + \left(
\mu+\frac{V_1}{2}\right)^4+...\right)\label{19s}\ee and for the
density \be \rho_q \equiv\frac{N_q}{V} = \frac{n_f}{\pi^2} \left\{
\left( \mu-\frac{V_1}{2}\right)^3 + \left(
\mu+\frac{V_1}{2}\right)^3\right\}.\label{20s}\ee The critical
$\mu$ at small $T$ has an expansion \be \mu_c (T\to 0)
=\frac{V_1(T_c)}{2} + (48)^{1/4} T^{(0)} \left(1-\frac{\pi^2}{2}
\frac{T^2}{(\mu_c-\frac{V_1(T_c)}{2})}+\right). \label{21s}\ee

For $V_1(T_c) =0.5$  GeV and $n_f=2$ one has:{  $\mu_c(T=0) \cong 0.6$ GeV}.

The critical density from (\ref{20s}), (\ref{21s}) is

\be \rho_q(crit) \cong \frac{2n_f m_c}{\pi^2} \left(
\mu-\frac{V_1^2}{2}\right)\approx 0.12~ {\rm GeV}^3=15~{\rm
fm}^{-3}\label{22s}\ee so that baryon number density is \be \rho^{crit}_{Bar} =
\frac{\rho_q}{3} \cong 5
 ~fm^{-3}~~\label{23s}\ee

 This is 30 times normal nuclear density, $ n_0=\frac16 ~fm^{-3}$ and the
 pressure is very high,
$ P^q_{crit}\cong 125 P_{nucl} (2n_0)$.

 It is clear that nuclear matter cannot exist as usual
 baryon matter for density of 5 baryons in  $fm^3$.



\section{Density effect on confinement potential below phase transition}

Consider a white  system of a quark and  a heavy antiquark $\bar
Q$, which in baryon can be replaced by the string junction. Then
in the partition function $Z= \int  DA D\psi \exp (\bar \psi (\hat
D+ m_q) \psi)$ one can average ever $DA$ and keep the quadratic in
$A_\mu$ term in the exponent, which yields $Z=\int D\psi \exp \int
\bar \psi (x) \psi (x) \bar \psi(y) \psi(y)J(x,y) dxdy.$. Here
$J(x,y)$  is proportional to the quadratic correlator $D^E(x) $, $
J(x,y) = \int^x_0 du\int^y_0 dv D^E(u-v).$

In case of confinement $(D^E\neq 0), J(x,x)\sim x$ at large $x$ and as shown in
\cite{12*} this property yields simultaneously confinement and chiral symmetry
breaking (CSB). As was shown in \cite{12*}, the  effective quark mass
$M^\circ_s$ (growing at large $x$ due to confinement) and quark propagator $S$
are found   from the system of equations (we neglect emission of pions) \be
M_s^{(0)} (x,y) = \bar J_{\mu\mu} (x,y) Tr S_q (x,y)\label{3.1}\ee

\be  i S_q (x,y) = \lan x | (\hat \partial + m_q +
M_s^{(0)})^{-1}|y\ran.\label{3.2}\ee

 Solutions  of (\ref{3.1}), (\ref{3.2}) are different for light or heavy quark $q$. For heavy
 quark, $m_q\to \infty$, one neglects $M_s^{(0)}$ in $S_q$ and has

 \be  S_q(x,y) (m_q\to \infty) \sim \delta^{(3)} (\vex -\vey)\label{3.3}\ee
 \be M_s^{(0)} (x,y) \to \bar J (\vex, \vex) \sim \sigma |\vex-\vex
 (\bar Q)|\label{3.4}\ee

 For light quark $q$ one solves the system using relativistic WKB
 for Dirac particles and again obtains CSB and   confinement for
 large $x$. That was done in \cite{12*} for the case of zero density
 (vanishing $\mu_q$).

Now we turn to the case of nonzero $\mu_q\equiv \mu$. As will be
seen, for heavy quarks nothing happens -linear confinement
at all $r$, unless $\mu \sim m_q$ (for standard nuclear density
$\mu \cong 0.3 $ GeV.).

Let us study now the case of light quarks, following the paper
\cite{13*}. The effective light quark mass can be written as \be \bar M (\vex,
\vey; \mu) =  J (\vex, \vey) \gamma_4
 \Lambda (\vex, \vey; \mu)\label{3.5}\ee
 where $\gamma_4\Lambda(\vex, \vey, \mu)$ is the time-independent quark Green's
 function, cf. (\ref{3.1}). It can be written in terms of one-particle
 eigenfunctions $\psi_n (\bar x)$ as

 $$\Lambda (\vex,\vey;\mu) =\sum_n \psi_n (\vex) \mathrm{sign}
 (\varepsilon_n -\mu) \psi^+_n (\vey)=$$
 \be =\Lambda_0 (\vex, \vey) -\Delta \Lambda (\vex, \vey)\label{3.6}\ee

\begin{figure}
\caption{Interactions $\bar M_{scal} (r)$ (solid line) and $\bar
M_{vect}(r)$ (dashed line) as functions  of distance $r$, with $b=
\mu/\sigma$. Dotted line and dashed line show the qualitative
smoothed form of both terms respectively.}
\label{fig:interactions}
\begin{center}
\includegraphics[height=5cm]{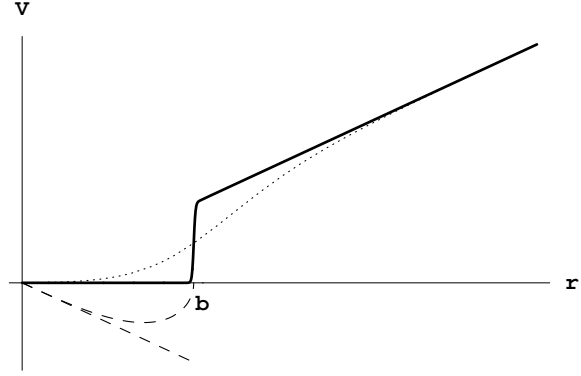}
\end{center}
\end{figure}

 In \cite{13*} $\psi_n(\vex) $ are found in the relativistic WKB method and
 the approximate summation over $n$ can be performed, yielding for $\Lambda_0
 (\vex,\vey) $ the smeared $\delta$-function form for large $|\vex|, |\vey|$
 and $|\vex-\vey|\ll |\vex|, \vey|$. In \cite{13*} also the term $\Delta
 \Lambda (\vex,\vey)$ in (\ref{3.6}) was calculated in the same way, which is
 nonzero for $\mu\neq 0$, and distorts the linear confinement at distances
 $r\leq\mu/\sigma$, where $\sigma =0.18$ GeV$^2$ is the standard string
 tension. The resulting picture for the averaged $\bar M(r)=\int d^3 \ver' \bar
 M (\ver,\ver',\mu)$ is given in Fig \ref{fig:interactions}, where $\bar M(r)$ is split into a
 Lorentz scalar and vector parts, \be \bar M_{scal} =\sigma r \theta (\sigma r
 -\mu),~~ \bar M_{vect} (r) =- 2\sigma r \theta (\mu-\sigma r). \label{3.7})\ee

\vspace*{0.5cm}


In MQS we shall for simplicity consider all quarks attached to the
same string junction (this is not crucial for quarks in the
modified confining potential, since near the string junction the
scalar part vanishes, but the vector part can be rewritten in the
form of pairwise potentials \cite{14*}). Consider now a system of
$N$ quarks around one string junction in the nuclear medium of
ordinary nucleons. Using the local limit for interactions of a
light quark with string junction, Eq. (\ref{3.7}), one can
calculate the change of masses of white states due to density.

Assuming that multiquark states located at different string
junction points do not overlap (see below the note on states
radii), one can associate the parameter $\mu$ for a given quark
with the largest value of $\varepsilon_n$, occupied by other
quarks, belonging to the same string junction point. In what
follows we shall show, that in some situations it will be
advantageous for two or more nucleons (with different string
junctions) to coalesce into a common  state with one string
junction, building in this way the contracted potential
(\ref{3.7}).

Note, that for one nucleon (3 quarks) the contraction mechanism
does not work and $\mu$ can be  taken at zero value, since no
forbidden states exist for each quark of a given color. This,
again, is true, provided one-quark states do not overlap at the
given density. The situation changes, however, for two nucleons,
since e.g. one can add to the first nucleon, i.e. 3 quarks in
S-states, 3 quarks in  P-states relative to the same string
junction point. The latter will move in the scalar part of the
potential $\Lambda(\vex, \vey)$, with $\mu$ being equal to the
S-state Dirac eigenvalue $\varepsilon_S$, thus the energy
$\varepsilon_P (\mu=\varepsilon_S)$ may become  less than
$\varepsilon_S(\mu=0)$: \be \varepsilon_P(\mu=\varepsilon_S)
<\varepsilon_S(\mu=0)\label{mt48}\ee The same type of inequality
may occur  for higher $L$ states, when more nucleons coalesce  to
the same Multiquark State. This would imply instability of nuclei
with respect to transition to MQS matter.

Note, that $\mu_B\neq 3\mu$, since $\mu$ is attributed to MQS
string junction and does not grow with $\mu_B$ unless different
MQS start to overlap. Hence for $\mu_B>0$ (\ref{mt48}) should be
replaced by
\begin{equation}
\varepsilon_P (\mu =\varepsilon_S)< \varepsilon_S(\mu=0) +
\varepsilon_B (\mathrm{Fermi}). \label{mt48a}
\end{equation}
To check (\ref{mt48a}) in \cite{13*} $\varepsilon_S$ and
$\varepsilon_P$ were computed numerically with the potential
(\ref{3.7}), and in perturbation theory, considering the change of
potential in (\ref{3.7}) as a small perturbation. The results are
given in Tables \ref{table2s},\ref{table2p}. The calculated
quantities are $\Delta M_S = 3 \Delta \varepsilon_n (S)$ and
$\Delta M_P (3q) =2\Delta \varepsilon_n (S) +
\Delta\varepsilon(P)$, where $\Delta\varepsilon_n = \varepsilon_n
(\mu) -\varepsilon_n(0)$.  One should note at this point, that the
vector part of interaction written in (\ref{3.7}) for the case of
light quark with heavy antiquark, and for the case of 3 quarks
interacting with string junction the vector part is transformed as
an addition to the color Coulomb potential with coefficient 1/2
(see \cite{14*} for explicit derivation of scalar and vector
interaction in the nucleon). This prescription was used for our
calculations.

\begin{table}
\caption{Mass  shifts of  $3q$  system with all quarks in the
$S$-state (purely numerically, and perturbatively, for comparison)
and the $S$-wave average radius, due to nonzero $\mu_q$}
\label{table2s}
\begin{center}
\begin{tabular}{|c|c|c|c|}
\hline \hline $\mu$, MeV & $\Delta M_{\text{exact}}$, MeV & $\Delta
M_{\text{pert.}}$, MeV & $\langle r\rangle_S$, fm \\ \hline \hline
100 & -5.8 & -5.7 & 0.505 \\ \hline 200 & -75 & -70 & 0.487 \\
\hline
300 & -290 & -265 & 0.460 \\ \hline 400 & -630 & -603 & 0.464 \\
\hline 500 & -1008 & -1020 & 0.501 \\ \hline \hline
\end{tabular}
\end{center}
\end{table}

\begin{table}
\caption{Mass  shifts of  $3q$  system with two quarks in the
$S$-state and one quark in the $P$-state (purely numerically, and
perturbatively, for comparison), and the $P$-wave average radius,
due to nonzero $\mu_q$} \label{table2p}
\begin{center}
\begin{tabular}{|c|c|c|c|}
\hline \hline $\mu$, MeV & $\Delta M_{\text{exact}}$, MeV & $\Delta
M_{\text{pert.}}$, MeV & $\langle r\rangle_P$, fm  \\
\hline \hline
100 & -4.0 & -3.9 & 0.613 \\ \hline 200 & -53 & -50 & 0.605 \\
\hline
300 & -221 & -199 & 0.558 \\ \hline 400 & -530 & -481 & 0.473 \\
\hline 500 & -917 & -873 & 0.448 \\ \hline \hline
\end{tabular}
\end{center}
\end{table}

From these numbers one can obtain $\varepsilon_S (\mu_q=0)\approx 0.52$ GeV,
$\varepsilon_P(\mu_q=0) =0.77$ GeV, $\varepsilon_P (\mu_q=0.52$ GeV) = 0.57
GeV. This is close to the $\varepsilon_S (\mu_q =0)$, i.e. one can expect the
formation of MQS for $\varepsilon _B$ (Fermi)=50 MeV, which corresponds to
critical density $\rho\cong 3 \rho_0$.

Also, the average radii of $S$- and $P$-wave quark states were
computed (see Tables \ref{table2s},\ref{table2p} ) and it is found
the radius is decreasing with a rise of $\mu$, at least for the
moderate $\mu$ values, so it is reasonable to neglect in the first
approximation the possible overlapping of multiquark state
wavefunctions attached to different string junction points.

As we shall discuss below, this might be important for new
heavy-ion colliders (FAIR, NICA,...), and crucial for neutron
stars and supernovae. However, we have still neglected interaction
between baryons in nuclear matter. In the next sections we show
that this interaction can be described in terms of MQS formation
in the intermediate states, and a new theory of nuclear matter
based on MQS can be formulated.

\section{Quark Compound Bag model}
\setcounter{equation}{0}

In this section we describe the general two-channel formalism of the
nucleon-nucleon interaction that provides dynamical framework for the
$P$-matrix formalism and is used in Sect. VI  to build the relativistic
formalism and to construct in Sect. X the Bethe-Goldstone $G$ operator for
nucleons in nuclear matter.

For simplicity we consider a hadronic channel consisting of two spinless nonrelativistic
hadrons, which is coupled to another channel, where any number of quarks and gluons are present.
We denote the hadronic channel with a subscript $h$ and  the quark channel
with a subscript $q$. The wave function is a column with two components $(\Psi_h, \Psi_q$),
while the interaction term in the Hamiltonian is a $2\times 2$ matrix.
\be
\hat V=\left(
\begin{array} {ll}
V_{hh} & V_{hq}\\
V_{gh} & V_{gq}
\end{array}
\right).
\label{1q}
\ee
We do not specify the quark dynamics and even the type of variables, on which $V_{qq}$ is
acting. At this point we can take the description of the quark-gluon system as
general as possible and we shall use only the property of confinement, i.e. that the
eigenstates of the quark Hamiltonian are discrete states confined to a finite region
in space whatever the binding energy is.

The  coupled dynamical equations for $\Psi_h, \Psi_q$ are
\begin{eqnarray}
(T_h+ V_{hh} -E) \Psi_h &=&- V_{hq} \Psi_q, \label{2q} \\
L_q \Psi_q + V_{qh} \Psi_h&=& E\Psi_q,      \label{3q}
\end{eqnarray}
where $L_q$ is some quark-gluon operator (its exact form is irrelevant for our
purposes) and $T_h$ is the kinetic operator for hadrons and $E$ is
the full energy, $E=\sqrt{s};~ V_{hh}$ is in general a nonlocal
energy-dependent operator. $T_h$ in general is a relativistic
operator.

One can solve formally equation (\ref{3q}) to express $\Psi_q$ in
terms of $\Psi_h$:
\be
\Psi_q = - G_q V_{qh} \Psi_h
\label{3}
\ee
where the quark Green function has a spectral representation
\be
G_q = \sum_\nu \frac{\Psi_q^\nu \bar \Psi^\nu_q}{E_\nu-E}.
\label{4}
\ee

The quark eigenstates $\Psi^\nu_q$ and eigenvalues $E_\nu$ belong
only to a discrete spectrum because of the confinement. Using
(\ref{3}) one can exclude the quark channel and rewrite the
hadronic equation  (\ref{2q}) as
\be
(T_h +V_{hh} + V_{hqh} )\Psi_h = E\Psi_h
\label{5q}
\ee
where
\be
V_{hqh} =- V_{hq} G_q V_{qh} =- \sum_\nu \frac{V_{hq} \Psi^\nu_q\bar \Psi^\nu_q
V_{qh}}{E_\nu-E}.
\label{6q}
\ee

Let us discuss the structure of the quark-induced hadronic
interaction  $V_{hqh}$.
\begin{enumerate}
    \item  It does not contain any quark or gluon degrees of
    freedom, which are present in $\Psi^\nu_q$ and $V_{hq}$.
    Indeed, in (\ref{6q}) the summation over all quark degrees of
    freedom is implied by the repeated subscript $q$, so that the
    only remaining coordinates  are hadronic relative distances,
    entering via  $V_{hq}$ and $V_{qh}$.
\item  $V_{hqh}$ is separable Hermitian interaction; if we
    denote
\be
f_\nu (r)\equiv V_{hq} \Psi^\nu_q,
\label{7q}
\ee
    $V_{hqh}$ has the form
\be
V_{hqh} =\sum_\nu\frac{f_\nu(r) f^*_\nu (r')}{E-E_\nu}.
\label{8q}
\ee
\item The numerator of (\ref{8q}) depends  in general case  on energy $E$
    through the quark-hadron interaction $V_{qh}$. The most
    characteristic energy  dependence enters through the
    denominator in (\ref{8q}); for $E$ near $E_\nu$ the
    quark-induced potential $V_{hqh}$ can be infinitely large.
\item  Since  quarks are not allowed outside the bag the
    function $f_\nu(r)$ should be nonzero only inside the bag.
    Hence it can be easily approximated there by any series on a
    finite interval of $r$.
\end{enumerate}

We assume now that the solution of the hadronic problem (\ref{5q})
without $V_{hqh} $ is known; the corresponding regular at origin
scattering wave functions and the Green function we denote by
$X(r)$ and $G_h$ respectively. In terms of these pure hadronic
quantities we can rewrite the full coupled quark-hadron equation
(\ref{5q}) as
\be
\Psi_h(r) = X(r) +\sum_\nu \frac{(G_h f_\nu)(f_\nu
\Psi_h)}{E_\nu-E}
\label{9q}
\ee
where we have used the notation
\begin{eqnarray}
(G_h f_\nu) &\equiv& \int G_h (r, r') f_\nu (r') dr'  \nonumber \\
(f_\nu \Psi_h)&\equiv& \int \Psi_h (r) f^*_\nu (r)dr. \label{10q}
\end{eqnarray}

The constants $(f_\nu \Psi_h)$ are to be found from  the system
of linear algebraic equations
\be
(f_\nu \Psi_h) = (f_\nu X) +\sum_\mu \frac{(f_\nu G_h f_\mu)
(f_\mu \Psi_h)}{E_\mu-E}
\label{11q}
\ee
where
\[
(f_\nu G_h f_\mu ) \equiv \int\int f^*_\nu (r) G_h (r,r') f_\mu (r') dr dr'.
\]

The resulting picture is particularly simple when $E$ is close to
some eigenvalue $E_\mu$ and all other terms in the sum in
(\ref{9q}) with $\nu\neq \mu$ can be neglected. In this case we
have
\be
\Psi_h (r) = X(r) +\frac{G_h f_\mu)(f_\mu X)}{E_\mu- E- (f_\mu
G_h f_\mu)}.
\label{12q}
\ee

At this point it is possible to discuss the general properties of
our solution $\Psi_h {(r)}$.  First of all,  multiplying  both
sides of (\ref{12q}) with $f^*_\mu(r)$ and integrating over $dr$,
we obtain
\be
(f_\mu \Psi_h) =0,~~ {\rm for}~~E=E_\mu.
\label{13q}
\ee

This property does not mean that $\Psi_h$ vanishes inside  the
bag, where $f_\mu(r)$ is nonzero. Indeed, $\Psi_h(r), $ being
complex, may have no zeros at all inside the bag, still the
condition (\ref{13q}) can be satisfied. One example of this kind
obtains if we take $f_\mu (r) = \theta(R-r)$ and $V_{hh} \equiv
0$. No definite consequences can be deduced for the observables in
that case. At this point we make an ansatz, that hadrons are
coupled to the QCB degrees of freedom only on the surface of the
compound bag. As we shall see below, this assumption immediately
leads to  explicit expressions for the wave function and the
observables. It is crucial for all results of the paper. Moreover
we show below that this assumption leads naturally to the $P$
matrix formalism of Jaffe and Low \cite{JALO} and therefore
experimental evidence in favor of that formalism presented in
\cite{JALO} justifies our  assumption. Moreover in Refs. \cite{SIMO81,NARO94}
additional experimental evidence is presented  in favor of the
stated ansatz.

 From theoretical point of view the assumption, that the hadrons
 effectively are not coupled  with the bag constituents inside the
 bag,
 but only at the surface, is not new. In the framework of the
 nuclear cluster models, (see the  book \cite{19}) it is well known
 that the region near the surface of the compound nucleus plays
 the crucial role.

In \cite{20} an additional  support  is given to the assumption
in terms of  quark  cluster models. The general character of these
considerations allows to apply it also to the quark systems.
Accordingly we assume the following simplified form for
$f_\mu(r)$:
\be
f_\mu (r) = c_\mu \delta (r-b)
\label{14q}
\ee
where
$b$ is connected to the QCB radius $R$ and $c_\mu$ is some
constant. With the choice (\ref{14q}) the solution $\Psi_h(r)$
simplifies:
\be
\Psi_h (r) =X(r) +\frac{G_h(r,b)
X(b)}{\frac{E_\mu-E}{\gamma_\mu} - G_h (b,b)},
\label{15q}
\ee
where $\gamma_\mu\equiv |c_\mu|^2.$

From (\ref{15q}) follows an important property of $\Psi_h(r)$,
namely
\be
\Psi_h (r) \equiv 0,~~ {\rm for} ~~ r\leq b ~~{\rm
and}~~ E =E_\mu,
\label{16q}
\ee
since
\be
G_h(r,r') =X(r_<) Y(r_>)
\label{17q}
\ee
where $Y(r)$ is a nonvanishing at $r=0$
solution of the purely hadronic  equation (\ref{2q})  with $V_{hq}=0$.
It is convenient to choose $X(r), Y(r)$ with the asymptotics
$X(r)_{r\to \infty} \sim \frac{\sin (kr +\delta_0(k))}{k}$
\be
Y(r)_{r\to\infty} \sim 2\mu \exp [i (kr +\delta_0
(k))],
\label{18q}
\ee
where $\mu$ is the reduced mass (reduced energy in relativistic
case).

Let us define the factor $F(E)$ as follows
\be
|\Psi_h(r)|^2 = F(E) |X(r)|^2, ~~ r\leq b.
\label{19q}
\ee
$F(E)$ shows how much $\Psi_h(r)$ decreases inside the bag because
of the presence of
the quark states; exact form of $F(E)$ is:
\be
F(E) =\left| \frac{E_\mu-E}{E_\mu-E-\gamma_\mu G_h (b,b)}\right|^2.
\label{20q}
\ee

 From
(\ref{15q}) we immediately conclude that the logarithmic
derivative of $\Psi_h(r)$ at $r=b$ should be meromorphic in $E$,
namely
\be
\frac{\Psi'_h(b)}{\Psi_h(b)}= \frac{X'(b)}{X(b)}+
\frac{2\mu\gamma_\mu}{E-E_\mu}.
\label{21q}
\ee

To obtain (\ref{21q}) we use the property of the solution $X,Y$:
$ X(r) Y'(r) - X'(r) Y(r)=-2\mu.$

An important quantity  entering (\ref{20q}) and (\ref{15q}),
$G_h(b,b)$, can be expressed through $P_0 \equiv
\frac{X'(b)}{X(b)}$ (the latter can  be  extracted from the
experimental data, see \cite{SIMO81}). Assuming that there is no
interaction beyond $r=b$, we can use (\ref{18q}) as the exact
values of $X(r)$ and $Y(r)$ for $r\geq b$, and obtain  from
(\ref{17q}):
\be
G_h(b,b) =\frac{2\mu}{k} \sin^2 (kb+\delta_0) (\cot (kb
 +\delta_0) +i).
\label{22q}
\ee

On the other hand we have $ P_0=k\cot(kb+\delta_0)$ so that \be G_h (b,b) =
\frac{2\mu}{P_0-ik}. \label{23q} \ee

It is instructive to evaluate $\gamma_\nu G_h(b,b)$ which enters (\ref{15q}).
For $NN$ scattering from the analysis of \cite{SIMO81} we have $P_0 \approx 0.4
$ GeV, $\gamma_\nu \approx  0.1$ GeV. Therefore  e.g. at  $E=0.1$ GeV,
$\gamma_\nu  G_h (b,b) = (0.15+i0.12)$ GeV.

The ratio $F(E)/F(0)$ is effectively zero for positive energy, e.g. for
$E=0.05$ GeV this ratio is $\sim 0.01$. Let us now consider many quark
eigenstates instead of one.

In this case the general result (\ref{9q}) for $\Psi_h$ is valid
where $(f_\nu \Psi_h)$ are to be  found from the solution of the
system of equations (\ref{11q}). The results simplify very  much
if we use the ansatz analogous to (\ref{14q}):
\be
f_\nu (r) = c_\nu \delta (r-b),~~ \nu=1,2,...N
\label{24q}
\ee
with the same range $b$ for all quark eigenstates.

The ansatz (\ref{24q}) can be  approximately true for those bag
eigenstates where the number of quarks and energy eigenvalues
do not change much.

Taking into account (\ref{24q}) we obtain $(f_\nu g_hf_\mu) =
c^*_\nu G_h (b,b)$ and $(f_\nu\Psi_h) = c^*_\nu \Psi_h(b)$.
Inserting these values we rewrite Eq. (\ref{9q}) as follows:
\be
\Psi_h(r) = X(r) +\frac{G_h(r,b) X(b) \sum_\nu
\frac{\gamma_\nu}{e_\nu-E}}{1-G_h(b,b) \sum_\nu
\frac{\gamma_\nu}{E_\nu-E}}
\label{26q}
\ee
where  $\gamma_\nu= |c_\nu|^2$.

From (\ref{26q}) one can see that whenever $E$ approaches some
eigenvalue $E_\nu$, the hadronic wave function vanishes for $r\leq
b$. In this way the property (\ref{16q}) generalizes to the many
quark eigenstates case: the form of the wave function (\ref{26q})
is similar to the one-state case (\ref{15q}) if in the latter one
replaces  $\frac{E_\nu-E}{\gamma}$ in the denominator by the
quantity $\left(\sum_\nu\frac{\gamma_\nu}{E_\nu-E}\right)^{-1}$.
Under the same condition as in (\ref{24q}) we obtain the following
expression for $P$ matrix \be P=k\cot (kb + \delta_0 (k)) +
\sum_\nu \frac{2\mu \gamma_\nu}{E-E_\nu} \label{27q} \ee and all
arguments about the correspondence of the quark eigenstates (``the
primitives'' the terminology of Jaffe and Low \cite{JALO}) and
poles of the $P$ matrix  remain true. Here $\delta_0(k)$ is the
phase due to interaction $V_{hh}$ which is assumed to vanish for
$r>b$.

In practical calculations one is using the potential of the form \cite{SIMO81,20}
\be V= V_{hqh} + V_{hh} \theta (r-b)\label{29q}\ee and $V_{hqh}  $ for
arbitrary $NN$ angular momentum $l$ is \be V_{hqh} (\ver, \ver') =\frac{1}{rr'}
\sum_m Y^*_{jmls} \frac{f^*_\nu(r,E) f_\nu(r', E)}{E-E_\nu}
Y_{jmlS},\label{30q}\ee while $f_\nu(r,E)$ can be represented as \be f_\nu(r,E)
=- c_\nu \delta (r-b) + x_\nu (E_\nu-E) \eta_\nu (r), \label{31q} \ee where
$\eta_\nu(r)$ describe the relative $NN$ motion inside the $MQS'$ and are
chosen in the form \be \eta_\nu (r) = N_\nu j_l (\beta_\nu
 r) \beta_\nu r,~~ \int \eta^2_\nu (r) dr =1.
\label{32q} \ee Finally, all coefficients are to be fitted to the experimental
phases, usually in the interval [ $0;~500\div 800$ MeV ]. As a first
approximation we shall assume, that $V_{hh} \equiv 0$ and only   $V_{hqh}$ is
retained; as is shown in \cite{SIMO81,NARO94} in this way one obtains a good
fit to experimental $NN$ $S$ phases for energies up to $\sim 500$ MeV. In Fig.
\ref{fig:S} comparison to the experimental phases is given for the $^1S_0$ and
$^3S_1$ states obtained from relativistic-expressions of Sect. VII.

We thus introduced in the $NN$ system the interaction due to MQS
(\ref{30q}), which can alone (or in combination with usual OBEP)
describe $NN$ force, and the same in $3N$, $4N$... systems. At
this stage $f_n, E_n$ are found phenomenologically from $NN$
phases, and due to \cite{14} are changing with matter density. One
can use $V_{hqs}$ changing with  matter density. One can use
$V_{hqs}$ both to reactions. At the next stage one should
calculate $f_n, E_n$ microscopically from QCD Lagrangian (e.g. in
the framework of the FCM).

\section{Effective Lagrangian of Quark Compound Bag model}
\setcounter{equation}{0}

In this section, we construct effective relativistic QCB Lagrangian for the description
of nucleon-nucleon interaction.

Two-nucleon currents entering the effective Lagrangian
are constructed as bilinear combinations of the nucleon wave functions:
\begin{equation}
J(x)=\bar{\Psi}_{c}(x)O\Psi (x).  \label{CURR}
\end{equation}
The nucleon wave functions, $\Psi (x)$, carry bispinor and isospin indices,
$\Psi_{c}(x)=C\bar{\Psi}^{T}(x)$ is $C$-conjugated wave function,
$\bar{\Psi}_{c} \equiv \overline{\left( \Psi _{c}\right) }=-\Psi ^{T}C$,
and $C=i\tau ^{2}i\gamma ^{2}\gamma ^{0}$ in the standard representation \cite{BJDR}.
The currents (\ref{CURR}) annihilate dibaryons. The creation currents
have the form
\begin{equation}
J^{+}(x)=\bar{\Psi}(x)\bar{O}\Psi _{c}(x).
\end{equation}
The matrix $O$ admits expansion over the Dirac $\gamma $-matrices and
the Pauli $\tau ^{\alpha }$-matrices.
Many terms of the formal expansion vanish, since $\Psi (x)$ anticommute.
Under the permutation of two nucleon fields $J(x)$
transforms identically provided
\begin{equation}
C^{T}O^{T}C=-O.  \label{OMAT}
\end{equation}
Transformation properties of the operators $O$ and the nucleon fields
under the $C$-conjugation are given in Appendix B.

Matrices $O$ entering the effective Lagrangian are therefore
antisymmetric (odd) under the $C$-conjugation. In Table \ref{tab:1} we show
elementary even and odd matrices and the composite odd ones which obey (\ref{OMAT}).
They are combined further with the first-order differential operators
$i\overleftrightarrow{\partial }_{\mu } = i(-\overleftarrow{\partial} + \overrightarrow{\partial }%
)_{\mu }$ (odd one) and $(\overleftarrow{\partial }+\overrightarrow{\partial
})_{\mu }$ (even one).


\begin{table}[t]
\centering
\caption
{Even and odd operators $O$.
}
\label{tab:1}
\begin{tabular}{|c|c|c|c|}
\hline \hline
Even                      & \multicolumn{3}{|c|}{Odd}                       \\ \hline
 even                     & odd               & even-odd       & odd-odd-odd  \\ \hline \hline
1 & $\tau ^{\alpha}$ & $\tau ^{\alpha}i\gamma _{5}$ & $\tau ^{\alpha}i\gamma _{5} \sigma _{\mu \nu }i\overleftrightarrow{\partial }_{\nu }$ \\
$i\gamma _{5}$ & $\gamma _{\mu }$ & $i\gamma _{5}i\overleftrightarrow{\partial }_{\mu }$ &  \\
$\gamma _{5}\gamma _{\mu }$ & $\sigma _{\mu \nu }$ & $(\overleftarrow{\partial }+\overrightarrow{\partial })_{\nu }\sigma _{\mu \nu }$ &  \\
$(\overleftarrow{\partial }+\overrightarrow{\partial })_{\mu }$ & $i\gamma_{5}\sigma _{\mu \nu }$ & $(\overleftarrow{\partial }+\overrightarrow{\partial
})_{\nu }i\gamma _{5}\sigma _{\mu \nu }$ &  \\
& $i\overleftrightarrow{\partial }_{\mu }$ &  & \\
\hline \hline
\end{tabular}
\end{table}


The matrices $\tau ^{\alpha}\gamma_{5}\gamma _{\mu }$ and $\tau ^{\alpha}\sigma _{\mu \nu }i%
\overleftrightarrow{\partial }_{\nu }$ are not shown, since currents they produce
with free nucleons represent full derivatives:
\begin{eqnarray}
\bar{\Psi}_{c}\tau ^{\alpha}\gamma _{5}\gamma _{\mu }\Psi &=&\frac{%
1}{m}\partial _{\mu }(\bar{\Psi}_{c}\tau ^{\alpha}i\gamma _{5}\Psi ), \label{iden1} \\
\bar{\Psi}_{c}\tau ^{\alpha}\sigma _{\mu \nu }i\overleftrightarrow{%
\partial }_{\nu }\Psi &=&2\partial _{\mu }(\bar{\Psi}_{c}\tau ^{\mathrm{%
\alpha }}\Psi ). \label{iden2}
\end{eqnarray}
Full derivatives shifted to the dibaryon vector fields give vanishing contributions
to the Lagrangian.

We restrict ourselves with scalar and vector dibaryon currents of positive
and negative parity. This is sufficient to provide the phenomenological
description of the $NN$ scattering in $J=0^{\pm },1^{\pm }$ channels
($S$- and $P$-wave scattering).


\begin{table}[h]
\centering
\caption
{
Two-nucleon currents, associated dibaryon fields, and their quantum numbers.
}
\label{tab:2}
\begin{tabular}{|c|c|c|}
\hline \hline
Current & Dibaryon & $(I,J^{P})$ \\ \hline \hline
$\bar{\Psi}_{c} \tau ^{\alpha} \Psi$            & $\varphi _{-}^{\alpha}$    & $(1,0^{-})$ \\
$\bar{\Psi}_{c} \tau ^{\alpha}i\gamma _{5}\Psi$ & $\varphi _{+}^{\alpha}$    & $(1,0^{+})$ \\
$\bar{\Psi}_{c}i\overleftrightarrow{\partial }_{\mu }\Psi$ & $\chi _{+\mu }$ & $(0,1^{+})$ \\
$\partial _{\nu }(\bar{\Psi}_{c}\sigma ^{\mu \nu }\Psi )$  & $\chi _{+\mu }$ & $(0,1^{+})$ \\
$\bar{\Psi}_{c}i\gamma _{5}i\overleftrightarrow{\partial }_{\mu }\Psi$       & $\chi_{-\mu }$ & $(0,1^{-})$ \\
$\partial _{\nu }(\bar{\Psi}_{c}i\gamma _{5}\sigma ^{\mu \nu }\Psi )$        & $\chi_{-\mu }$ & $(0,1^{-})$ \\
$\bar{\Psi}_{c}\tau ^{\alpha}i\gamma _{5}\sigma ^{\mu \nu }i\overleftrightarrow{\partial }_{\nu }\Psi$ & $\chi _{-\mu }^{\alpha}$ & $(1,1^{-})$  \\
\hline \hline
\end{tabular}
\end{table}


The list of possible two-nucleon currents is shown in Table \ref{tab:2}.

A third vector current $\bar{\Psi}_{c}\gamma _{\mu }\Psi$ with quantum numbers ($0,1^{+})$
can be expressed in terms of the two ones listed in Table \ref{tab:2} using
the Gordon's expansion
\[
\bar{\Psi}_{c}\gamma _{\mu }\Psi =\frac{1}{2m}\bar{\Psi}_{c}i%
\overleftrightarrow{\partial }_{\mu }\Psi +\frac{1}{2m}\partial _{\nu }
(\bar{\Psi}_{c}\sigma ^{\mu \nu }\Psi ).
\]

The vector dibaryons are characterized by two coupling constants.
Recall that photon couplings with the nucleons are characterized by
two coupling constants also.

In the channel $(0,1^{+})$, the pseudovector coupling with the current
$\bar{\Psi}_{c}i\overleftrightarrow{\partial }_{\mu }\Psi$
gives vanishing
contribution at two-nucleon threshold, so the $S$-wave $NN$ scattering length
is determined completely by the pseudotensor coupling $(0,1^{+})$.
In what follows, we neglect by the pseudovector coupling.

In the channel $(0,1^{-})$, a linear combination of two currents enters
the $P$-wave $NN$ scattering length, so we set the vector coupling equal
to zero and redefine the tensor coupling.

Finally, every dibaryon is coupled to one two-nucleon current.

The dibaryon fields $\varphi _{P}^{\alpha}$, $\chi _{P \mu }$ and
$\chi _{P\mu }^{\alpha}$ carry (suppressed) indices
of radial excitations,
index $P=\pm 1$ stands for the parity,
$\alpha$ is isospin index, and $\mu $ is the Lorentz index:
\begin{eqnarray*}
\varphi ^{\alpha}=\left(
\begin{array}{l}
\varphi _{1-}^{\alpha} \\
\varphi _{1+}^{\alpha} \\
\varphi _{2-}^{\alpha} \\
\varphi _{2+}^{\alpha} \\
\vdots
\end{array}
\right) ,\;\;\;\chi _{\mu }=\left(
\begin{array}{l}
\chi _{1+\mu } \\
\chi _{1-\mu } \\
\chi _{2+\mu } \\
\chi _{2-\mu } \\
\vdots
\end{array}
\right) ,\;\;\;\;\chi _{\mu }^{\alpha}=\left(
\begin{array}{l}
\chi _{1+\mu }^{\alpha} \\
\chi _{1-\mu }^{\alpha} \\
\chi _{2+\mu }^{\alpha} \\
\chi _{2-\mu }^{\alpha} \\
\vdots
\end{array}
\right) .
\end{eqnarray*}
The coupling constants are defined accordingly:
\begin{eqnarray}
g=\left(
\begin{array}{l}
g_{1-} \\
g_{1+} \\
g_{2-} \\
g_{2+} \\
\vdots
\end{array}
\right) ,\;\;h=\left(
\begin{array}{l}
h_{1+} \\
h_{1-} \\
h_{2+} \\
h_{2-} \\
\vdots
\end{array}
\right) ,\;\;\;\;h^{\prime }=\left(
\begin{array}{l}
h_{1+}^{\prime } \\
h_{^{\prime }1-} \\
h_{2+}^{\prime } \\
h_{2-}^{\prime } \\
\vdots
\end{array}
\right) .
\label{COUP}
\end{eqnarray}

The effective Lagrangian splits into free and interaction parts:
\begin{equation}
\mathcal{L} = \mathcal{L}^{[0]} + \mathcal{L}^{[1]}_{int} + \mathcal{L}^{[2]}_{int}.
\label{LAGRANGIAN}
\end{equation}
$\mathcal{L}^{[1]}_{int}$ describes interaction of dibaryons with
nucleons, while $\mathcal{L}^{[2]}_{int}$ describes a four-fermion contact interaction.

The free part has the form
\begin{eqnarray}
\mathcal{L}^{[0]} &=& \bar{\Psi}(i\hat{\nabla}-m)\Psi  \label{LAGR} \\
&+&\partial _{\mu }\varphi ^{\alpha +} \partial _{\mu }\varphi ^{\alpha } - \varphi ^{\alpha +}M_{10}^{2}\varphi^{\alpha }  \nonumber \\
&-&\partial _{\nu }\chi _{\mu }^{+}\partial _{\nu}\chi _{\mu } + \chi _{\mu }^{+} M_{01}^{2}\chi _{\mu}  \nonumber \\
&-&\partial _{\nu }\chi _{\mu }^{\alpha +}\partial _{\nu}\chi _{\mu }^{\alpha} + \chi _{\mu}^{\alpha +}M_{11}^{2}\chi _{\mu}^{\alpha }.  \nonumber
\end{eqnarray}

The interaction parts are as follows
\begin{eqnarray}
&&\mathcal{L}^{[1]}_{int} = \varphi ^{\alpha +}\bar{\Psi}_{c}\tau ^{\alpha}\Gamma \Psi g + g^{+}\bar{\Psi}\tau ^{\alpha}\Gamma \Psi_{c}\varphi ^{\alpha}  \label{LAGR1}  \\
&&\;\;\;\;\;\;\;\;-\;\partial _{\nu}\chi _{\mu}^{+}\bar{\Psi}_{c}\Gamma \sigma _{\mu \nu}\Psi \frac{h}{2m}-\frac{h^{+}}{2m}\bar{\Psi}%
\Gamma \sigma _{\mu \nu}\Psi _{c}\partial _{\nu}\chi _{\mu}  \nonumber \\
&&+\;\;\chi _{\mu}^{\alpha+}\bar{\Psi}_{c}\tau ^{%
\alpha}\Gamma \sigma ^{\mathrm{\mu \nu }}i\overleftrightarrow{%
\partial }_{\nu}\Psi \frac{h^{\prime }}{2m} +\frac{h^{\prime +}}{2m}\bar{\Psi}%
\tau ^{\alpha}\Gamma \sigma ^{\mathrm{\mu \nu }}i%
\overleftrightarrow{\partial }_{\nu}\Psi _{c}\chi _{\mu}^{\alpha},  \nonumber \\
&&\mathcal{L}^{[2]}_{int} =  \lambda_{10} (\bar{\Psi}\tau ^{\alpha}\Gamma \Psi_{c})( \bar{\Psi}_{c}\tau ^{\alpha}\Gamma \Psi) \label{LAGR2}  \\
&&\;\;\;\;\;\;\;\;+\; \lambda_{01} (\bar{\Psi} \Gamma \sigma _{\mu \nu}\Psi _{c} )(\bar{\Psi}_{c}\Gamma \sigma _{\mu \nu}\Psi)             \nonumber      \\
&&\;\;\;\;\;\;\;\;+\; \lambda_{11} (\bar{\Psi}\tau ^{\alpha}\Gamma \sigma ^{\mathrm{\mu \nu }}i%
\overleftrightarrow{\partial }_{\nu}\Psi _{c}) (\bar{\Psi}_{c}\tau ^{\alpha}\Gamma \sigma ^{\mathrm{\mu \nu }}i\overleftrightarrow{\partial }_{\nu}\Psi).  \nonumber
\end{eqnarray}

Here, $m$ is the nucleon mass, $M_{\mathrm{IJ}}$ are dibaryon masses
(matrices) diagonal in the radial and parity quantum numbers and
\[
\Gamma =\left(
\begin{array}{lllll}
1 & 0 & 0 & 0 & \cdots \\
0 & i\gamma _{5} & 0 & 0 & \cdots \\
0 & 0 & 1 & 0 & \cdots \\
0 & 0 & 0 & i\gamma _{5} & \cdots \\
\vdots & \vdots & \vdots & \vdots & \ddots
\end{array}
\right) .
\]
The coupling constants $g,$ $h$, $h^{\prime }$, and $\lambda_{IJ}$ depend on
the radial and parity quantum numbers as indicated e.g. by Eq.~(\ref{COUP}).
$\mathcal{L}$ is Hermitian by the construction.

There exists six various vertices for the Lorentz structures: $\{1,\gamma _{%
\mu},\sigma _{\mathrm{\mu \nu }},i\gamma _{5},\gamma _{5}\gamma _{%
\mu},i\gamma _{5}\sigma _{\mathrm{\mu \nu }}\}$. The vector and
tensor ones correspond to two spin-1 fields, vector and
pseudovector ones. So, we have at most two spin-$0$ fields with
parity $P=\pm 1$ and two spin-$1$ fields with parity $P=\pm 1$,
altogether four fields. The isospin degrees of freedom provide two
additional structures, so we might have eight dibaryons. However,
not all of them are coupled to the $NN$ channel. The system of two
nucleons may carry quantum numbers shown in Table \ref{tab:3}.


\begin{table}[h]
\addtolength{\tabcolsep}{3pt}
\centering
\caption
{
Two-nucleon states with lowest quantum numbers.
}
\label{tab:3}
\begin{tabular}{|cc|}
\hline \hline
$^{2S + 3}L_J$ & $(I,J^P)$ \\ \hline \hline
$^{3}$S$_1$ & $(0,1^+)$ \\
$^{1}$S$_0$ & $(1,0^+)$ \\
$^{3}$P$_1$ & $(1,1^-)$ \\
$^{1}$P$_1$ & $(0,1^-)$ \\
$^{3}$P$_0$ & $(1,0^-)$ \\
\hline \hline
\end{tabular}
\end{table}


We thus have five possible dibaryons $(I,J^{P})= (1,0^{\pm}),(0,1^{\pm}),(1,1^{-})$.
Missing are three dibaryons with the exotic quantum numbers $(0,0^{\pm}),(1,1^{+})$,
that do not correspond to two-nucleon states. From other hand, states
$(0,0^{\pm})$ have the vanishing couplings with the $NN$ channel,
since $\bar{\Psi}_{c}\Psi = \bar{\Psi}_{c}i\gamma _{5}\Psi =0$.
In the $(1,1^{+})$ channel one has, in addition, $\bar{\Psi}_{c}\tau ^{%
\alpha}\gamma _{\mu}\Psi =\bar{\Psi}_{c}\tau ^{\mathrm{%
\alpha }}\sigma _{\mathrm{\mu \nu }}\Psi = 0$.

The $(1,1^{-})$ channel is not exotic. It is included into the
effective Lagrangian through the vertex $\bar{\Psi}_{c}\tau ^{\mathrm{\alpha
}}i\gamma _{5}\sigma _{\mathrm{\mu \nu }}\overleftrightarrow{\partial }_{%
\nu}\Psi $. The vertex $\bar{\Psi}_{c}\tau ^{\alpha%
}\sigma _{\mathrm{\mu \nu }}\overleftrightarrow{\partial }_{\nu%
}\Psi $ is full derivative according to Eq.~(\ref{iden2}).

\section{In-medium nucleon and dibaryon propagators}
\setcounter{equation}{0}

We consider symmetric nuclear matter with equal proton and neutron fractions.

\subsection{Nucleon propagators in ideal Fermi gas}

The in-medium nucleon propagator is defined by
\begin{equation}
iS_{F}(x)=<T\Psi (x)\bar{\Psi}(0)>.  \label{SFC}
\end{equation}
The propagator depends on the Fermi momentum $p_F$.
In the momentum representation,
\begin{equation}
S_{F}(p) = \int dxe^{ipx}S_{F}(x)
\label{FOTR}
\end{equation}

We need the in-medium nucleon propagator constructed
out of the $C$-conjugated nucleon fields
\begin{equation}
iS_{F}^{c}(x) = <T\Psi _{c}(x)\bar{\Psi}_{c}(0)>.
\end{equation}
The momentum representation of $S_{F}^{c}(x)$ is derived from Eq.~(\ref{FOTR}).
The propagators are related by
\begin{eqnarray}
S_{F}^{c}(x)&=&C^{T}S_{F}^{T}(-x)C, \label{COMC1} \\
S_{F}^{c}(p)&=&C^{T}S_{F}^{T}(-p)C. \label{COMC2}
\end{eqnarray}
The $C$-conjugation matrix is defined in Appendix A.

In the momentum representation, the propagators of the ideal gas have the form
\begin{eqnarray}
S_{F}(p) = S_{F}^{c}(p) = \frac{1}{\hat{p} - m}.
\end{eqnarray}
The signs of imaginary parts of the pole positions on the complex energy plane should be specified separately.

The poles of $S_{F}(p)$ approach the real axis from the upper half of the complex $p_{0}$-plane
for $\Re p_{0} < \mu = + \sqrt{m^{2} + p_F^2}$ and
from the lower half of the complex $p_{0}$-plane for $\Re p_{0} > \mu$.
$S_{F}^{c}(p)$ has poles shifted to the upper half-plane for $\Re p_{0} < - \mu$
and to the lower half-plane otherwise.

$S_{F}(x)$ describes therefore propagation
of nucleons in presence of the Fermi sphere made up of nucleons.
$S_{F}^{c}(x)$ describes propagation of nucleons in presence of the Fermi
sphere, however, made up of antinucleons, the Fermi momentum being the same.

The plane waves expansion of the in-medium propagators has the form
\begin{widetext}
\begin{eqnarray}
iS_{F}(x) &=& \int \frac{d\mathbf{p}}{(2\pi )^{3}}\left[
e^{-ipx}\Lambda _{+}(\mathbf{p})\left( \theta (x_{0})\theta (|\mathbf{p}%
|-p_{F})-\theta (-x_{0})\theta (p_{F}-|\mathbf{p}|)\right) +e^{ipx}\Lambda
_{-}(\mathbf{p})\theta (-x_{0})\right], \label{PLAW1} \\
iS_{F}^{c}(x) &=&\int \frac{d\mathbf{p}}{(2\pi )^{3}}
\left[ e^{ipx}\Lambda _{-}(\mathbf{p})\left( \theta (-x_{0})\theta
(|\mathbf{p}|-p_{F})-\theta (x_{0})\theta (p_{F}-|\mathbf{p}|)\right) + e^{-ipx}\Lambda _{+}(\mathbf{p})\theta (x_{0}) \right],
\label{PLAW2}
\end{eqnarray}
\end{widetext}
where $p=(E(\mathbf{p}),\mathbf{p}),$ $E(\mathbf{p})=+\sqrt{m^{2}+\mathbf{p}^{2}}$, and
\begin{eqnarray}
\Lambda _{\pm }(\mathbf{p})=\frac{ \pm \hat{p} + m}{2E(\mathbf{p})}
\end{eqnarray}
are projection operators that obey
\begin{eqnarray}
C^{T} \Lambda _{ \pm }(\mathbf{p})^{T}C = \Lambda _{ \mp }(\mathbf{p}).
\end{eqnarray}

\subsection{Dibaryon propagators}

The dibaryon propagators with quantum numbers
$(1,0^{\pm})$, $(0,1^{\pm})$, $(1,1^{-})$ are defined, respectively, by
\begin{eqnarray}
i\Delta^{\prime \mathrm{\alpha \beta }}(x-y) &=&<T\varphi ^{\mathrm{%
\alpha }}(x)\varphi ^{\mathrm{\beta }+}(y)>, \\
iD_{\mathrm{\mu \nu }}^{\prime }(x-y) &=&<T\chi _{\mu}(x)\chi _{%
\nu}^{+}(y)>, \\
iD_{\mathrm{\mu \nu }}^{\prime \mathrm{\alpha \beta }}(x-y) &=&<T\chi _{%
\mu}^{\alpha}(x)\chi _{\nu}^{\mathrm{\beta }%
+}(y)>.
\end{eqnarray}
The radial indices are suppressed. The coupling to the $NN$
channel produces mixing of the radial excitations.

The free propagators are given by
\begin{eqnarray}
\Delta^{\alpha \beta} (p) &=& \frac{1}                                    {p^{2} - M_{10}^{2}} \delta ^{\alpha \beta }, \\
D_{\mu \nu}(p)                &=& \frac{-g_{\mu \nu } + p_{\mu} p_{\nu}/p^{2}}{p^{2} - M_{01}^{2}},   \\
D_{\mu \nu}^{\alpha \beta}(p) &=& \frac{-g_{\mu \nu } + p_{\mu} p_{\nu}/p^{2}}{p^{2} - M_{11}^{2}} \delta ^{\alpha \beta}.
\end{eqnarray}

Tensor structures of the dressed propagators can be factored out
\begin{eqnarray}
\Delta^{\prime \alpha \beta} (p) &=& \delta ^{\alpha \beta } \Delta^{\prime}(p), \\
D_{\mu \nu}^{\prime}(p)          &=& (-g_{\mu \nu } + p_{\mu} p_{\nu}/p^{2})D^{\prime}(p),   \\
D_{\mu \nu}^{\prime \alpha \beta}(p) &=& \delta ^{\alpha \beta} (-g_{\mu \nu } + p_{\mu} p_{\nu}/p^{2}) D^{\prime}_1(p).
\end{eqnarray}

\subsection{In-medium dispersion law for interacting nucleons}

In systems with interaction, self-energy operators of fermions
appear. They can be expanded over the $\gamma $-matrices:
\begin{eqnarray}
\hat{\Sigma}(p) &=& \hat{\Sigma}_{V}(p)+\Sigma _{S}(p), \\
\hat{\Sigma}^{c}(p) &=& \hat{\Sigma}_{V}^{c}(p)+\Sigma _{S}^{c}(p),
\end{eqnarray}
where $\hat{\Sigma}_{V}(p)=\Sigma _{V}^{\mu }(p)\gamma _{\mu }$ and $\hat{\Sigma}_{V}^{c}(p)=\Sigma^{c \mu}_{V}(p)\gamma_{\mu }$.
The $C$-conjugation gives (cf. (\ref{COMC2}))
\begin{eqnarray}
\hat{\Sigma}^{c}(p) = C^{T}\hat{\Sigma}^{T}(-p)C.
\end{eqnarray}
We thus obtain
\begin{eqnarray}
\Sigma^{c \mu}_V (p) &=& - \Sigma^{\mu}_V (- p), \label{SYMM1}\\
\Sigma^{c    }_S (p) &=&   \Sigma      _S (- p), \label{SYMM2}
\end{eqnarray}
and
\[
\hat{\Sigma}^{c}(p) = \gamma _{5}\hat{\Sigma}(-p)\gamma _{5}.
\]

\begin{figure}[!htb]
\begin{center}
\includegraphics[angle = 0,width = 4.96 cm]{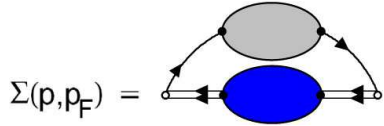}
\caption{
Nucleon self-energy operator.
The double solid line with the shaded blob shows the dressed dibaryon propagator,
the single solid line the shaded blob shows the dressed nucleon propagator.
}
\label{fig2}
\end{center}
\end{figure}

The diagram representation of the self-energy operator
arising due to the dibaryon-induced nucleon-nucleon interaction
is shown in Fig. \ref{fig2}. Its non-relativistic version
is $U_2(\mathbf{p},\rho)$ defined by Eq.~(\ref{4.17}).
Equation (\ref{4.18}) is the analogue
of the one-loop approximation shown in Fig. \ref{fig2}.

The in-medium propagators have the form
\begin{eqnarray*}
S_{F}^{\prime }(p) &=&\frac{1}{\hat{p}-m-\hat{\Sigma}(p)}=\frac{\hat{p} + m + \hat{\Sigma}^{c}(-p)}{\phi (p)}, \\
S_{F}^{\prime c}(p) &=&\frac{1}{\hat{p}-m-\hat{\Sigma}^{c}(p)}=%
\frac{\hat{p}+m+\hat{\Sigma}(-p)}{\phi_{c}(p)},
\end{eqnarray*}
where
\begin{eqnarray*}
\phi (p) &=&(p-\Sigma _{V}(p))^{2}-(m+\Sigma _{S}(p))^{2}, \\
\phi _{c}(p) &=&(p-\Sigma^{c} _{V}(p))^{2}-(m+\Sigma^{c} _{S}(p))^{2}.
\end{eqnarray*}
Using Eqs.~(\ref{SYMM1}) and (\ref{SYMM2}) we find
\begin{eqnarray}
\phi_{c}(p) = \phi(-p).
\end{eqnarray}

The dispersion laws of elementary excitations
of nucleons are determined from equation
\begin{eqnarray}
\phi (p) = 0.
\label{DISPLAW}
\end{eqnarray}
This equation gives two solutions $p_{0} = \pm E_{\pm }(\mathbf{p})$ with positive
and negative energies of the excitations.
Respectively, positive and negative energy
solutions of equation $\phi_{c} (p) = 0$ are $p_{0} = \pm E_{\mp }( \mathbf{p})$.
In the ideal Fermi gas, $E_{\pm }(\mathbf{p}) = +\sqrt{m^2 + \mathbf{p}^2}$.

The plane waves expansion of the in-medium propagators has the form
\begin{widetext}
\begin{eqnarray}
iS_{F}^{\prime }(x) &=& \int \frac{d\mathbf{p}}{(2\pi )^{3}}\left[e^{-ip_{+}x}\Lambda _{+}^{\prime}(\mathbf{p})\left( \theta (x_{0})\theta (|\mathbf{p}%
|-p_{F})-\theta (-x_{0})\theta (p_{F}-|\mathbf{p}|)\right)
+e^{ip_{-}x}\Lambda _{-}^{\prime}(\mathbf{p})\theta (-x_{0})\right] , \\
iS_{Fc}^{\prime }(x) &=&\int \frac{d\mathbf{p}}{(2\pi )^{3}}\left[
e^{ip_{+} x}\Lambda _{+}^{\prime c}(\mathbf{p})\left( \theta (-x_{0})\theta (|%
\mathbf{p}|-p_{F})-\theta (x_{0})\theta (p_{F}-|\mathbf{p}|)\right)
+e^{-ip_{-} x}\Lambda _{-}^{\prime c}(\mathbf{p})\theta (x_{0})\right] ,
\end{eqnarray}
\end{widetext}
where $p_{\pm}=(E_{\pm}(\mathbf{p}),\mathbf{p})$ and
\begin{eqnarray}
\Lambda _{ \pm }^{\prime}(\mathbf{p})   &=&\frac{ \pm \hat{p}_{\pm } + m + \hat{\Sigma}^{c}(\mp p_{\pm })}{\pm \phi     ^{\prime }(\pm p_{ \pm })}, \\
\Lambda _{ \mp }^{\prime c}(\mathbf{p}) &=&\frac{ \pm \hat{p}_{\mp } + m + \hat{\Sigma}    (\mp p_{\mp })}{\pm \phi _{c}^{\prime }(\pm p_{ \mp })},
\end{eqnarray}
with $\phi^{\prime }(p) = {d \phi (p)}/{dp_{0}}$ and $\phi^{\prime }_{c}(p) = {d \phi_{c}(p)}/{dp_{0}}$.
In the ideal Fermi gas, Eqs.~(\ref{PLAW1}) and (\ref{PLAW2}) are recovered.

\subsection{Dyson equation for nucleon propagator}

$\Sigma (p)$ is calculated to one loop, contributions of
the antinucleons are neglected.
The loops are formed thereby by nucleons and nucleon holes in the Fermi sphere.
Such approximation is close to the mean-field approximation.

\begin{figure}[h]
\begin{center}
\includegraphics[angle = 0,width = 14 cm]{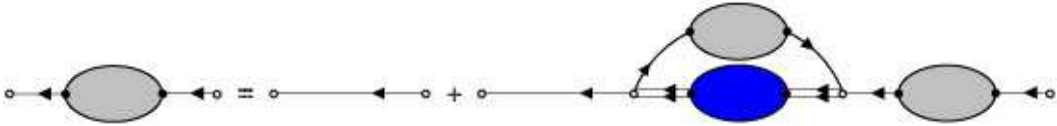}
\caption{
System of Dyson equations for nucleon propagator $S_F^{\prime}(p)$.
The single and double solid lines show the nucleon and dibaryon
propagators. The shaded blobs indicate the dressed propagators. Summation over
dibaryons in the loop is assumed.
}
\label{fig1}
\end{center}
\end{figure}

Shown in Fig. \ref{fig1} is the system of Dyson equations,
\begin{equation}
S_{F}^{\prime }=S_{F}+S_{F}\Sigma S_{F}^{\prime },
\end{equation}
for the in-medium nucleon propagator. The virtual dibaryons
in the loop contribute to the self-energy operator. The four-fermion
interaction will be included later.

Starting from Lagrangian (\ref{LAGRANGIAN}) one gets
\begin{widetext}
\begin{eqnarray}
\frac{1}{4}\hat{\Sigma}(p) &=&\int \frac{d^{4}p_{c}}{(2\pi )^{4}}g^{+} \tau ^{\alpha}\Gamma
\Delta^{\prime \alpha \beta}(p-p_{c})
iS_{F}^{\prime c}(p_{c})\tau ^{\beta }\Gamma g  \nonumber \\
&-&\int \frac{d^{4}p_{c}}{(2\pi )^{4}} \frac{h^{+}}{2m} i\sigma _{\mu \tau }(p_{c}-p)_{\tau }\Gamma
D_{\mu \nu }^{\prime}(p-p_{c})
iS_{F}^{\prime c}(p_{c})\Gamma i\sigma _{\nu \sigma }(p_{c}-p)_{\sigma }\frac{h}{2m}  \nonumber \\
&+&\int \frac{d^{4}p_{c}}{(2\pi )^{4}}\frac{h^{\prime +}}{2m} i\sigma _{\mu \tau }(p+p_{c})_{\tau }\tau ^{\alpha}i\gamma
_{5}D_{\mu \nu }^{\prime \alpha \beta }(p-p_{c})iS_{F}^{%
\prime c}(p_{c})\tau ^{\beta }i\gamma _{5}i\sigma
_{\nu \sigma }(-p_{c}-p)_{\sigma }\frac{h^{\prime }}{2m}.
\label{SIGMA}
\end{eqnarray}
\end{widetext}
In Eq.~(\ref{SIGMA}) summation over the radial numbers and parity states ($\Gamma = 1,i\gamma_5$)
is assumed. The integrals over the timelike component of $p_{c}$
can be calculated assuming $\phi(p)$ has simple zeros in the complex energy plane
at $p_{0} = \pm E_{\pm}(\mathbf{p})$.
The negative energy states are neglected. The remaining three-dimensional
integrals describe contributions
to the self-energy operators from the nucleon holes.

The self-energy can be expanded over the different channels
\begin{eqnarray}
\hat{\Sigma}(p) = \sum_{IJ^{P}} \hat{\Sigma}_{IJ^{P}}(p).
\end{eqnarray}
The separate contributions for
$(I,J^{P})= (1,0^{\pm}),(0,1^{\pm}),(1,1^{-})$ are given in Appendix B.


\subsection{Bethe-Brueckner equation for dibaryon propagator}

\begin{figure}[h]
\begin{center}
\includegraphics[angle = 0,width = 4.96 cm]{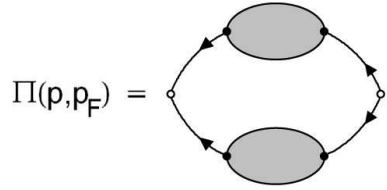}
\caption{
Dibaryon self-energy operator.
The single solid lines with the shaded blobs are the dressed nucleon
propagators. The loop produces mixing between dibaryons with
different radial quantum numbers.
}
\label{fig4}
\end{center}
\end{figure}

Dibaryon self-energy operator to one loop is determined by two-body
nucleon Green function, its diagram representation is shown in Fig. \ref{fig4}.
The system of equations of Fig. \ref{fig4} is equivalent
to Eqs.~(\ref{4.1}) which determine two-nucleon scattering in the vacuum ($T$-matrix)
and in the medium ($G$-matrix). The potential $V_2$ is the analogue of the
dibaryon propagator in the vacuum, $G_2(W)$ is the analogue of the dressed dibaryon propagator,
$Q_2 e_2^{-1}$ is the dibaryon self energy shown in Fig. \ref{fig4}.

The corresponding analytical expressions have the form
\begin{widetext}
\begin{eqnarray*}
\Pi ^{\mathrm{\beta \alpha }}(p,p_{F}) &=&-2i\int \frac{d^{4}p_{c}}{(2\pi)^{4}}g
\mathrm{Sp[}\tau ^{\mathrm{\beta }}\Gamma S_{F}^{\prime}(p+p_{c},p_{F})\tau ^{\alpha}\Gamma S_{Fc}^{\prime }(p_{c},p_{F})\mathrm{]}g^{+}, \\
\Pi _{\mathrm{\nu \mu }}(p,p_{F}) &=&-2i\int \frac{d^{4}p_{c}}{(2\pi )^{4}} \frac{h}{2m}%
\mathrm{Sp[}(-i\sigma _{\mathrm{\nu \sigma }}p_{\mathrm{\sigma }%
})\Gamma S_{F}^{\prime }(p+p_{c},p_{F})i\sigma _{\mathrm{\mu
\tau }}p_{\mathrm{\tau }}\Gamma S_{Fc}^{\prime }(p_{c},p_{F})\mathrm{]}\frac{h^{+}}{2m}, \\
\Pi _{\mathrm{\nu \mu }}^{\mathrm{\beta \alpha }}(p,p_{F}) &=&-2i\int \frac{%
d^{4}p_{c}}{(2\pi )^{4}}\frac{h^{\prime }}{2m}\mathrm{Sp[}(-\tau ^{\mathrm{%
\beta }}i\gamma _{5}i\sigma _{\mathrm{\nu \sigma }}(p+2p_{c})_{\mathrm{%
\sigma }})S_{F}^{\prime }(p+p_{c},p_{F}) i\sigma _{\mathrm{\mu
\tau }}(p+2p_{c})_{\mathrm{\tau }}\tau ^{\alpha}i\gamma
_{5}S_{Fc}^{\prime }(p_{c},p_{F})\mathrm{]} \frac{h^{\prime +}}{2m}\mathrm{.}
\end{eqnarray*}
\end{widetext}
Here, $\mathrm{Sp[...]}$ is taken over isospin and bispinor indices.
The self-energy operators are diagonal in $(I,J^{P})$ and represent matrices
in the radial numbers.

The system of Dyson equations for the dibaryon propagators has the form:
\begin{eqnarray*}
\Delta ^{\prime \alpha \beta }(p) &=&\Delta^{\alpha \beta } (p)+ \Delta^{\alpha \sigma } (p)\Pi ^{\sigma \gamma }(p)\Delta ^{\prime \gamma \beta }(p),  \\
D_{\mu \nu }^{\prime }(p) &=& D_{\mu \nu }(p) + D_{\mu \tau}(p) \Pi _{\tau \sigma }(p)D_{\sigma \nu }^{\prime }(p),  \\
D_{\mu \nu }^{\prime \alpha \beta }(p) &=&D_{\mu \nu}^{\alpha \beta }(p)+D_{\mu \tau }^{\alpha \delta }(p)\Pi _{\tau \sigma }^{\delta \gamma}(p)D_{\sigma \nu }^{\prime \gamma \beta }(p).
\end{eqnarray*}
These equations are depicted in Fig. \ref{fig3}.

\subsection{Link to optical potential model}

Probing particles acquire new decay channels in the medium, connected
to collisions with the surrounding particles, and become unstable.
Phenomenologically, the in-medium spin-1/2 particles are described
by the Dirac equation
\begin{equation}
(i\hat{\nabla}-m_{1}+\frac{i}{2}\Gamma )\Psi =0,
\end{equation}
where $\Gamma $ is collision width
\begin{equation}
\Gamma =\rho \sigma v\gamma .
\end{equation}
Here, $v$ is velocity and $\gamma $ Lorentz factor. The cross section can be
expressed in terms of the forward scattering amplitude with the help of the
optical theorem
\begin{equation}
\sigma =\frac{4 \pi }{p^{*}} \mathrm{Im} f(\theta =0).
\end{equation}

The self-energy operator and the forward scattering
amplitude are analytical functions. Given
the imaginary part is known, the real part can be restored using the
dispersion relations. If, however, the scattering amplitude is known,
one can write
\begin{equation}
\Sigma (p)=-\rho \frac{2\pi \sqrt{s}}{m_{1}m_{2}}f(\theta =0).
\label{OPTI}
\end{equation}
where $m_{2}$ is mass of the particles of the medium.

Near the threshold, one can expand the amplitude over the partial waves
and keep the lowest order $L=0$ terms:
$f(\theta =0)\approx e^{i\delta (s)}\sin \delta (s)/p^{*}$ where $\delta(s)$
is the $S$-wave phase shift.
If probing particle is identical with particles of the medium,
one has $m_{1}=m_{2} = m$, $f(\theta =0)\approx 2e^{i\delta
(s)}\sin \delta(s)/p^{*}$ and
\begin{equation}
\Sigma (p)=-\rho \frac{8\pi }{mp^{*}}e^{i\delta (s)}\sin \delta (s).
\label{OPTICA}
\end{equation}

The symmetric nuclear matter consists of the two types of the particles,
so the statistical factors modify Eq.~(\ref{OPTICA}).

Let a spin-up proton scatters on the in-medium protons in the $(1,0^+)$ channel.
The in-medium protons have spin down. The statistical weight of the configuration $\uparrow \downarrow $
in the $S=0$ spin wave function is $1/2$. The proton-proton collisions
contribute to the self-energy operator with the weight
${1}/{2} \times \rho/4$, where $\rho/4 $ is the spin-down proton density.
A spin-up proton scatters on the in-medium neutrons also,
with the weight factor ${1}/{2}$ for the $I=1$ channel
and the weight factor ${1}/{2}$ for the spin-down neutrons.
The spin-isospin factor in the $I=1$ channel becomes ${1}/{4} \times {1}/{2} \times {1}/{2} = {1}/{16}$.
The in-medium protons and neutrons contribute altogether with the factor
${1}/{8} + {1}/{16} = {3}/{16}$. We therefore replace
$\rho \rightarrow {3} \rho /{16} $ in Eq. (\ref{OPTICA}).
The $(0,1^+)$ channel can be analyzed in the similar fashion.

 From other hand, in the low-density limit of the symmetric matter,
Eqs.(\ref{SIGMA}) give
\begin{eqnarray*}
\Sigma_{IJ^P} (p) = - \rho \frac{3\pi }{2mp^{*}}e^{i\delta _{IJ^P}(s)}\sin \delta _{IJ^P}(s)\frac{\gamma _{0}+1}{2} + \ldots,
\end{eqnarray*}
\newline
where $(I,J^P)=(1,0^+)$ and $(0,1^+)$ in agreement with the optical potential model.

It is instructive to compare the nucleon self-energy with the
Relativistic Mean Field (RMF) model \cite{Walecka:1974qa,Chin:1977iz},
according to which
\begin{equation}
\Sigma (p) = \rho \frac{g_{\omega }^{2}}{m_{\omega }^{2}}\gamma
^{0}-\rho \frac{g_{\sigma }^{2}}{m_{\sigma }^{2}}.
\label{RMF}
\end{equation}
The first term describes the $\omega $-meson exchange between the nucleons,
the second one originates from the $\sigma $-meson exchange.
In the RMF models $\Sigma (p) \sim \gamma ^{0}-1$ (numerically).
The strong cancellation of the large vector and the large scalar mean fields
is the typical feature of the RMF models.

In our model, the nucleon self-energy in the low density limit
is proportional to $\gamma_{0}+1$.
The first term can be interpreted as the vector mean field, the second one
as the scalar mean field. The mean fields are of the same sign and of the same
order.

\section{Dibaryons in phase shifts analysis of nucleon-nucleon scattering}
\setcounter{equation}{0}

In order to describe the in-medium properties of nucleons and dibaryons,
one has to fix parameters entering the effective Lagrangian.
Those parameters determine the $NN$ scattering phase shifts
which are known experimentally.

\subsection{Dibaryon self-energy operator in the vacuum}

The dibaryon self-energy operators are matrices
proportional $gg^{+}$, $hh^{+}$, and $h^{\prime }h^{\prime +}$.
Such a circumstance allows to factorize the radial numbers
and the parity structures. Also, the isospin and Lorentz structures
of the self-energy operators can be factored out:
\begin{eqnarray}
\Pi ^{\beta \alpha }(p) &=&\delta ^{\beta \alpha} g \Pi _{10^{P}}(p)g^{+}, \\
\Pi _{\nu \mu}      (p) &=&(- g_{\nu \mu} + \frac{ p_{\nu } p_{\mu} }{p^{2}}) h \Pi_{01^{P}}(p)h^{+}, \\
\Pi _{\nu \mu }^{ \beta \alpha }(p) &=& \delta^{\beta \alpha}(- g_{\nu \mu } + \frac{p_{\nu}p_{\mu }}{p^{2}}) h^{\prime }\Pi _{11^{-}}(p)h^{\prime +}, \;\;\;
\end{eqnarray}
where
\begin{widetext}
\begin{eqnarray}
\Pi _{IJ^P}(p)=\frac{1}{(2\pi )^{4}}\int d^{4}q_{c}d^{4}p_{c}\delta^{4}(p-q_{c}-p_{c})\frac{\varphi _{IJ^P}(q_{c},p_{c})}
{(q_{c}^{2}-m^{2}+i0)(p_{c}^{2}-m^{2}+i0)}.
\end{eqnarray}
The functions $\varphi _{IJ^P}(q_{c},p_{c})$ have the form
\begin{eqnarray*}
\varphi _{10^P}(q_{c},p_{c}) &=&-\frac{2i}{3}\mathrm{Sp[}\tau ^{%
\alpha}\Gamma (\hat{q}_{c}+m)\tau ^{\alpha}\Gamma (-\hat{%
p}_{c}+m)\mathrm{]}, \\
\varphi _{01^P}(q_{c},p_{c}) &=&-\frac{2i}{3}\frac{1}{4m^{2}}(-g_{%
\mathrm{\nu \mu }}+\frac{p_{\nu}p_{\mu}}{p^{2}})\mathrm{%
Sp[}\Gamma (-i\sigma _{\mathrm{\nu \sigma }}p_{\mathrm{\sigma }})(\hat{q}%
_{c}+m)i\sigma _{\mathrm{\mu \tau }}p_{\mathrm{\tau }}\Gamma (-\hat{p}_{c}+m)%
\mathrm{]}, \\
\varphi _{11^-}(q_{c},p_{c}) &=&-\frac{2i}{9}\frac{1}{4m^{2}}(-g_{%
\mathrm{\nu \mu }}+\frac{p_{\nu}p_{\mu}}{p^{2}})\mathrm{%
Sp[}\tau ^{\alpha}i\gamma _{5}(-i\sigma _{\mathrm{\nu \sigma }%
}(q_{c}-p_{c})_{\mathrm{\sigma }})(\hat{q}_{c}+m)i\sigma _{\mathrm{\mu \tau }%
}(q_{c}-p_{c})_{\mathrm{\tau }}\tau ^{\alpha}i\gamma _{5}(-\hat{p}%
_{c}+m)\mathrm{]}.
\end{eqnarray*}


\begin{figure}[h]
\begin{center}
\includegraphics[angle = 0,width = 14 cm]{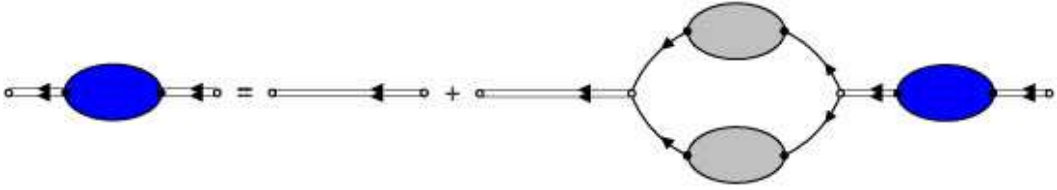}
\caption{
Systems of Dyson equations for the dibaryon propagators shown by the double solid lines.
The single solid lines show the nucleon propagator. The shaded blobs indicate
dressed propagators.
}
\label{fig3}
\end{center}
\end{figure}


The discontinuity over the cut $(4m^{2},+\infty )$ can be found to be
\[
\Pi _{IJ^P}(s+i0)-\Pi _{IJ^P}(s-i0)=-\frac{1}{4\pi ^{2}}\Phi
_{2}(s)\varphi _{IJ^P}(q_{c},p_{c}),
\]
where $q_{c}$ and $p_{c}$ are the on-mass-shell nucleon momenta and
$\Phi_{2}(s)$ is two-body phase space
\begin{eqnarray*}
\Phi _{2}(s) &=&\frac{\pi p^{*}(s)}{\sqrt{s}},
\end{eqnarray*}
where $p^{*}(s) = \sqrt{s/4 - m^{2}}$. In what follows, $p^{*}(s+i0)\equiv k>0$ above the threshold and
$p^{*}(s-i0)=-k$.

The functions $\varphi _{IJ^P}(q_{c},p_{c})$ are given in Table \ref{tab:6}.
They show the correct threshold behavior $\sim k^{2L}$. The imaginary part of $\Pi _{IJ^P}(s)$
is negative and corresponds to positive transition
probabilities.


\begin{table}[h]
\addtolength{\tabcolsep}{3pt}
\centering
\caption
{
$\varphi _{IJ^P}(q_{c},p_{c})$ and $\kappa_{IJ^P}$ is different channels.
}
\label{tab:6}
\begin{tabular}{|cccc|}
\hline \hline
$^{2S + 3}L_J$ & $(I,J^P)$ & $\varphi _{IJ^P}(q_{c},p_{c})$ & $\kappa_{IJ^P}$ \\ \hline \hline
$^{3}$S$_1$ & $(0,1^+)$ &  $8is(1 + p^{*2}/(3m^{2}))$       &   $\pi b/(2m)   $  \\
$^{1}$S$_0$ & $(1,0^+)$ &  $8is                      $      &   $\pi b/(2m)   $  \\
$^{3}$P$_1$ & $(1,1^-)$ &  $64i p^{*2}/3              $     &   $\pi m b^3/12 $  \\
$^{1}$P$_1$ & $(0,1^-)$ &  $16i p^{*2}s/(3m^{2})       $    &   $\pi m b^3/12 $  \\
$^{3}$P$_0$ & $(1,0^-)$ &  $32i p^{*2}                  $   &   $\pi m b^3/18 $  \\
\hline \hline
\end{tabular}
\end{table}


\subsection{$S$ matrix}

We consider first the scalar dibaryons.
The dressed propagator can be written as follows
\[
\Delta^{\prime \alpha \beta}(p) = \delta ^{\alpha \beta }\Delta^{\prime }(p).
\]
In the one-loop approximation, the system of Dyson equations can
be solved to give
\[
\Delta ^{\prime }(p)=\Delta (p)+\frac{\Delta (p) g \Pi _{10^P}(p)g^{+}\Delta (p)}{1-\Lambda _{10^P}^{-1}(p)
\Pi _{10^P}(p)}.
\]
Here, $\Delta (p)$ is the bare dibaryon propagator diagonal in the radial numbers,
\begin{eqnarray}
\Lambda _{10^P}^{-1}(s)=g^{+}\Delta (p)g&=&g^{+}\frac{1}{s-M_{10^P}^{2}}g \label{LAMB} \\
&=& \sum_{r}g_{r}^{*}\frac{1}{s-M_{r10^P}^{2}}g_{r},                      \nonumber
\end{eqnarray}
where $s=p^{2}$, the sum is taken over the radial numbers.

The model described in Sect. V is the relativistic extension of
the Quark Compound Bag model \cite{SIMO81}. $\Lambda _{01^P}^{-1}(s)$ is, in particular,
the function defined in Eq.~(5) of Ref. \cite{SIMO81}.
The poles of $\Lambda _{01^P}^{-1}(s)$ correspond to the compound $6q$ states i.e. dibaryons,
whereas poles of $\Lambda _{01^P}(s)$ are the Castillejo-Dalitz-Dyson (CDD) poles \cite{CDD}.
The model described in Sect. V can also be considered
as a version of the Dyson model \cite{DYSO}, which is, in turn,
a version of the Lee model \cite{LEE}.

The four-point Green function of the nucleons is shown in Fig. \ref{fig5}.
The scattering amplitude is proportional to
\[
g^{+}\Delta ^{\prime }(p)g = \frac{1}{\Lambda _{10^P}(p)-\Pi _{10^P}(p)}.
\]

Extension to the vector dibaryons is straightforward.
In general case, the $S$-matrix can be written as follows:
\begin{equation}
S_{IJ^P}=e^{2i\delta _{_{IJ^P}}(s)}=\frac{\Lambda _{IJ^P}(s)-\Pi _{IJ^P}(s-i0)}{\Lambda _{IJ^P}(s)-\Pi _{IJ^P}(s+i0)}.  \label{SMAT}
\end{equation}

\subsection{$P$ matrix. Dibaryons as $P$-matrix poles}

The experimental searches of exotic MSQ did not give conclusive results. Jaffe and Low \cite{JALO}, therefore,
proposed to identify exotic MSQ with so-called "primitives", i.e., the poles of the $ P $ matrix, not $ S $ matrix.
The primitives manifest themselves in the form of zeros of $ D $ functions on the unitary cut. Analysis of
the phase shifts of meson-meson scattering and nucleon-nucleon scattering \cite{JASH}
showed the presence of primitives with masses close to those expected in the MIT bag model.

The method of Jaffe and Low is basically heuristic and can
hardly be extended to modified external conditions e.g.
at finite temperature or density. In this sense, the dynamical $P$ matrix
developed by Simonov \cite{SIMO81} is perfectly suited for studying
collective properties of nuclear matter. Earlier, it has
been applied successfully for description of nucleon-nucleon
interaction in the vacuum \cite{SIMO81,SIMO84,BHGU,FALE,NARO94}.

Since the $P$-matrix formalism is phenomenologically successful,
it is desirable for any model to have a link with it.
As noticed by Simonov \cite{SIMO81}, the $P$ matrix has poles
at primitive masses provided the interaction between
nucleons and dibaryons is restricted to the surface of a radius $b$.
In the coordinate space, the form factor entering the $\bar{N}\bar{N}D$ vertex
represents a delta-function $\delta (r-b)$.

In the momentum representation, we have to use therefore, in the partial $L$-wave,
\begin{equation}
\mathcal{F}_{L}(s)= \left( \frac{s}{4m^2} \right) ^{1/4} \frac{(2L+1)!!j_{L}(kb)}{(kb)^{L}}.  \label{FF}
\end{equation}
The partial wave scattering amplitudes of the pointlike nucleons and dibaryons
discussed so far already have the correct threshold behavior.
The denominator $(kb)^{L}$ in Eq.~(\ref{FF}) is thus introduced not to violate
that feature. The additional dependence on $s/(4m^2)$ is introduced to simplify
calculations of the dibaryon self-energy operators. The common factor is chosen
to fulfill
\[
\mathcal{F}_{L}(s=4m^{2})=1.
\]
In the nonrelativistic limit, $\mathcal{F}_{L}(s) \equiv 1$ for pointlike interaction with $b=0$.


\begin{figure}[!htb]
\begin{center}
\includegraphics[angle = 0,width = 14 cm]{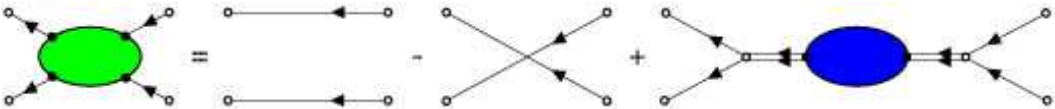}
\caption{
Four-point Green function of nucleons. }
\label{fig5}
\end{center}
\end{figure}


Let us consider function
\[
f_{L}(z)=zj_{L}(z)h_{L}^{(1)}(z),
\]
where $j_{L}(z),$ $y_{L}(z),$ and $h_{L}^{(1)}(z)=j_{L}(z)+iy_{L}(z)$ are,
respectively, the 1-st, 2-nd and 3-rd spherical Bessel functions defined
as Ref.~\cite{ABRA}. One has, for $z>0,$%
\begin{eqnarray*}
j_{L}(z)-j_{L}(-z) &=&\left\{
\begin{array}{ll}
0, & \mathrm{even\;}L, \\
2j_{L}(z), & \mathrm{odd\;}L,
\end{array}
\right. \\
y_{L}(z)-y_{L}(-z) &=&\left\{
\begin{array}{ll}
2y_{L}(z), & \mathrm{even\;}L, \\
0, & \mathrm{odd\;}L
\end{array}
\right.
\end{eqnarray*}
and, as a consequence,
\begin{equation}
f_{L}(z)-f_{L}(-z)=2zj_{L}^{2}(z).  \label{FUND}
\end{equation}

The form factor modifies the discontinuity of $\Pi _{IJ^P}(s)$:
\begin{widetext}
\begin{eqnarray}
\varphi _{IJ^P}(q_{c},p_{c})  &\rightarrow&  \varphi _{IJ^P}(q_{c},p_{c}) \mathcal{F}_{L}^{2}(s)  \nonumber \\
 &=& \varphi _{IJ^P}(q_{c},p_{c}) \left( \frac{s}{4m^2}\right) ^{1/2} \left( \frac{(2L+1)!!}{(kb)^{L}}\right) ^{2} \frac{f_{L}(kb)-f_{L}(-kb)}{2kb}.
\label{DPFF}
\end{eqnarray}
\end{widetext}
$\Pi _{IJ^P}(s)$ can be reconstructed using the dispersion integral.

In order to simplify calculations, we replace the ratio
$\varphi _{IJ^P}(q_{c},p_{c})/(kb)^{2L}$ by its threshold value.
Such a procedure can be viewed as a redefinition of the form factor $\mathcal{F}_{L}(s)$
which is a phenomenological element of the model.
One can easily verify that the discontinuity has
the correct form for the expression
\begin{equation*}
\Pi _{IJ^P}(s) = -\frac{1}{16 \pi b m} \varphi _{IJ^P}(q_{c},p_{c}) \left( \frac{(2L+1)!!}{(kb)^{L}}\right)^{2}f_{L}(s),
\end{equation*}
where $f_{L}(s) \equiv f_{L}(p^{*}(s))$.
$\Pi _{IJ^P}(s)$ defined this way is analytical function
in the complex $s$-plane with the cut $(4m^{2},+\infty )$.
On the first Riemann sheet, $\Pi _{IJ^P}(s) \to 0$ for $s \to \infty$.
Equation for $\Pi _{IJ^P}(s)$ can be written as
\begin{equation}
\kappa_{IJ^P} \Pi _{IJ^P}(s) = -if_{L}(s),
\label{KAPPA}
\end{equation}
with coefficients $\kappa_{IJ^P}$ listed in Table~\ref{tab:6}.

$P$ matrix is defined as the logarithmic derivative of
two-nucleon radial wave function:
\begin{equation}
P = \left. \frac{d\ln \chi _{L}(r)}{d\ln r}\right| _{r=b}.
\label{PMAT}
\end{equation}
Here, we suppress indices of the channel except for $L$.

One can express $S$ matrix in terms of $P$ matrix and $\Lambda(s)$,
$P$ matrix in terms of $S$ matrix and $\Lambda(s)$, and
$\Lambda(s)$ in terms of $S$ and $P$ matrices.

Firstly, using Eq.~(\ref{PMAT}) and Eqs.~(\ref{SMAT}) and (\ref{KAPPA}) one gets
\begin{eqnarray}
S &=& - \frac{h_{L}^{(2)}(z)P-(zh_{L}^{(2)}(z))^{\prime }}
             {h_{L}^{(1)}(z)P-(zh_{L}^{(1)}(z))^{\prime }}  \label{PJPI} \\
  &=&   \frac{\kappa \Lambda - izj_{L}(z)h_{L}^{(2)}(z)}
             {\kappa \Lambda + izj_{L}(z)h_{L}^{(1)}(z)},   \label{PJPI2}
\end{eqnarray}
where $z=kb$ (see e.g. \cite{FLUG}, problem 82).

Secondly, $P$ matrix can be found to be
\begin{eqnarray}
P &=& \frac{(zh_{L}^{(2)}(z))^{\prime } + S (zh_{L}^{(1)}(z))^{\prime }}
           {  h_{L}^{(2)}(z) + S h_{L}^{(1)}(z)}  \label{PJPI3}\\
  &=& P_{free} + \kappa^{-1} \Lambda ^{-1},       \label{PJPI4}
\end{eqnarray}
where
\begin{eqnarray}
P_{free}= \frac{(zj_{L}(z))^{\prime }}{j_{L}(z)}.
\label{FREEPMAT}
\end{eqnarray}
Equation (\ref{PJPI4}) is equivalent to Eq.~(\ref{27q}) with
$\delta_0 = 0$. $P$ matrix can be modified in general by adding to $P$ matrix
of the noninteracting particles new poles with positive residues.
The negative ones are not allowed since they bring the corresponding
$D$ function out of the class of generalized $R$ functions.
The free $P$ matrix has residues
\[
r_{p} = 8\left( \frac{a_{L}}{b}\right) ^{2} ,
\]
where $a_{0}=\pi$, $a_{1}=4.49$ etc. are zeros of the Bessel functions.

Equation (\ref{PJPI4}) can be used for the extraction from
experimental phase shifts the information on masses of the compound states
which show up as $P$-matrix poles on the real axis of the complex $s$-plane.

Thirdly, $\Lambda$ is expressed through $S$ and $P$ matrices:
\begin{eqnarray}
\kappa \Lambda &=& - izj_{L}(z)\frac{Sh_{L}^{(1)}(z) + h_{L}^{(2)}(z)}{S-1} \\
        &=& (P-P_{free})^{-1}.
\end{eqnarray}
Poles of $\Lambda$ are the CDD poles.

\subsection{$P$ matrix and Castillejo-Dalitz-Dyson poles}

Castillejo, Dalitz and Dyson \cite{CDD} introduced the so-called CDD poles to parameterize
the uncertainty of the amplitudes which satisfy the correct analytic properties in the complex
energy plane, the unitarity relation and are solutions of the scattering Low's equation.
The physical meaning of the CDD poles has been clarified by Dyson \cite{DYSO} who showed,
using one of the earlier versions of the Lee model \cite{LEE}, that CDD poles correspond
to bound states and resonances. New type of CDD pole is associated with primitives \cite{MIK}.

The CDD poles are poles of $\Lambda$ and $D$ functions defied by the denominator
of Eq.~(\ref{SMAT}):
\begin{equation}
D(s)=\Lambda (s)-\Pi (s),  \label{DFUN}
\end{equation}
where the indices $(I,J^P)$ are suppressed.
The CDD poles are localized between zeros of $\Lambda (s)$, i.e.,
between $M_{rIJ^P}^{2}$ and $M_{r + 1IJ^P}^{2}$.

$D$ function we discuss is the generalized $R$ function
\cite{CDD}. It has no zeros on the first Riemann sheet of the complex
$s$-plane. It has no zeros and on the real half-axis $(-\infty ,s_{0})$,
corresponding to bound states, also provided $D(s_{0})<0$ and
$s_{0} < M_{ rIJ^P }^{2}$.

Resonances are simple roots of equation
\begin{equation}
D(s)=0,
\label{DNULL}
\end{equation}
located on the second Riemann sheet below the unitary cut. In the limit of small $g_{\alpha }$,
roots of Eq.~(\ref{DNULL}) are localized in the neighborhood of $s = M_{ rIJ^P }^{2}$. $\Re s$ gives the
resonance mass, $\Im s$ determines the decay width $\Gamma_{\alpha} = g_{\alpha}^2 \Im D(M_{\alpha }^2)/M_{\alpha }$.

At the CDD poles $\delta (s)=0 \mod(\pi)$, the phase shift crossed the level
with a positive slope. Indeed, let $s_{\gamma}$ be a CDD pole. Equations (\ref{LAMB}) and (\ref{SMAT})
imply $\Lambda^{-1}(s_{\gamma}) = 0$ and $\Lambda^{-1}(s_{\gamma})^{\prime}
< 0$. Expanding the $D$ function around $s = s_{\gamma}$ and using Eq.~(\ref{SMAT}%
), one finds
\[
\delta(s_{\gamma})^{\prime} = - \Im D(s_{\gamma})
\Lambda^{-1}(s_{\gamma})^{\prime} > 0.
\]
The same behavior is implied by the Breit-Wigner formula, according to which
on isolated resonances phase shift increases by $\pi$. In the theory of
potential scattering, growth of phase is associated with attraction.

Models such as the Dyson-Lee model~\cite{LEE,DYSO} apply directly to systems where the phase shifts
increase with energy.

Nucleon-nucleon interaction, on the contrary, points toward the dominance of repulsion, that
results in decrease of phase shifts with increasing energy. It may seem that the Dyson-Lee models
are not applicable to the problem of nucleon-nucleon scattering.
However, this is not true. A delicate generalization allows one to extend the CDD methodology of constructing
analytical unitary amplitudes to systems with repulsion:

In the earlier studies it was always assumed that $\Im D(s)$ is strictly positive~\cite{CDD,DYSO}.
Simonov \cite{SIMO81} showed that the softening of this constraint to $\Im D(s) \geq 0$
is sufficient to ensure that systems with repulsive interactions got into the class of Dyson-Lee models.

The scattering of two nucleons through compound 6QS, dibaryons, can be described with a form factor $\mathcal{F}_{L}(s)$
that has simple zero(s) at $s = s_{p} > s_{0} = 4m^{2}$ like in Eq.~(\ref{FF}).
As a consequence, $\Im D(s)\sim (s-s_{p})^{2}$ in the vicinity of $s_{p}$.

If $\Re D(s_{p}) \neq 0$, the phase touches at $s=s_{p}$ one of the levels
$\delta (s)=0 \mod(\pi )$ without crossing.

However, if the real and imaginary parts of the equation (\ref{DNULL}) vanish simultaneously at one point,
the phase crosses one of the levels \textit{with negative slope}. One can verify this by expanding
the $D$ function in the neighborhood of $s = s_{p}$. Given the equation (\ref{SMAT}) and the conditions
$\Im D(s_{p})^{\prime \prime } > 0$ and $\Re D(s_{p})^{\prime } >0$, we obtain the following inequality
\[
\delta (s_{p})^{\prime }=-\frac{\Im D(s_{p})^{\prime \prime }}{2\Re
D(s_{p})^{\prime }} < 0.
\]
In the theory of potential scattering, negative slope of phase is associated with repulsion.

The compound MQS show up as poles of the $P$ matrix. By their physical characteristics,
poles of $P$ matrix can be divided into two groups:

The first group is a group of bound states and resonances:

Provided that $ D (s_{0})> 0 $ there is always one bound state. Other bound states
are generated by the compound states with masses $M_{\alpha }<\sqrt{s_{0}}$.

A distinctive feature of resonance is the condition $\mathcal{F}_{L}(s)
\neq 0$ in the neighborhood of $s=M_{ rIJ^P }^{2}$ on the second sheet of the Riemann surface.
Equation (\ref{DNULL}) can then be used to find the simple poles of the scattering matrix.

The roots of Eq.~(\ref{DNULL}), which lie on the real half-axis $(-\infty ,s_{0})$
of the second sheet of the Riemann surface, define virtual states. Their position
is determined by the masses of the compound states and interaction.

The poles of the second group are associated with the roots of Eq.~(\ref{DNULL})
\textit{on the unitary cut} in the neighborhood of $s=M_{ rIJ^P }^{2}$.
Such poles do not lead to resonant behavior of the scattering cross sections,
they do not generate poles in the $ S $-matrix and are not resonances.
Jaffe and Low have proposed to call such states "primitives" \cite{JALO}.
When resonance hits the unitary cut, its singular effect on the $ S $ matrix disappears.
In contrast to the resonances, primitives drive phase shifts down and mimic repulsion.

Resonances and primitives do not exist as asymptotic states. In the diagram technique
of Feynman, propagators of primitives are multiplied by form factors of the vertices.
Such a combination has no poles at $s=M_{ rIJ^P }^{2}$. Primitives, thus, do not propagate,
although produce a direct impact on the dynamics.

In the general case, therefore, CDD poles correspond to bound states, resonances,
and primitives. The last ones are $ P $-matrix poles associated with the zeros of
$ D $ functions on the unitary cut, which do not manifest themselves as poles of $ S $ matrix.
The primitive-type CDD poles arise in systems with repulsion.

\subsection{Nucleon-nucleon scattering in the vacuum}

Techniques developed in Ref.~\cite{CDD} can be used to
parameterize the nucleon-nucleon scattering amplitudes by
functions with correct analytical properties within the QCB model.

\subsubsection{$^{3}$S$_{1}$ channel}

The most striking feature of the $^{3}S_{1}$ nucleon-nucleon phase shift
is the vanishing of the phase shift at $T_{lab}=354$ MeV. $S$-matrix
according to Eq.(\ref{SMAT}) is unit in two cases: $\Lambda (s) = \infty$
(a CDD pole) or $\Im \Pi(s)=0$.

At the CDD poles $\delta (s)=0 \mod(\pi )$ with a positive slope,
whereas the slope at $T_{lab}=354$ MeV is negative. The CDD poles
generate resonances for $\Im \Pi(s) < 0$.

Let us consider the primitive-type CDD poles.
The imaginary part of $\Pi (s)$ is related to dibaryon decay width,
so it must be nonpositive. According to Eq.(\ref{KAPPA}),
\begin{equation}
\Im \kappa \Pi (s) = -zj_{L}(z)^{2} \leq 0.  \label{POSI}
\end{equation}
$\Im \kappa \Pi (s)$ has second order zeros, connected to the simple
zeros of $j_{L}(z)$. If $\Lambda (s)-\Re \Pi (s)$ is finite,
then $\delta(s)=0 \mod(\pi )$ but the phase does not cross the horizontal axes
$0,\pm \pi ,...$. If, however, $\Lambda
(s)-\Re \Pi (s)=0$ and $s$ is a simple zero, the phase crosses the
horizontal axes $\delta (s)=0 \mod(\pi )$ \textit{with a negative slope}.

A negative slope of the phase is associated with the repulsion.
If we want to describe the repulsion, we have to fulfill Eq.~(\ref{DNULL})
both for real and imaginary parts at real $s = s_{p} > s_{0} = 4m^{2}$.
The condition $\Im \Pi (s) \leq 0$ gives third equation
\[
\Im \Pi (s_{p})^{\prime }=0,
\]
which is, however, fulfilled automatically.

$T_{lab}=354$ MeV corresponds to $k=408$ MeV. Equation $\Im D(s) = 0$
simplifies to $kb=\pi $. We thus fix $b=1.52$ fm. This value
is close to $b=1.41-1.44$ fm obtained in Ref. \cite{SIMO81}.
The zero of the phase alone thus determines $\Pi (s)$.

In the $S$-wave,
\[
\kappa \Pi (s+i0)=-\frac{\sin z}{z}e^{iz},
\]
and we observe that $\Re \Pi (s)=0$ for $s = s_{p}$. Eq.~(\ref{DNULL})
simplifies therefore to $\Lambda (s) = 0$. The inverse $\Lambda $ function
has a pole at $s = s_{p}$. Equation $\Lambda (s) = 0$ allows to find the
dibaryon mass $M = 2\sqrt{k^{2} + m^{2}}=2047$ MeV. This value is
close to the dibaryon mass $M=2071-2081$ MeV found in Ref. \cite{SIMO81}.
It is related to the primitive-type CDD pole at $M = 3203$ MeV as shown below.

We set
\[
\gamma =-\Re \kappa \Pi (s=s_{d})>0,
\]
where $s_{d}=M_{d}^{2}$ is the deuteron pole. In the case of one
dibaryon, one can write
\begin{equation}
\kappa ^{-1} \Lambda ^{-1}=c_{p}(\frac{r_{p}}{s-s_{p}}-\frac{r_{p}}{s_{d}-s_{p}})-\frac{1}{\gamma },
\label{INVLAMB}
\end{equation}
where $c_{p}$ is a free positive definite parameter such that $\kappa c_{p}r_{p}= g^{2}$.

The constant term entering $\Lambda ^{-1}$ describes the four-fermion contact
interaction between the nucleons.

The parameterization (\ref{INVLAMB}) ensures the existence
of zero of $D$ function and of a pole of $S$ matrix at $s=s_{d}$,
respectively.

To eliminate unphysical zeros of the $D$ function, one has to
constrain $c_{p}$. We write the $D$ function in form of the
dispersion integral
\[
D(s)=\Lambda (s)-\frac{1}{\pi }\int_{4m^{2}}^{+\infty }\frac{\Im \Pi
(s^{\prime })}{s^{\prime }-s-i0}ds^{\prime }.
\]
The imaginary part of $D$ can be represented as follows
\[
\Im D(s) = \Im s \left( \frac{\Lambda^2 (s)}{|s-s_{p}|^2}
\kappa c_{p} r_{p} - \frac{1}{\pi } \int_{s_{0}}^{+\infty }
\frac{\Im \Pi(s^{\prime })}{|s^{\prime }-s|^{2}}ds^{\prime }\right) .
\]
Since $\Im \Pi (s^{\prime }) \leq 0 $, the bracket is positive for
$c_{p}>0$, in which case $D(s)$ has no complex roots in the complex
$s$-plane modulo the half-axis $(-\infty ,s_{0}).$

It remains to check the half-axis $(-\infty ,s_{0})$. The derivative
\[
D(s)^{\prime }= \frac{\Lambda^2 (s)}{(s-s_{p})^2} \kappa c_{p}r_{p}-%
\frac{1}{\pi }\int_{s_{0}}^{+\infty }\frac{\Im \Pi (s^{\prime })}{%
(s^{\prime }-s)^{2}}ds^{\prime }
\]
is positive below the threshold. $D(s)$ crosses the real axis
at $s=s_{d}<s_{0}.$ This is the unique root of the $D$ function
provided $\Lambda (s)$ has no poles at $s<s_{0}$.
Let us investigate therefore roots of $\Lambda ^{-1}(s)$.
Since $\kappa ^{-1} \Lambda^{-1}(s_{d})=-1/\gamma <0$, $\Lambda ^{-1}(s)^{\prime } <0$
and $\Lambda^{-1}(s)$ has no poles at $s<s_{p}$ by the construction,
the condition $\Lambda ^{-1}(-\infty )<0$ is sufficient to exclude
unphysical roots below the threshold. The parameter $c_{p}$ should finally
be subjected to the constraint
\begin{equation}
0<c_{p}<c_{p}^{\max }=\frac{s_{p}-s_{d}}{ \gamma r_{p}}.  \label{CONS}
\end{equation}

In Fig. \ref{fig:S} (a) we show the $^{3}S_{1}$ phase shift versus
kinetic energy of the proton in the laboratory system for $c_{p}=0.9c_{p}^{\max }$.
In Fig. \ref{fig:S} (c) $\Re D(s)$ and $\Im D(s)$ as functions of $T$ are shown.
We recall that for $pn $ system $s = s_{0} + 2m_{n} T_{lab}$ where $m_{n}$ is
the neutron masses. $\Re D(s)$ has one zero below the two-nucleon threshold,
which corresponds to the deuteron.

If the number of dibaryons is greater than one, the constraint \ref{CONS} reads
\[
\sum_{p} c_{p}\frac{r_{p}}{s_{p}-s_{d}}<\frac{1}{\gamma },
\]
where $c_{p}$ are positive.

\begin{figure} [h] 
  \centering
  \includegraphics[angle =   0,width=0.382\textwidth]{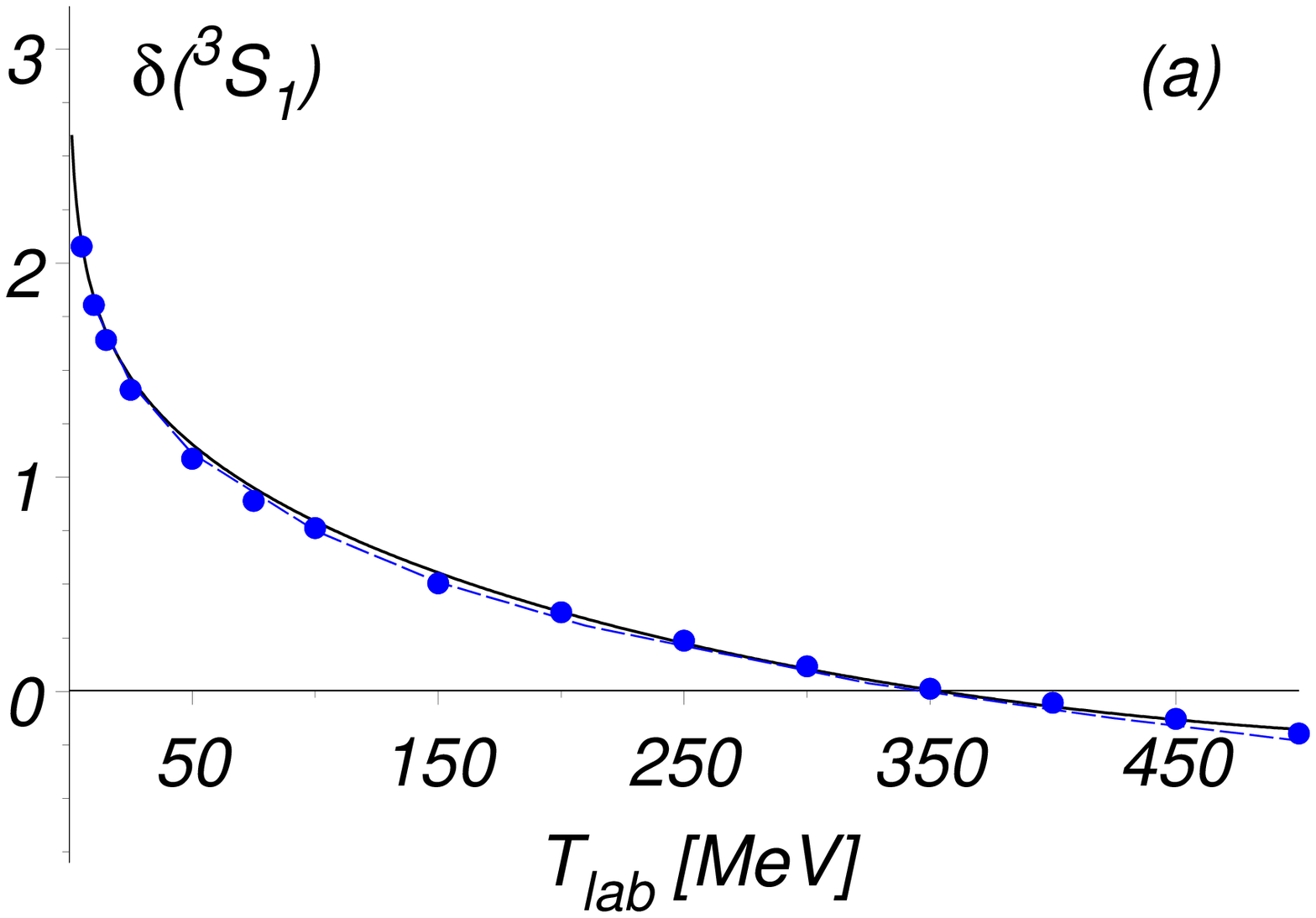}~~~~~~~~~~~~~~~~~~~~~
  \includegraphics[angle =   0,width=0.382\textwidth]{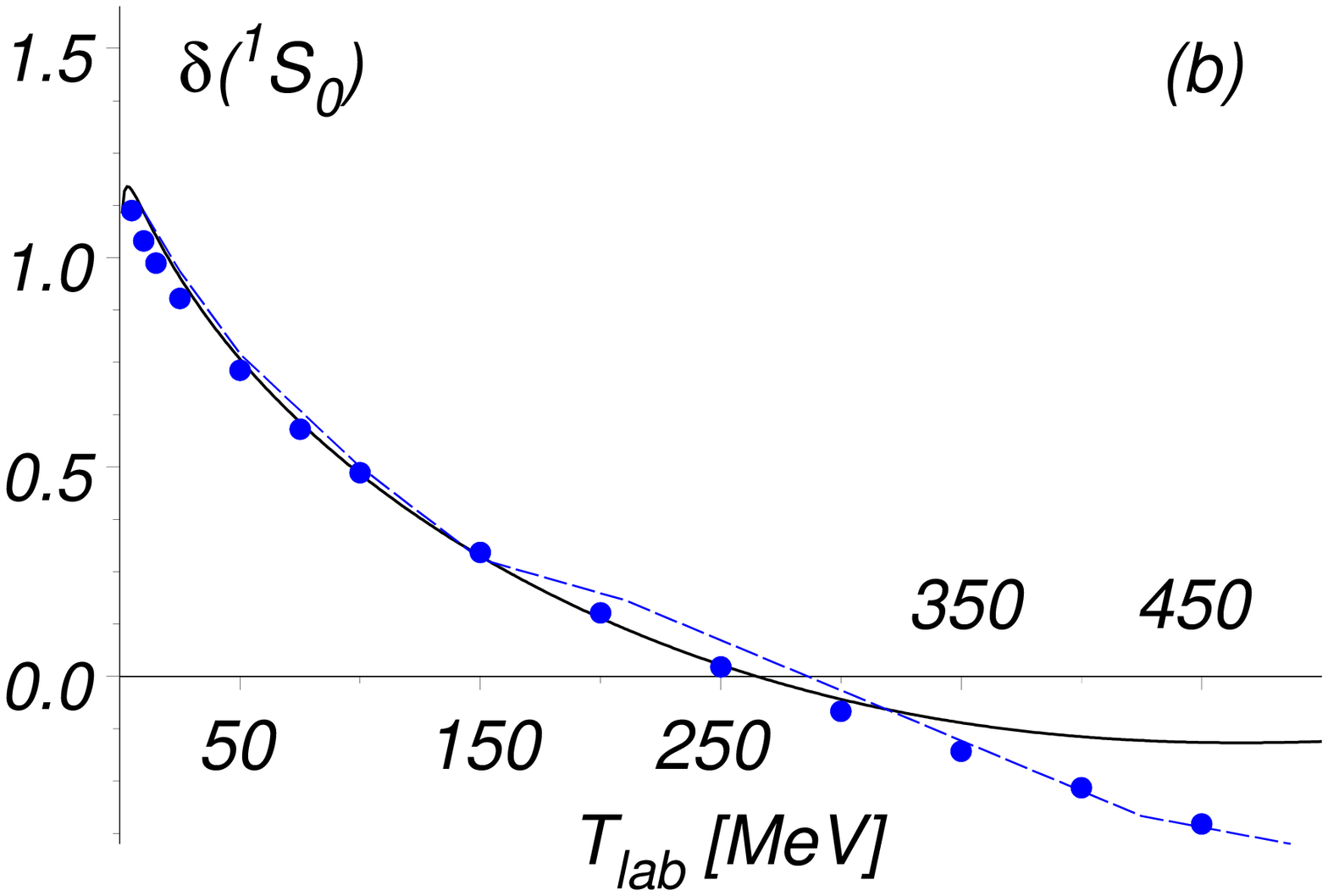} \\
  \includegraphics[angle =   0,width=0.413\textwidth]{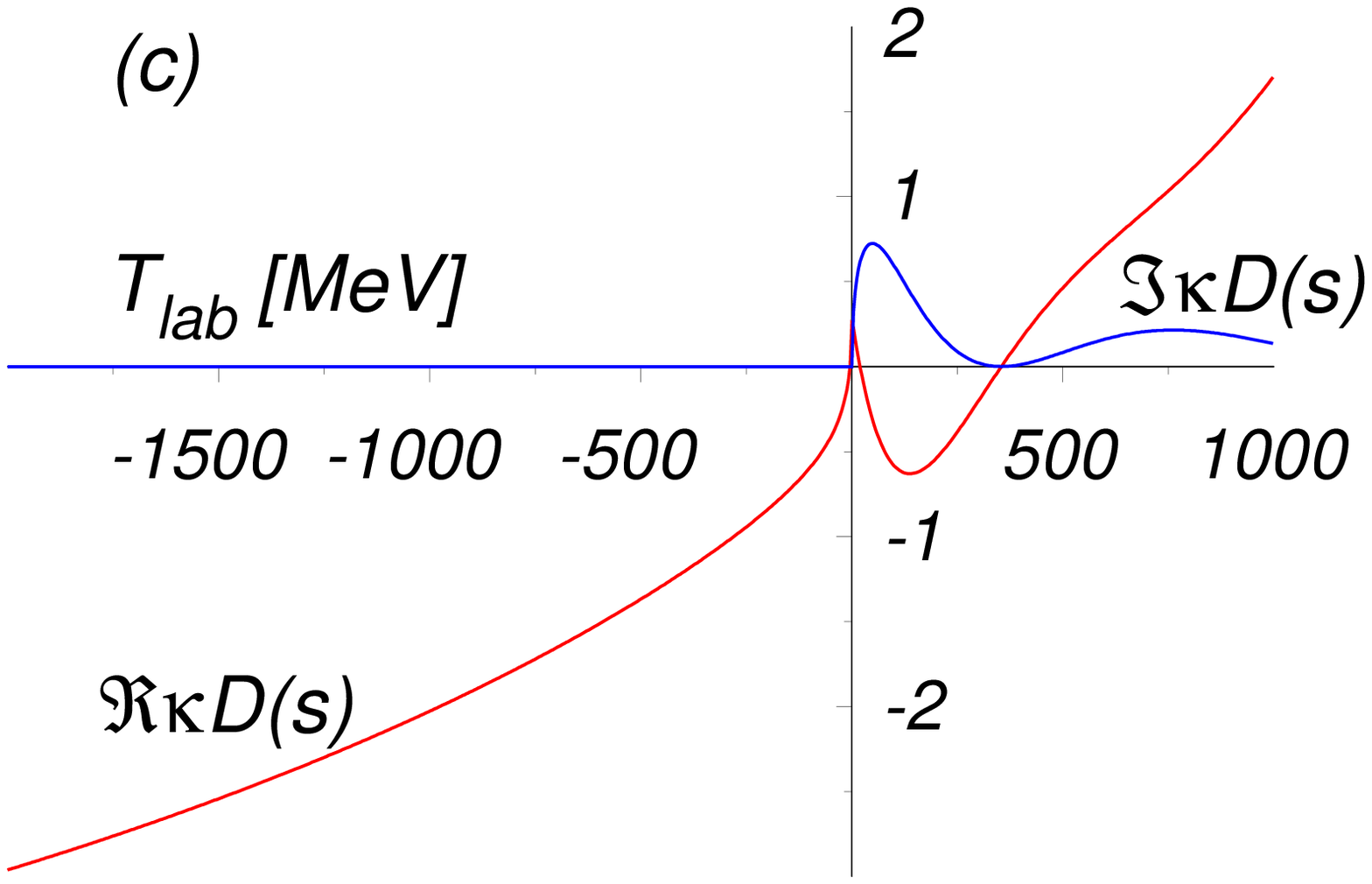}~~~~~~~~~~~~~~~~
  \includegraphics[angle =   0,width=0.413\textwidth]{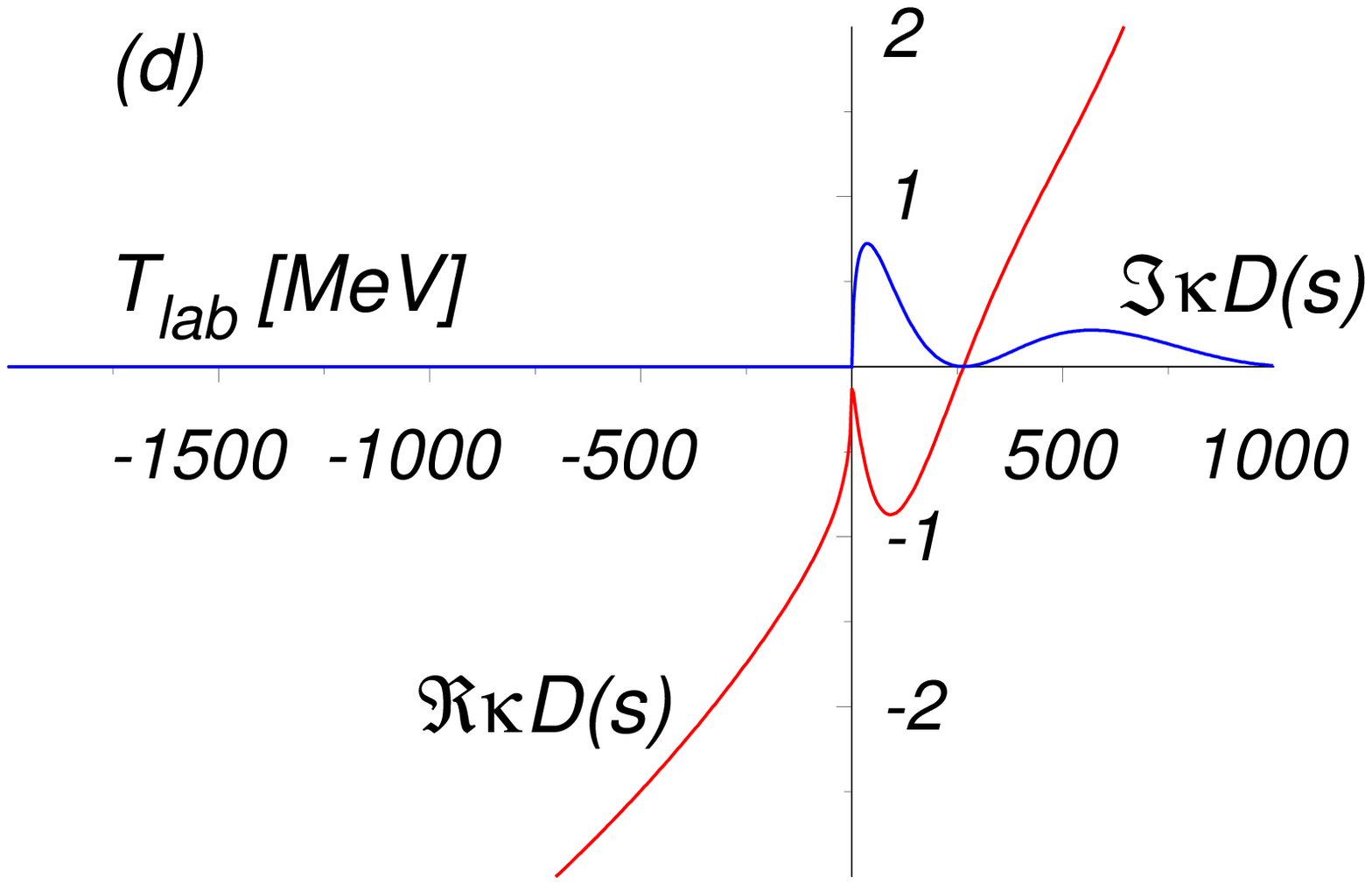}~~~~
\caption{ (color online)
$^{3}$S$_{1}$ and $^{1}$S$_{0}$ scattering phase shifts in radians (upper panel) and real and imaginary parts of
the $D$ functions (lower panel) versus the proton kinetic energy. The long-dashed curves are parameterizations
of the QCB model from Ref. \cite{SIMO81}, the solid curves are parameterizations
of the relativistic QCB model. The experimental phase shifts \cite{PHAS} are shown by the circles.
}
\label{fig:S}
\end{figure}

\subsubsection{$^{1}$S$_{0}$ channel}

The scattering phase vanishes at $T_{lab}=265$ MeV. Arguments
similar to the ones used for the $^{3}$S$_{1}$ channel lead us to
the estimates $b=1.76$ fm and $M=2006$ MeV. The numbers are close
to those obtained in Ref. \cite{SIMO81} $b=1.44$ fm and $M=2112$ MeV.
The CDD pole is localized at $M = 2916$ MeV.

The inverse $\Lambda $ function has the form of Eq.(\ref{INVLAMB})
with $s_{d}$ replaced by $s_{0} = 4m^{2}$, where $m$ is the average
mass of proton and neutron. Near two-nucleon threshold,
$\kappa D(s)=-\gamma +1+ikb+...$ From other hand, $D(s)\sim 1-i\delta
(k)+...=1-ika+...,$ where $a=23.56$ fm is the scattering length.
One has to require
\begin{eqnarray}
\gamma = - \kappa \Pi (s_{0}) + \frac{b}{a},
\end{eqnarray}
with $\Pi (s_{0})=-1$.

$D$ function of the $^{1}S_{0}$ channel has no roots for the complex values
of $s$ and its derivative is positive on the real half-axis below
two-nucleon threshold. In order to avoid the unphysical poles of $S$ matrix,
it is sufficient to require
\[
\kappa D(-\infty )=\frac{1}{-\frac{c_{p}r_{p}}{s_{0}-s_{p}}-\frac{1}{\gamma }}%
<\kappa D(s_{0})=1-\gamma <0.
\]
Since $\gamma >1$, the second inequality is fulfilled. The first one gives
\begin{equation}
c_{p}<\min (c_{p}^{\max },\frac{c_{p}^{\max }}{\gamma -1}),  \label{INEQ}
\end{equation}
where $c_{p}^{\max }=(s_{p}-s_{0})/(\gamma r_{p}).$ Since $b\ll a,$ it
reduces to $c_{p}<c_{p}^{\max }$.

In Fig. \ref{fig:S} (b) we show out fit of the $^{1}S_{0}$ phase shift
versus the proton kinetic energy with $c_{p}=0.9c_{p}^{\max }$.
In Fig. \ref{fig:S} (d), real and imaginary parts of the $D$
function are shown on the real $T_{lab}$-axis. $\Re D(s)$ has no zeros
below the threshold. $\Im D(s)$ is positive definite above two-nucleon
threshold. At $s=s_{p}$ both the real and imaginary parts of the $D$ function
vanish.

\subsubsection{$^{3}$P$_{1}$ and $^{1}$P$_{1}\ $channels}

The $^{3}P_{1}$ and $^{1}P_{1}$ phase shifts dot not cross the real axis.
So, we have to use other methods to estimate the parameter $b$ and
the dibaryon mass.

\begin{figure} [h] 
  \centering
  \includegraphics[angle =   0,width=0.382\textwidth]{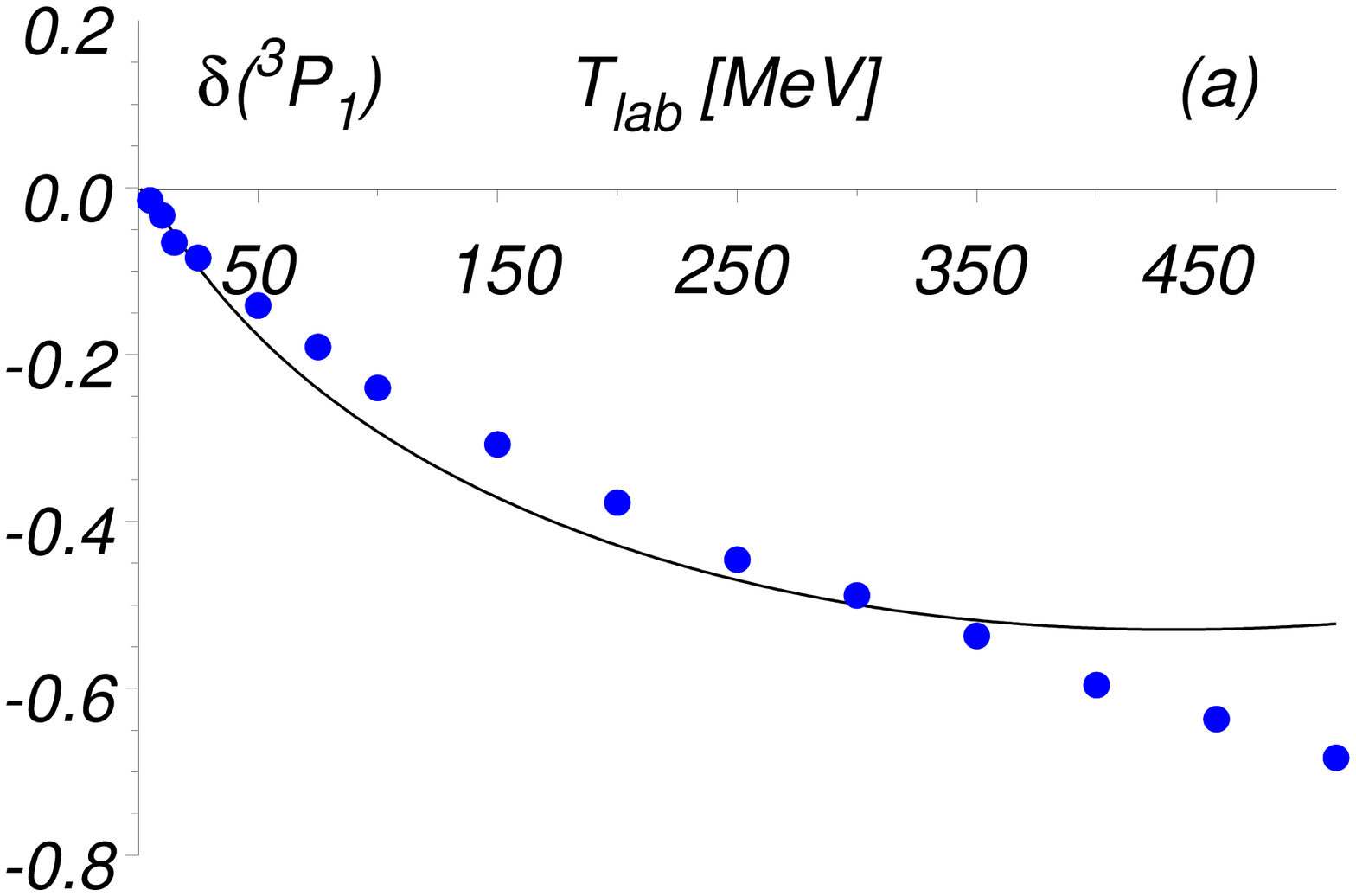}~~~~~~~~~~~~~~~~~~~~~
  \includegraphics[angle =   0,width=0.382\textwidth]{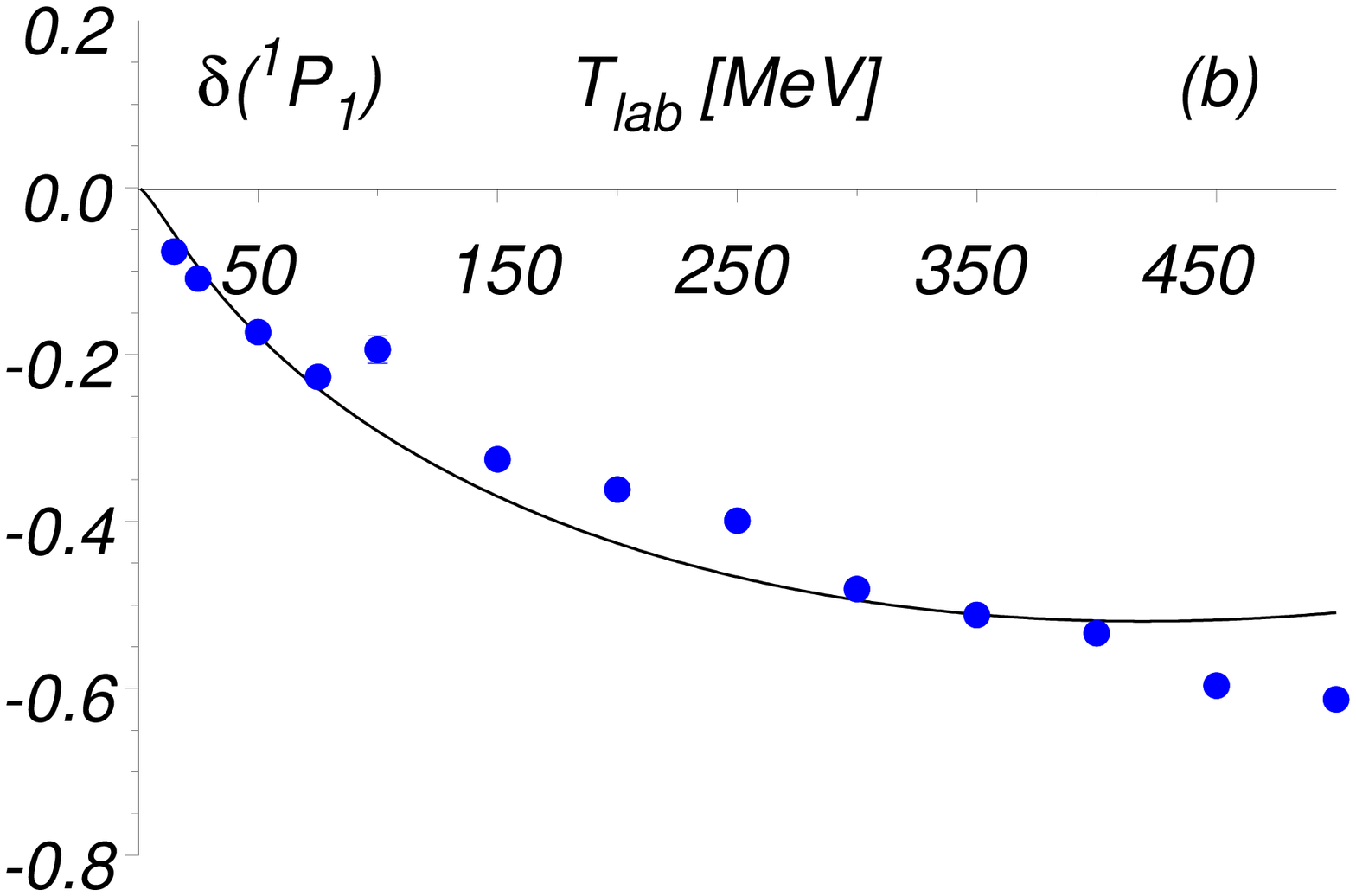} \\
  \includegraphics[angle =   0,width=0.413\textwidth]{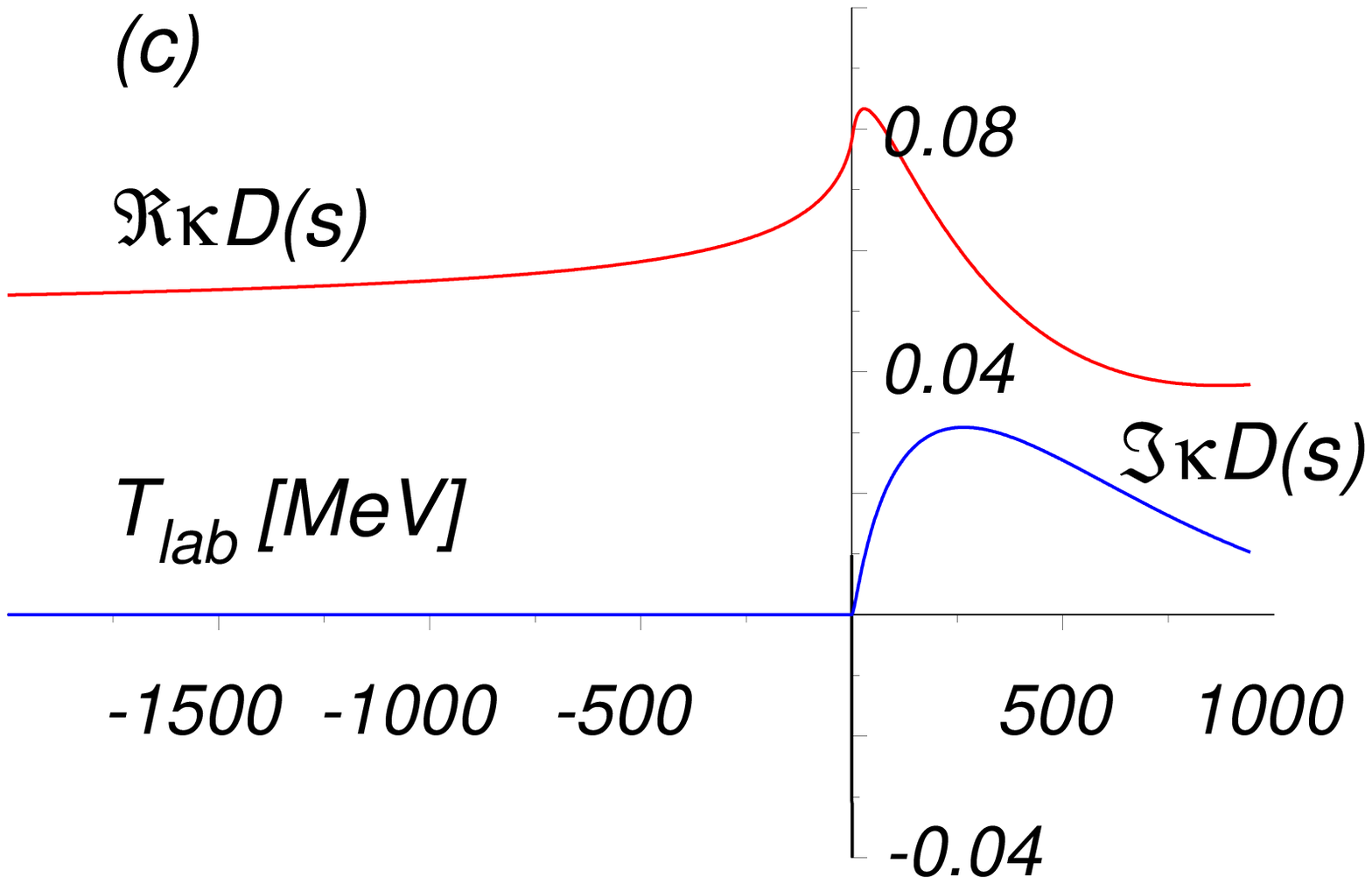}~~~~~~~~~~~~~~~~
  \includegraphics[angle =   0,width=0.413\textwidth]{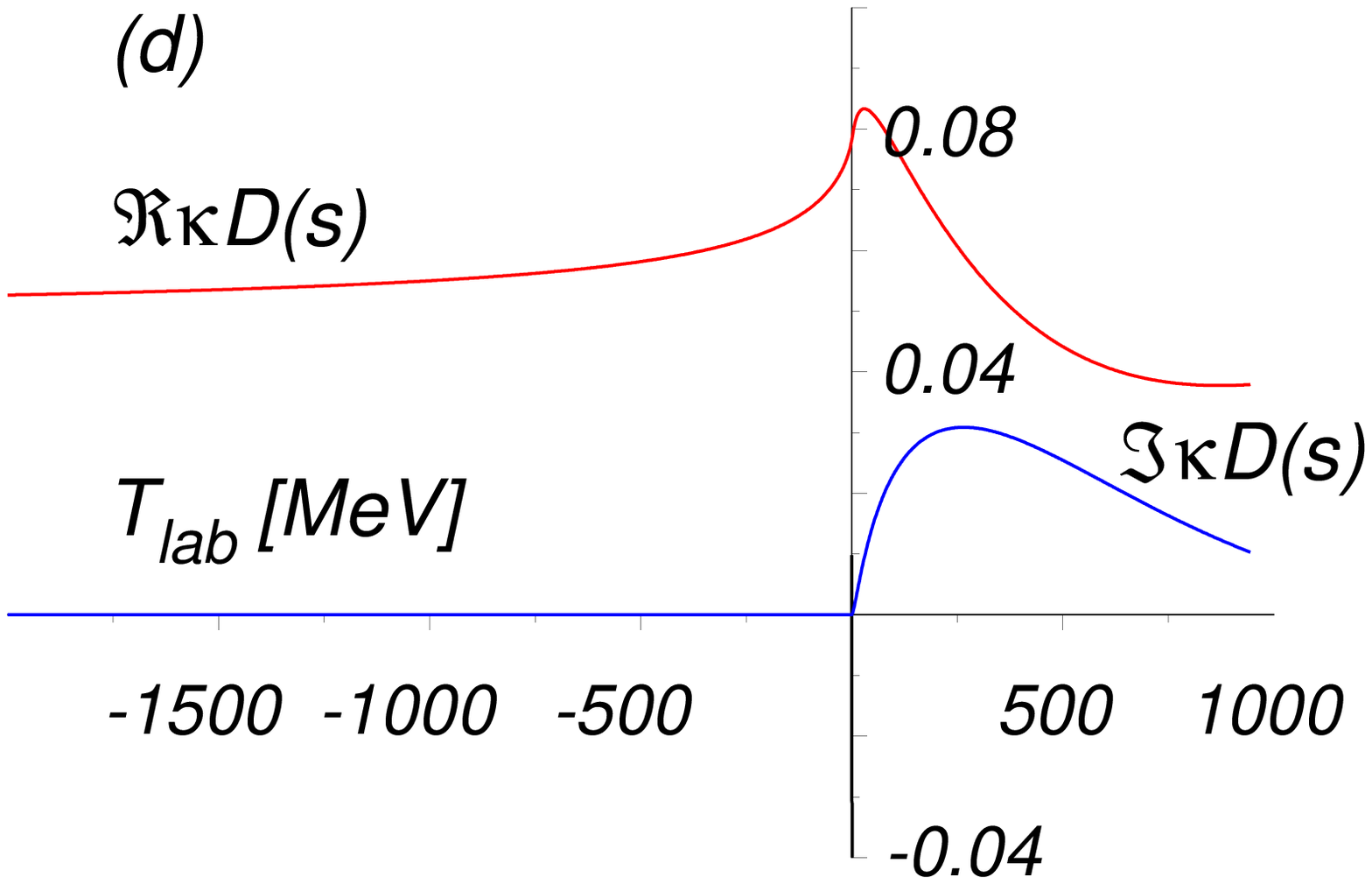}~~~~
\caption{ (color online) $^{3}$P$_{1}$ and $^{1}$P$_{0}$
scattering phase shifts in radians (upper panel) and real and
imaginary parts of the $D$ functions (lower panel) versus the
proton kinetic energy. The solid curves are parameterizations
within the relativistic QCB model. The experimental phase shifts
\cite{PHAS} are shown by circles. } \label{fig:3P11P1}
\end{figure}

The inverse $\Lambda $ function can be taken in the form of
Eq.(\ref{INVLAMB}) with $s_{d}$ replaced by $s_{0}=4m^{2}$.
At the threshold, $\kappa D(s) = -\gamma - \kappa \Pi(s_{0}) + \frac{i}{9}k^{3}b^{3}+...$.
 From other hand, $D(s)\sim 1-i\delta (k)+...=1-ik^{3}a+...,$ where $a=-1.5\times
10^{-7}$ MeV$^{-3}$ for the $^{3}P_{1}$ channel and $a=-3\times 10^{-7}$ MeV$^{-3}$
for the $^{1}P_{1}$ channel. To provide correct threshold behavior
one has to require
\begin{eqnarray}
\gamma = - \kappa \Pi (s_{0})+\frac{b^{3}}{9a}.
\label{gamm}
\end{eqnarray}
Since $a$ is negative, $\kappa D(s_{0})=-\kappa \Pi (s_{0})-\gamma >0.$
In order to remove unphysical poles, it is sufficient to require
\begin{equation*}
0< \kappa D(-\infty )=\frac{1}{\frac{c_{p}r_{p}}{s_{p}-s_{0}}-\frac{1}{\gamma }}%
< \kappa D(s_{0})=- \kappa \Pi (s_{0})-\gamma .
\end{equation*}
In the case of $\gamma <0$, this equation provides no constraints on $c_{p}$.
In the case $0<\gamma <- \kappa \Pi (s_{0})$, $\kappa ^{-1} \Lambda ^{-1}(s_{0})= -\frac{1}{\gamma }<0$,
$\kappa ^{-1} \Lambda ^{-1}(-\infty )= \frac{c_{p}r_{p}}{s_{p}-s_{0}}-\frac{1}{\gamma }>0,$ and
$\Lambda ^{-1}(s)^{\prime }<0.$ There is therefore a pole in the $D$ function,
and zero of the $D$ function on the real half-axis $(-\infty ,s_{0})$.
The case of $0<\gamma <- \kappa \Pi (s_{0})$ is thus not useful.

Let us assume $\gamma =0.$ The value of $b$ for the $^{3}P_{1}$ channel can
be found to be $b=1.51$ fm. The dibaryon of mass $M=2214$ MeV that makes
$\Lambda (s_{p})$ to vanish ensures the crossing of the level $\delta (s)=-\pi $
at $T_{lab}=732$ MeV. In such a case, the repulsion is largely
overestimated. The linear interpolation of the experimental data e.g.
gives $\delta (s_{p})\sim -1$. If $M \neq 2214$ MeV, the phase remains negative and
touches one of the levels $\delta (s)=0 \mod(\pi )$ at $s=s_{p}$. A
better description is possible for lower value of $b$ and greater
value of $M$, respectively. However, lower values of $b$ require
positive $\gamma $ which is not acceptable.

\begin{figure} [h] 
  \centering
  \includegraphics[angle =   0,width=0.382\textwidth]{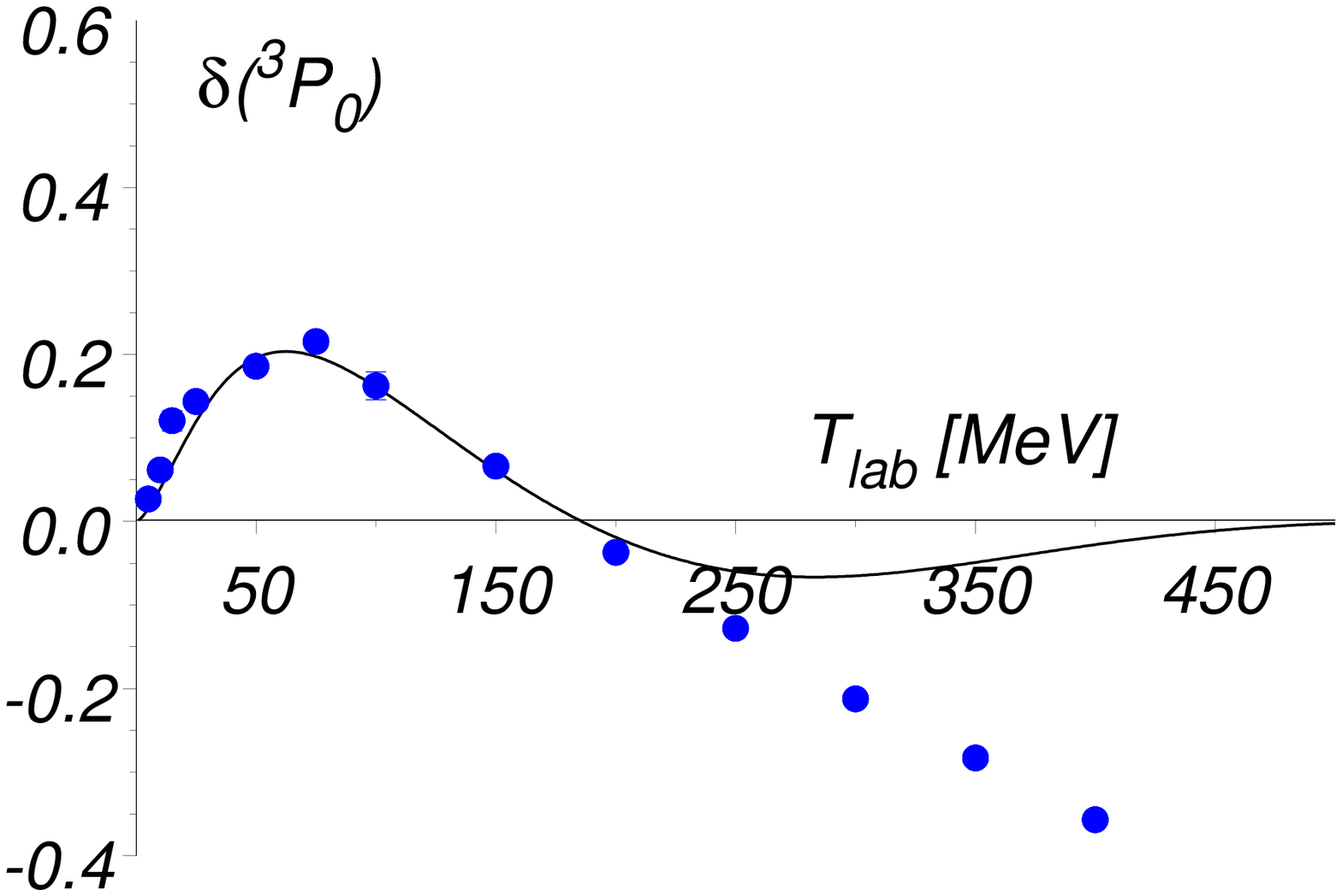} \\
  \includegraphics[angle =   0,width=0.413\textwidth]{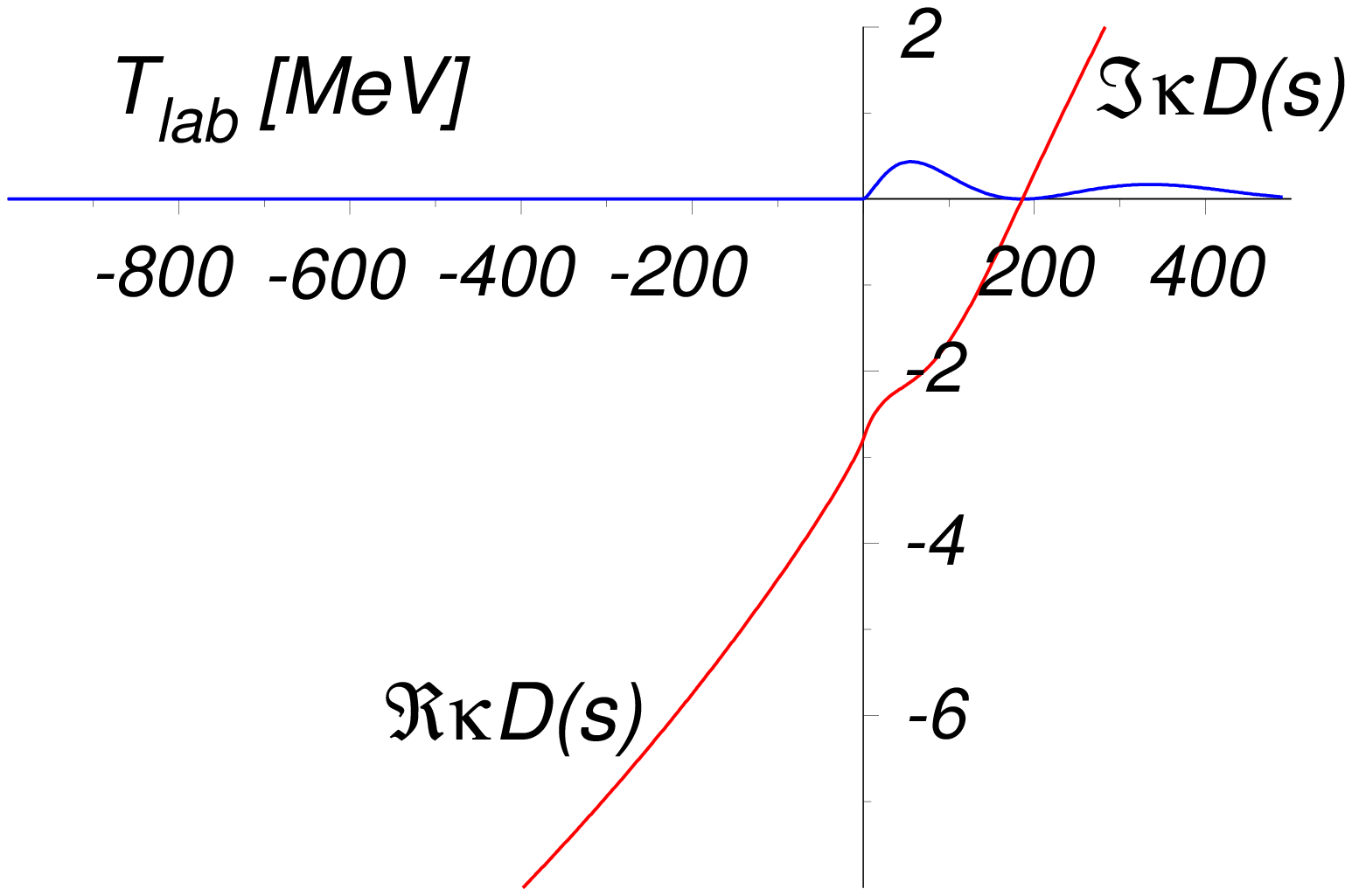}~~~~
\caption{ (color online) $^{3}$P$_{0}$ scattering phase shifts in
radians (upper panel) and real and imaginary parts of the $D$
functions (lower panel) versus the proton kinetic energy. The
solid curves are parameterizations within the relativistic QCB
model. The experimental phase shifts \cite{PHAS} are shown by
circles. } \label{fig:3P0IT}
\end{figure}
We thus consider more general expression for the imaginary part of
$\Pi (s)$:
\begin{eqnarray}
\kappa \Im \Pi (s) &=&-zj_{l}(z)^{2}\frac{\varkappa ^{2}}{\varkappa ^{2}+z^{2}},
\end{eqnarray}
where $\varkappa $ is a parameter. $\Pi (s)$ is restored from the dispersion integral.
The explicit analytical expression for $\Pi (s)$ can be found e.g. with the help of MAPLE.
For $\varkappa =\infty$, we recover previous formulas. A finite $\varkappa$
leads to reduced values of $\Pi (s_{0})$ in the absolute value and the parameter $b$, respectively, keeping $\gamma $ negative.
For $L=1$,
\begin{eqnarray*}
\kappa \Pi (s_{0})=\frac{1}{6\varkappa ^{3}}(-3+3\varkappa ^{2}-2\varkappa ^{3}+3\left(
1+\varkappa \right) ^{2}e^{-2\varkappa }).
\end{eqnarray*}
One can notice that $0\leq - \kappa \Pi (s_{0})\leq 1/3$.

The same arguments as before lead us to the conclusion of absence of
unphysical poles of the $S$-matrix in the complex $s$-plane for negative
$\gamma $.

A reasonable fit in the $^{3}P_{1}$ channel can be obtained without the CDD poles and primitives.
$c_{p}=0$ implies that the only important interaction is
the contact four-fermion interaction described by $\gamma $. We use $\gamma =-1/20$,
$b=0.93$ fm and $\varkappa^{2}=1/3$ which corresponds to $\kappa \Pi (s_{0})=-0.028$.

In Fig. \ref{fig:3P11P1} (a) we show phase shift of the $^{3}P_{1}$ channel.
Shown in Fig. \ref{fig:3P11P1} (c) are real and imaginary parts of the $D$ function.

The phase shift in the $^{1}P_{1}$\ channel behaves similar to the phase shift
in the $^{3}P_{1}$ channel. We use therefore precisely the same set of the parameters.
The scattering length appears to be a factor of two lower that the experimental one,
however, the overall description of the data is better than with the increased
scattering length. In Fig. \ref{fig:3P11P1} (b) the phase shift in the $^{1}P_{1}$
channel is shown. In Fig. \ref{fig:3P11P1} (d) the real and imaginary parts of the $D$ function
are shown.

\subsubsection{$^{3}$P$_{0}\ $channel}

The parameterization of the phase shift in the $^{3}P_{0}\ $channel is similar to
that in the $^{1}S_{0}$ channel. Here, we do not modify $\Im \Pi (s)$ and set
$\varkappa = \infty$. The values of $b=3$ fm and the dibaryon mass $M=1969$ MeV are
found from the position $T_{lab}=186$ MeV of zero of the phase shift.
The primitive-type CDD pole appears at $M = 2650$ MeV. The parameter (cf. Eq.~(\ref{gamm}))
\begin{equation}
\gamma =\frac{1}{3}+\frac{b^{3}}{9a}
\end{equation}
can be fixed by the scattering length $a=2.8\times 10^{-7}$ MeV$^{-3}.$ We consider, however,
the scattering length as a fitting parameter. The general agreement with the
data is better when slope of the phase shift at $T_{lab}=0$ is two times lower than the
experimental one. Fig. \ref{fig:3P0IT} (a) shows the phase shift as compared to the
experimental data as functions of the proton kinetic energy. Shown
in Fig. \ref{fig:3P0IT} (b) are the real and imaginary parts of the $D$ function.


\subsection{Discussion}

In Sects. V-VII we constructed and analyzed the effective Lagrangian to describe the two-nucleon forces
based on their $s$-channel dibaryon exchange mechanism. Dibaryons in our approach are the primitives,
and not resonances. They manifest themselves as poles of $P$ matrix or, equivalently, as zeros of $D$
functions on the unitary cut. Resonances with the energy-dependent width have similar properties,
when the width vanishes at the resonance mass, while far from the mass of the resonance the width is finite.
The effective Lagrangian was formulated for the nucleons, which are well-defined asymptotic states,
and for the primitives, which are decoupled of the two-nucleon channel on the mass shell, and, apparently,
are not asymptotic states. Note that although the resonances are not asymptotic states, they are often
included in effective Lagrangians. Despite the unusual properties of the primitives,
there is no intrinsic difficulty in applying the diagram technique for calculation of various processes
involving primitives, based on their effective Lagrangian.

In Sect. V we gave a complete list of relativistic two-nucleon
currents with the angular momentum $J=0,1$, through which the
primitives are coupled to two nucleons in the low-$J$ partial
waves. We found five primitives with the lowest quantum numbers
$(I,J^P)=(0,1^{\pm}),(1,0^{\pm})$, and $(1,1^-)$ and discussed
their effect on the nucleon-nucleon interaction in Sect. VII. In the
various versions of the QCB model, the $S$-wave nucleon-nucleon
phase shifts are fitted by the primitives excellently and we
confirm it again. We considered the $P$-wave scattering also. We
found good qualitative agreement with the experiment. The
description of the channels $^3 P_1$ and $^1 P_1$ requires the
contact four-fermion interaction only, without primitives, whereas
the channel $^3 P_0$ gives an evidence for the primitive. We
suppose more accurate parameterization of the experimental phases
can be achieved by means of the $s$-channel dibaryon exchange only
without invoking the $t$-channel meson-exchange mechanism.

$P$-matrix analysis has been successful in describing the various processes, including nucleon-nucleon interaction.
This formalism is built into the effective Lagrangian with an appropriate choice of nucleon-nucleon-dibaryon
form factors and four-fermion vertices.

In the limit of low density we reproduce results of optical potential model. Lorentz structure of
the nucleon self-energy operator is different from the OBE RMF models. In our approach, vector and scalar
parts of the self-energy operator are of the same order of magnitude.

Solutions of the self-consistent system of equations for the nucleon and dibaryon propagators
and $G$-matrix in nuclear matter will be presented elsewhere \cite{FUTU}.

\section{Models of nuclear matter}
\setcounter{equation}{0}

The nuclei are known to be stable under normal conditions on Earth, while the collection of nuclei
forms the ground state in QCD at finite baryon charge.

The range of densities and temperatures $\rho \sim 10^{39}$
cm$^{-3}$ and $T \sim 10^{12}$ K, where the phase transition in nuclear matter takes place
with chiral symmetry restoration and deconfinement, can be studied experimentally in heavy-ion collisions.
Supercold nuclear matter exists in the interiors of massive neutron stars.

Quark matter appears in compression of nuclear matter. New forms of nuclear matter can
exist at intermediate densities, such as hyperon matter, crystalline neutron matter,
pion and kaon condensation and a Bose condensation of dibaryons.

In the field of atomic physics, modifications of the properties of ordinary atoms at density
$\rho \sim 10^{10} - 10^{18}$ cm$^{-3}$ and temperature $T \sim 10^{4}$ K are studied since the mid-1930's (for a review see \cite{SOBE,ROPK}).

To ensure the stability of nuclear matter under conditions
$T \lesssim 10\div 20$ MeV and $\rho \sim 0.16 \div 0.30$ fm$^{-3}$, the $NN$ forces must satisfy necessary and sufficient saturation conditions
(see, e.g., the review of F. Calodgero and Yu. A. Simonov \cite{Cal+Sim}). For
local and velocity-independent potentials these conditions require a strong repulsive core at small distances, $r\la 0.4$ fm.

The quark-hadron phase transition in nuclear matter has been
discussed first by Zel'dovich \cite{ZELD65}. By comparing
pressures of the ideal Fermi gases of nucleons and quarks, he
arrived at a conclusion that the composite structure of the
nucleons requires a short-range repulsion between the nucleons.
His discussion was based on the Le-Shatelie - Brown principle.
Earlier he showed that the stiffest, consistent with causality
nuclear matter EoS corresponds to exchange by a massive vector
particle between the nucleons \cite{ZELD61}. The $\omega$-meson
exchange mechanism is the most typical and challenging feature of
the modern OBE models.

As discussed above, at short distances $NN$ force becomes nonlocal -- in OBE picture due to finite size of
nucleons and mesons -- and in the QCB picture due to formation of MQS.
It is remarkable that nonlocal $NN$ interaction, even when everywhere attractive,
easily passes the requirements of saturation, while in the local OBE case one needs a very large coupling of $\omega$-meson,
producing repulsive core.

The realistic OBE models of nuclear matter split into three groups:

Relativistic mean field (RMF) models are based on the mean-field approximation
to account for the meson-exchange interactions between nucleons, $\sigma$- and
$\omega$-mesons ensure the saturation property. The mean field models are the
phenomenological ones.

Dirac-Brueckner-Hartree-Fock (DBHF) approach is a field-theoretic scheme,
which is based on accurate experimental data of nucleon-nucleon interaction.
Nuclear many-body problem is solved microscopically by the summation of wide subclass
of Feynman diagrams. The scheme claims to be parameter-free description
of the nuclear matter EoS.

The variational approach is proposed in \cite{PAND}. This uses the potentials of the nucleon-nucleon
interaction, whose characteristics are extracted from the nucleon-nucleon scattering and descriptions
of light nuclei. Ground-state wave function of infinite nuclear matter is constructed with the use
of the variational principle. Nuclear EoS is claimed to be parameter-free as well.

The DBHF models and the variational approach are successful in the description
of nuclear matter at saturation starting essentially from the first principles,
however, with an additional tuning, e.g., of $3N$ forces in the variational approach.

In addition to the above three methods, the non-relativistic BHF calculations
\cite{muether00}, phenomenological theories with density dependent interactions
such as Gogny or Skyrme forces \cite{reinhard04} and various modifications of
the RMF scheme are widely used for modeling nuclear matter also.

A critical test of nuclear models is possible on the basis of astrophysical observations
on the properties of massive neutron stars.

The most massive pulsars reported in the literature are PSR B1516+02B with
the mass of $1.96^{+0.09}_{-0.12}~\mathrm{M_{\odot}}$ and PSR J1748-2021B with
the mass of $2.74 \pm 0.22~\mathrm{M_{\odot}}$ \cite{FREI07}.
The mass of rapidly rotating neutron star in the low mass X-ray binary 4U 1636-536 is found
to be $M=2.0 \pm 0.1 ~\mathrm{M_{\odot}}$ \cite{BARR05}. The mass of the X-ray
source EXO 0748-676 is limited to $M \geq 2.10 \pm
0.28~\mathrm{M_{\odot}}$.
These observations indicate clearly that the $\beta$-equilibrated nuclear matter EoS
is stiff and, furthermore, eliminate the soft EoS constructed, e.g., on the basis
of the classical Reid soft core model \cite{PAND}.

The new degrees of freedom lower the energy of matter and make EoS softer. Scenarios with phase transitions
shift the masses to the region of even smaller values. Existence of quark matter and other forms of exotic
matter in the cores of neutron stars was questioned \cite{OZEL06} (see, however, \cite{Ozel:2010fw}).

The so-called realistic models of neutron stars usually ignore hyperon channels, for example,
the reactions $\Sigma^- \to n + e + \bar{\nu}_e$. In the RMF models \cite{GLEN91,GLEN96,BURG03},
which take hypernuclear data into account, the inclusion of the $\beta$-equilibrium balance for hyperons reduces
the limiting mass of neutron stars by $0.5 \div 0.8~\mathrm{M_{\odot}}$.
These results are in agreement with recent calculations \cite{ISHI08,SCHA08,DAPO08}.
The $\beta$-equilibrium brings thus difficulties to reproduce the observed high masses of neutron stars.

The problem can be solved by assuming the existence of light weakly interacting bosons,
the possibility of the existence of which is discussed in some generalizations of the
Standard model \cite{Krivoruchenko:2009kx}. A more conventional and perhaps more realistic approach is to modify
the baryon-baryon interaction at short distances, taking the quark-gluon degrees of freedom
into account. It is based on the formalism of the $P$ matrix and QCB model discussed in the previous sections.

\subsection{Relativistic mean field models}

Development of many-body theory in nuclear physics was motivated by the problem of describing
heavy nuclei and, since 1970's, neutron stars. The change in mass of the nucleon
in nuclear matter has been introduced in the pioneering work of Walecka \cite{Walecka:1974qa}.
Mean-field (MF) and relativistic mean field (RMF) approximations were proposed to describe the
dense nuclear matter self-consistently to all orders in density, but in tree level of perturbation theory.
MF and RMF were very successful in describing the basic properties of finite nuclei and nuclear matter
near saturation density. RMF models are currently the most popular models of nuclear
matter, although they do not provide a strict connection with the phenomenology of nucleon-nucleon scattering.

Since mid-1970's, RMF models have been improved in many directions: by including nonlinear self-interactions of $\sigma$-, $\rho$-, and
$\delta$-mesons, introduction of the density-dependent coupling constants and vertices with higher derivatives,
in order to describe the energy, radius, surface thickness of nuclei, neutron skins, spin-orbit interaction of nucleons,
as well as restrictions on flows obtained in heavy-ion collisions, and fulfill constraints imposed by
astrophysical observations of neutron stars.

In RMF models a significant decrease of the effective mass of the nucleon was found at a density of saturation.
This effect was predicted purely phenomenologically. The development of QCD confirms the prediction and the effect
has attracted much attention, since the decrease of the nucleon mass is connected to the partial
restoration of chiral symmetry.

In the simplest version, the effective Lagrangian contains the nucleon fields
and $\sigma$- and $\omega$-mesons:
\begin{eqnarray}
\mathcal{L} &=& \bar{\Psi}(i\hat{\nabla} - m_N - g_{\sigma} \sigma - g_{\omega} \hat{\omega} ) \Psi  \nonumber \\
&+& \frac{1}{2}(\nabla_{\mu} \sigma \nabla_{\mu} \sigma - m_{\sigma}^2 \sigma^2) \nonumber \\
&-& \frac{1}{4}\mathcal{F}_{\mu \nu}\mathcal{F}_{\mu \nu} + \frac{1}{2}
m_{\omega}^2 \omega_{\mu} \omega_{\mu},
\end{eqnarray}
where $\mathcal{F}_{\mu \nu} = \nabla _{\nu} \omega_{\mu} - \nabla _{\mu}
\omega_{\nu}$, $\hat{\nabla} = \gamma_{\mu} \nabla_{\mu}$, etc.

Property of saturation is easily seen from the following argument: Suppose that low-density
attraction of the scalar field dominates the $\omega$ exchange. With increasing density,
the scalar charge of the nucleons, generating a scalar field, is suppressed by the Lorentz factor
(The scalar charge of a particle placed in a box disappears in the ultrarelativistic limit.
The vector charge, of course, remains constant.) In infinite-density limit, $\sigma$-meson decouples from the system
of nucleons, but vector fields survive. As a result, the attraction at low densities is replaced by
the $\omega$-induced repulsion
at high densities. In some intermediate-density regime, repulsion balances the attraction, this density is
the density of saturation where the substance is in equilibrium.
The coupling constants $g_{\sigma}$, $g_{\omega}$, and the $\sigma$-meson mass are determined from
fitting the properties of nuclear matter at saturation density.

RMF models offer low binding energy of nuclear matter at saturation due to cancellation
between the strong repulsive vector and strong attractive scalar potentials.
These two potentials are equal in magnitude to about a quarter of the nucleon rest mass.
The scalar part reduces the nucleon mass and increases the influence of small components
of the Dirac bispinors.

RMF models allow the most precise description of the nuclear properties with 5
- 15 parameters.

\subsection{Dirac-Brueckner-Hartree-Fock approximation}

DBHF approximation scheme is based on realistic meson-exchange models. As compared
with the RMF model, it takes additional requirements of fermionic statistics and in-medium
modification of two-body $ T $-matrix into account. DBHF successfully reproduces the properties
of nuclear matter at saturation mainly from first principles, using the interaction, based on data
from nucleon-nucleon scattering.

Shown in Fig. \ref{fig:dbhf} is the Bethe-Salpeter equation for the in-medium two-body scattering
$ T $-matrix (called reaction $ G $-matrix). The difference from the vacuum equations is associated with the change
of nucleon propagators in the loop and Pauli blocking, which restricts the admissible regions in the phase
space of nucleons.

\begin{figure}[!htb]
\begin{center}
\includegraphics[angle=0,width= 8 cm]{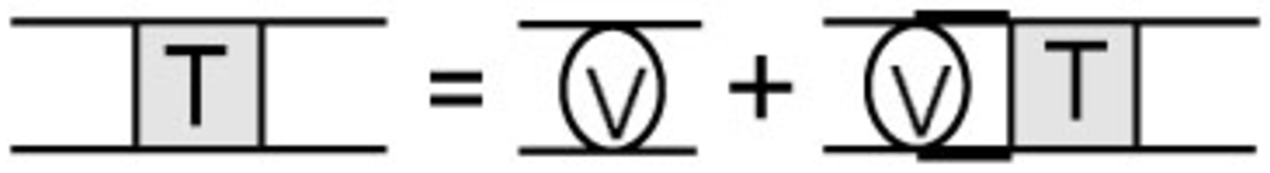}
\end{center}
\caption {System of equations for in-medium $T$-matrix. $V$ is meson-exchange potential.
The bold lines in the loop are in-medium nucleon
propagators. } \label{fig:dbhf}
\end{figure}

The nucleon self-energy can be found by solving equation shown in Fig.
\ref{fig:dbhf2}.

\begin{figure}[!htb]
\begin{center}
\includegraphics[angle=0,width= 7.5 cm]{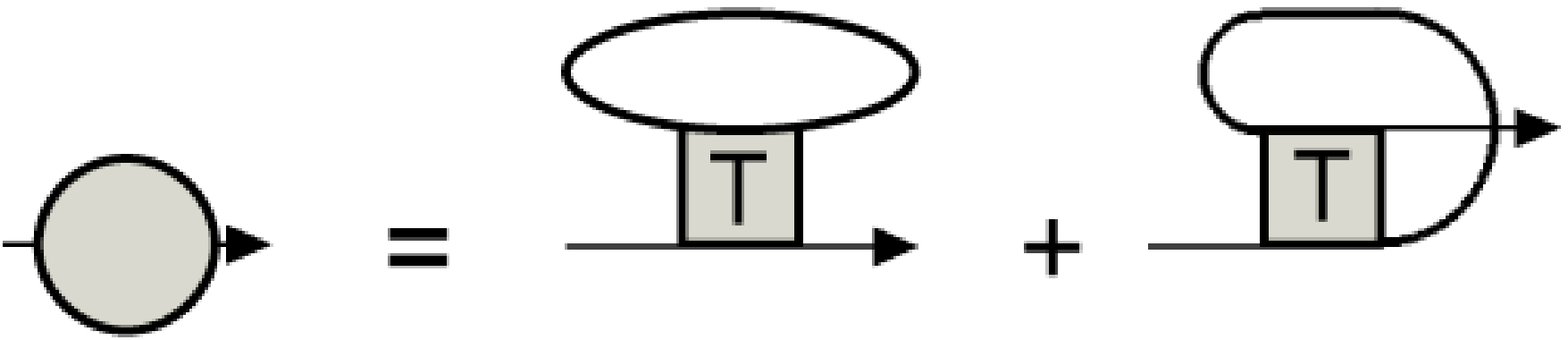}
\end{center}
\caption {System of equations for in-medium nucleon self-energy operator.
The bold lines in the loops are in-medium nucleon propagators. $T$ is
in-medium $T$-matrix of Fig. \ref{fig:dbhf}. }
\label{fig:dbhf2}
\end{figure}

Equations in Figs. \ref{fig:dbhf} and \ref{fig:dbhf2} constitute selfconsistent
system of equations of the DBHF approximation.

Examples of the DBHF calculations can be found in Refs.~\cite{boelting99,honnef,dalen04,vanDalen:2005ns}.
DBHF scheme is phenomenological only marginally: The ladder approximation is used to solve the relativistic
Bethe-Salpeter equation, negative-energy nucleons are neglected,
nucleon resonances are neglected also.

\subsection{Variational approach}

Estimates of $ G $-matrix with realistic potentials show that $ G $-matrix has a sharp peak with respect
to the difference of distances between the nucleons in the initial and final states. This means that in the process of scattering the distance
between nucleons is not much changed, and for the variational calculations the dependence of the trial wave
function on the relative momentum of the nucleon pairs can be neglected. The trial wave function has the form
\begin{equation}
\Psi = \prod_{i<j} \sum_{\ell} f^{\ell}(r_{ij}) \mathcal{P}^{\ell} \mathcal{A}
\prod_{k} \phi_{k}(r_{k}), \label{JAST}
\end{equation}
where $\mathcal{P}^{\ell}$ is projection operator to state of the pair $ij$
with the orbital momentum $\ell$, $\mathcal{A} $ is the antisymmetrization
operator, $\phi(r_{k})$ are uncorrelated wave functions of the nucleons.
The correlation function $f^{\ell}(r)$ is equal to one when the distance
between the nucleons increases, but less than unity, when $r$ is reduced.

The Jastrow wave function (\ref{JAST}) is used for the calculation
of the ground state energy from the cluster expansion
\begin{eqnarray}
E &=& \frac{<\Psi |H |\Psi>}{<\Psi| \Psi>} \nonumber \\
  &=& \sum_{i} C_{i} + \frac{1}{2} \sum_{ij} C_{ij} + \ldots, \label{PREN2}
\end{eqnarray}
where
\begin{eqnarray}
C_{k}   &=& <\phi_{k}| \frac{p^2_{k}}{2m} |\phi_{k}>, \nonumber \\
C_{ij}  &=& \frac{<\phi_{ij}| H_{2} |\phi_{ij}>}{<\phi_{ij}|\phi_{ij}>} - C_{i}
- C_{j}, \label{PREN3}
\end{eqnarray}
etc. Here, $H_{2}$ is two-body Hamiltonian, $\phi_{ij}$ is two-body wave function.

The smallness of the correlation effect in the two-body nucleon wave function,
\begin{equation}
\rho_0 \sum_{\ell}\int d \mathbf{r} (1 - f^{\ell}(r))^2 \sim 0.1,
\end{equation}
at the saturation density $\rho_0 = 0.16$ fm$^{-3}$ makes it possible to truncate the cluster expansion.

Recent examples of the variational calculations can be found in Refs.~\cite{akmal98,Mukherjee:2006rt}.

\subsection{In-medium modifications of hadrons}

In-medium modification of the properties of hadrons affects the nuclear matter EoS. In the RMF models,
vector mesons play a special role. The interaction mediated by vector mesons at high densities
leads asymptotically to the stiffest EoS of nuclear matter. The effect of the vector meson on the energy
density of nuclear matter can be estimated by averaging the Yukawa potential \cite{ZELD61}
\begin{equation}
\varepsilon_{I}=\frac{1}{2 V}\int d\mathbf{x}_{1}d\mathbf{x}_{2}\rho (\mathbf{x}_{1})%
\frac{g^{2}}{4\pi }\frac{e^{-\mu r}}{r}\rho (\mathbf{x}_{2}),  
\end{equation}
where $\rho (\mathbf{x}_{1})=\rho (\mathbf{x}_{1})\equiv \rho $ is baryon number density,
$r=|\mathbf{x}_{2}-\mathbf{x}_{1}|$, $g$ is coupling constant, $\mu $ is vector meson mass, and $V$ is normalization volume.
A simple integration gives
\begin{equation}
\varepsilon_{I} = \frac{g^{2}\rho ^{2}}{2\mu ^{2}}. \label{OMEGA}
\end{equation}

The RMF models are used to study the modifications of the properties of vector mesons
in nuclear medium \cite{Chin:1977iz,BGP,Mishra,Chen03}. Assuming that the
mass of mesons depends on the density, the contribution of the $\omega $-meson to pressure can be found to be
\begin{equation}
P_{I}=\frac{g^{2}\rho ^{2}}{2\mu ^{2}}\left( 1-\frac{2\rho }{\mu }\frac{%
\partial \mu }{\partial \rho }\right) .  \label{PRESS}
\end{equation}
This equation shows that the pressure decreases if the meson mass increases with density.
The problem of modifying the mass of vector mesons in nuclear matter was studied in experiments
on the dilepton production in heavy-ion collisions. Simulations of the dilepton production
using transport models and the subsequent comparison with experiment gives a clear indication
of the presence of collisional width, however, a noticeable shift of the masses of $\rho$- and $\omega$-mesons
is not registered \cite{SANT08}.

\section{Nuclear matter with multiquark baryons. Nonrelativistic treatment}
\setcounter{equation}{0}

For two-nucleon contribution to the nuclear matter one starts with the
Bethe-Brueckner  series which is summed up by the operator equation \cite{22}
\be G_2(W) =V_2-V_2Q_2 e^{-1}_2 G_2(W) \label{4.1} \ee where $G_2(W)$ is the
reaction matrix of two nucleons in nuclear matter $V_2$ is the two-body
interaction operator (potential), $Q_2$ is the projection operator on states
outside of the Fermi sphere, and finally $e_2$ is energy denominator \be
e_2|\vep\ven> = (E_p + E_n-W) |\vep\ven>.\label{4.2}\ee Eq. (\ref{4.1}) is an
equivalent of two-body $t$-matrix (Lippmann-Schwinger) equation for the case
when two nucleons are inside nuclear matter.

One can also write equations for the pair wave functions. For the plane wave
$\Phi_{\vep\ven}$ and  actual w.f. $\Psi_{\vep\ven}$ one has \be
\Psi_{\vep\ven}= \Phi_{\vep\ven}- Q_2e^{-1}_2 G_2 \Phi_{\vep\ven} \label{4.3}
\ee Separating out the c.m. motion, $\Psi_{\vep\ven}=\frac{e^{i\veK\veR}}{V}
\psi_{\vep\ven}$, and keeping only relative coordinates in $\psi_{\vep\ven},
\varphi_{\vep\ven}\equiv \exp (i \vek\ver)$, one has the Bethe -Goldstone
equation (for nonlocal in general interaction) \be \psi_{\vep\ven} (\ver)
=e^{i\vek\ver}- \int F_2 (\ver, \ver') V_2 (\ver', \ver^{\prime\prime}; E)
\psi_{\vep\ven} (\ver^{\prime\prime}) d\ver' d\ver^{\prime\prime} \label{4.4}
\ee where
\begin{widetext}
\begin{eqnarray}
F_2 (\ver, \ver') &\equiv& \int \frac{d\vek}{(2\pi)^3} Q_2 (\vek, \veK) [e_2
(\vek,\veK)]^{-1} \exp [i\vek
(\ver-\ver')],\label{4.5} \\
Q_2 (\vek, \veK) |\vep,\ven> &=& \left\{ \begin{array} {ll} |\vep,\ven>,& {\rm
for}~~
|\frac12 \veK\pm \vek |>k_F,\\
0,& {\rm otherwise,}\end{array}\right.\label{4.6} \\
e_2 |\vep\ven\ran &=& [ E(\frac{\veK}{2} +\vek) + E (\frac{\veK}{2} -\vek)-W]
|\vep,\ven\ran\equiv e_2 (\vek, \veK)| \vep,\ven\ran,\label{4.7}
\end{eqnarray}
\end{widetext}
and \be \veK=\vep+\ven,~~ \vek=\frac12 (\vep-\ven). \label{4.8} \ee From
$\psi_{\vep\ven}$ one easily finds the matrix $G$,\be \lan \vel,\vem| G_2|
\vek,\ven\ran =\frac{1}{V} \delta_{\veK\veK'}\lan \vek' |
G_2|\vek\ran\label{4.9}\ee \be \lan \vek' |G_2 |\vek\ran =\int e^{-i\vek'\ver}
V_2(\ver, \ver') \psi_{\vep\ven} (\ver') d\ver d\ver'.\label{4.10} \ee

Inserting the separable interaction $V_2 (\ver,\ver')=v_{hqh}=\sum_\nu
\frac{f_\nu (r) f_\nu (r')}{E-E_\nu}$, one obtains from (\ref{4.4}) the
following system of algebraic equations \be (f_\nu \psi ) = f_\nu (\vek)
-\sum_{\nu'}\lambda_{\nu\nu'} (f_{\nu'} \psi)\label{4.11}\ee where we have used
notations \be (f_\nu\psi) =\int d\ver f_\nu (\ver) \psi_{\vep\ver} (\ver),~~
f_\nu (\vek) =\int d\ver e^{i\vek\ver}f_\nu (\ver),\label{4.12}\ee \be
\lambda_{\nu\nu'} =\frac{ \int f_\nu (\ver) F_2 (\ver, \ver') f_{\nu'} (\ver')
d\ver d\ver'}{E-E_{\nu'}}\equiv
\frac{\Lambda^{(2)}_{\nu\nu'}}{E-E_{\nu'}}.\label{4.13}\ee

In a similar way one easily calculates $\lan \vek' |G|\vek\ran$, \be \lan \vek'
|G_2 | \vek\ran =\sum_\nu \frac{f_\nu(\vek')
(f_\nu\psi)}{E-E_\nu}.\label{4.14}\ee Keeping only one $MQS$ level in
(\ref{4.11}), (\ref{4.14}) one has \be (F_\nu \psi) =\frac{f_\nu (\vek)}{1+
\lambda_{\nu\nu}},\label{4.15}\ee \be \lan \vek'| G_2 |\vek\ran = \frac{ f_\nu
(\vek') f_\nu (\vek)}{E-E_\nu +\Lambda_{\nu\nu}^{(2)}}.\label{4.16}\ee
The case of more $MQS$ levels is discussed in the next sections.
Note, that $E$ in all equations in this paper denotes the total energy of
two-nucleon system, including its  masses, which  finally in (\ref{4.20})
coincides with $W$, and one-particle energies $E(\vep)$ also include rest mass
in kinetic term. To make (\ref{4.1}) selfconsistent, one should also define the
single-particle energies  $E_n, E_p$ in (\ref{4.2}) through the same matrix
operator $G_2$ \cite{22}. \be
 E(\vep)= T(\vep)+ U_2(\vep,\rho)\label{4.17}
\ee \be
 U_2(\vep,\rho) = \sum_{p'\leq p_F} \lan \vep\vep' | G_2 (W= E(\vep)
 + E (\vep')) | \vep\vep'\ran_a
\label{4.18} \ee where subscript $a$ implies antisymmetrization of state $
(\vep \vep'>$.

 To calculate the energy per nucleon one can use the BHF
 approximation
 \be
 \frac{E}{A}(\rho)= \frac{\lan T\ran}{A}+ \frac{\lan
 U_2\ran}{2A}
\label{4.19} \ee
 where we have defined
\be
 \lan U_2\ran = \sum_{p,p'\leq p_F} \lan \vep\vep'|G_2 (E(\vep) +
 E(\vep'))|\vep\vep'\ran_a.
\label{4.20} \ee

 We now turn to the contribution of higher $MQS$ states, and start
 with the case, when the only interaction present in nuclear
 matter is the $MQS$ composed of $3 N$ quarks, $N>2$, while all
 other interactions  with $N'\neq N$ are absent. Denoting the
 corresponding reaction matrix for $N$ nucleons $G_N$, one can
 write similarly to (\ref{4.1})
 \be G_N(W) = V_N-V_NQ_Ne^{-1}_N G_N (W)
\label{4.21} \ee
 where $Q_N,e_N$ are natural generalizations of $Q_2, e_2$ and
 $V_N$ is
 \be
 V_N=\sum_\nu\frac{f_\nu^{(N)} (\ver_1,... \ver_N)
 f_\nu^{(N)}(\ver'_1,... \ver'_n)}{E-E_\nu}
\label{4.22} \ee

 It is clear, that generalization of Eqs. (\ref{4.11}-\ref{4.16})
 to the case of $N$-nucleon interaction is straightforward if one
 replaces there one relative momentum $\vek$ by the  set of $N-1$
 relative momenta $\{\vek_\alpha\}, \alpha=1, ... N-1,$ and
 similarly generalize $Q_2 (\vek, \veK)$ and $e_2 (\vek, \veK)$ to
 $Q_N(\{ \vek_\alpha\} , \veK), ~~ e_N (\{ \vek_\alpha\} , \veK)$.
 For the function $F_N$ one defines $N-1$ relative coordinates $\{
 \ver_\alpha\}, \alpha=1, ... N-1,$ and one has instead of
 (\ref{4.5})
\begin{widetext}
 \be
 F_n(\{\ver_\alpha\},\{\ver'_\alpha\}) =\int
 \prod^{N-1}_{\alpha=1} \frac{d\vek_\alpha}{(2\pi)^3} Q_N(\{ \vek_\alpha\} , \veK) e_N^{-1} (\{ \vek_\alpha\} ,
 \veK)\exp (i\sum^{N-1}_\alpha
 \vek_\alpha(\ver_\alpha-\ver'_\alpha)).
\label{4.23} \ee
\end{widetext}

 The answer (\ref{4.16}) for the case of one $MQS$ made of $3N$
 quarks is
 \be
 \lan \{\vek'_\alpha\}| G_N | \{\vek_\beta\}\ran = \frac{f_\nu
 (\{\vek'_\alpha\}) f_\nu (\{ \vek_\beta\})}{E-E_\nu +
 \Lambda^{(N)}_{\nu\nu}}\label{4.24}\ee
 and $\Lambda^N_{\nu\nu'}$ is expressed through $F_N$ in the same
 way as in (\ref{4.13}) with the obtains replacement $\ver\to
 \{\ver_\alpha\}, \ver' \to \{\ver'_\beta\}$. Finally in
 (\ref{4.19}) one replaces $\frac{\lan U_2\ran}{2A}$ by $\frac{\lan
 U_N\ran }{N!A}$,where $\lan U_N\ran $ is
 \be \lan U_N\ran =\sum_{\{p_i\} \leq p_F} \lan \{\vep_i\}|G_N
 (W_N)|\{\vep_i\}\ran_a\label{4.25}\ee
 and the set of one-particle momenta $\{\vep_i\},~~ i=1,...N$ is
 inside the Fermi sphere, while $W_N= \sum^N_{i=1} E(p_i)$.

\section{$G$ matrix}
\setcounter{equation}{0}

Diagrams in Figs. \ref{fig1}, \ref{fig3}, and \ref{fig5} show
a closed system of equations that can be solved e.g. by iterations
starting from the vacuum solution.

The projection operator $Q_2$ on states outside of the Fermi sphere
influences scattering of the nucleons as described in Sect. IV.
In the relativistic theory, the dibaryon self-energy Fig. \ref{fig1}
is represented in the form of dispersion integral.
The phase space of two nucleons is restricted by $Q_2$ and weight of
the dispersion integral modifies, accordingly. The invariant part of the dibaryon
self-energy operator can be written in the form
\begin{equation}
\Pi_{IJ^P}(s,v) = \frac{1}{\pi}\int_{s_{0}}^{+\infty}B_F(s^{\prime},v)\frac{\Im \Pi_{IJ^P}(s^{\prime}) ds^{\prime}}{s^{\prime} - s - i0}
\label{9182}
\end{equation}
where $B_F$ is the Pauli blocking factor and $v$ is the dibaryon velocity.
In terms of the nucleon momenta, the imaginary
part of $\Pi_{IJ^P}(s)$ is the same as in the vacuum. The only modification
of $\Pi_{IJ^P}(s)$ comes from $B_F \neq 1$ and the in-medium nucleon mass
$m^{*} = m + \Sigma _{S}(p_{+})$ where $p_{+} = (\sqrt{m^{*2}+\mathbf{p}^2},\mathbf{p})$.

\subsection{Pauli blocking}

One can give explicit relativistic form of the blocking factor (\ref{4.6}).
It can be used with Eq.~(\ref{9182}) for evaluation of the in-medium
dibaryon self-energy.

\begin{figure} [h] 
  \centering
\includegraphics[angle = 0,width = 11 cm]{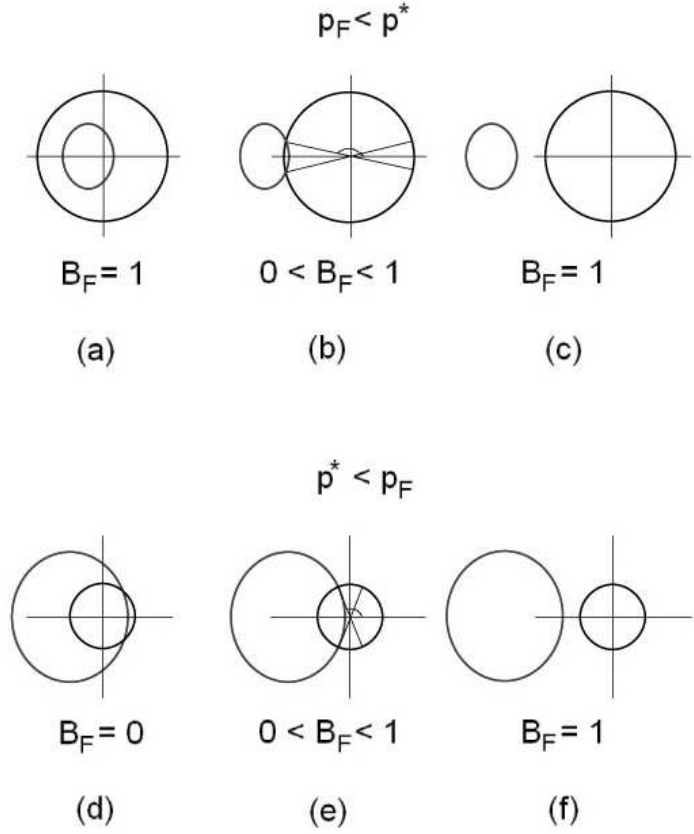}
\caption{
Possible configurations of the boost-deformed Fermi sphere and the decay sphere
in the rest frame of two nucleons.
\textit{Upper panel $p_{F} < p^{*}$}:
Condition $-p^{*} < -\gamma (p_{F}+vE_{F})$ is fulfilled in the case (a).
The Fermi sphere is inside of the decay sphere. The decays are allowed and $B_{F} = 1$.
The spheres intersect each other in the case (b) when $- \gamma (p_{F} + vE_{F}) \leq -p^{*} < \gamma (p_{F}-vE_{F})$
and $0 < B_{F} < 1$.
In the case (c), the relative velocity is high, $\gamma (p_{F}-vE_{F}) \leq  -p^{*}$,
the spheres do not intersect each other and the decays are allowed $B_{F} = 1$.
\textit{Lower panel $p^{*} < p_{F}$}:
In the case of $p^{*}  \leq  \gamma (p_{F}-vE_{F})$, the decay sphere is
inside of the Fermi sphere and the decays are forbidden $B_{F} = 0$.
The case (d) corresponds to intersection of the
spheres for $\gamma (p_{F}-vE_{F}) < p^{*}$ and $E^{*} < E_{F}/\gamma$
where $B_{F} = 0$ also.
In the case (e), $-p^{*} < \gamma (p_{F}-vE_{F}) < p^{*}$ and $E_{F}/\gamma < E^{*} < E_{F}$,
the decays are possible and $0 < B_{F} < 1$. In the case (f),
the relative velocity is high to provide $\gamma (p_{F}-vE_{F}) \leq -p^{*}$,
in which case the spheres do not intersect and the decays are allowed $B_{F} = 1$.
}
\label{fig:BF}
\end{figure}

In the rest frame, $K$, of the substance the Fermi sphere is described by
equation
\begin{equation}
p_{x}^{2}+p_{y}^{2}+p_{z}^{2} = p_{F}^{2}.
\end{equation}
In the co-moving frame, $K^{\prime }$, of the dibaryon the decay products
have momenta
\begin{equation}
p_{x}^{\prime 2}+p_{y}^{\prime 2}+p_{z}^{\prime 2}=p^{*2}.
\label{ds}
\end{equation}
We assume that the dibaryon moves in the positive direction of the $x$-axis.
Using boost along $x$, one can write equation for the Fermi sphere in $K^{\prime }$:
\begin{equation*}
\gamma^{2}(p_{x}^{\prime 2} + v E^{\prime})^{2} + p_{y}^{\prime 2} + p_{z}^{\prime 2} = p_{F}^{2},
\end{equation*}
where $v > 0$ and $p_{x}= \gamma(p_{x}^{\prime } + v E^{\prime})$,
$p_{y} = p_{y}^{\prime }$, $p_{z}=p_{z}^{\prime }$. Since $p^{\prime }$ belongs
to both the Fermi sphere and the decay sphere (\ref{ds}),
$E^{\prime } = E^{*} = \sqrt{m^{* 2}+p^{* 2}}$ and
\begin{equation*}
\gamma ^{2}(p_{x}^{\prime }+vE^{*})^{2}+p^{*2}-p_{x}^{\prime 2}=p_{F}^{2},
\end{equation*}
from which we derive
\begin{equation*}
p_{x\pm }^{\prime }=\frac{-E^{*}\pm E_{F}\sqrt{1-v^{2}}}{v},
\end{equation*}
where $E_{F}=\sqrt{m^{* 2}+p_{F}^{2}}$. Note that $p_{x-}^{\prime } < -p^{*}$.
There exists therefore only one intersection of the spheres.
The nucleon self-energy depends on the absolute values of the momenta,
so directions of the nucleon velocities have equal weights. The modified
dispersion law implies that the nucleon mass entering $E^{*}$ and $E_{F}$
is the effective nucleon mass $m^{*}$.

The critical values $p_{x+}^{\prime }=\pm p^{*}$ at which the Fermi sphere
touches the decay sphere give equations
\[
\pm p^{*}=\frac{-E^{*}+E_{F}\sqrt{1-v^{2}}}{v}.
\]

The blocking factor is defined by fraction of the area of the decay sphere outside of the Fermi sphere.
The possible cases are shown in Fig. \ref{fig:BF}. The blocking factor becomes
\begin{equation}
B_{F}=-\cos \theta _{+},
\label{BLOCK}
\end{equation}
where
\[
\cos \theta _{+}=\max (-1,\min (0,\frac{p_{x+}^{\prime }}{p^{*}})).
\]

\subsection{Heterophase nuclear matter with a Bose condensate of dibaryons}

Small perturbations of $s$-channel $NN$ interaction, as a rule, shift primitives from
the unitary cut and turn them into resonances. One can expect therefore that the effects
of the environment transform the primitives into the resonances. Such dibaryons can provide
deep modification of nuclear matter at densities above the critical density for the formation
of such dibaryons on mass shell. Before proceeding to discuss realistic models
it is worthwhile to consider a mixed dibaryon-nucleon phase in ideal gas approximation.

Suppose that we fill the box with neutrons, as shown in Fig.~\ref{boxx}. Pauli principle allows
neutrons with increasing density consistently occupy higher energy levels. We can continue this
process until the chemical potential of neutrons at the surface of the Fermi sphere will not be
higher than half of dibaryon mass. In the approximation of an ideal gas, the chemical potential
of fermions is equal to the Fermi energy, while the chemical potential of bosons is simply
the mass of boson $m_D$.

When the Fermi energy of neutrons becomes higher than half the mass of the boson, it gets energetically
advantageous for the two neutrons to merge and form a dibaryon. The critical density at which the process
of the dibaryon condensation starts is thus determined by the mass of the lightest dibaryon. Above the critical density,
the chemical potential of the nucleons $\mu _n$ is frozen at $\mu _n^{\max } = m_D/2$ due to chemical equilibrium
with respect to conversions $ NN \leftrightarrow D $.

\begin{figure} [htb!]
\includegraphics[width = 0.382\textwidth]{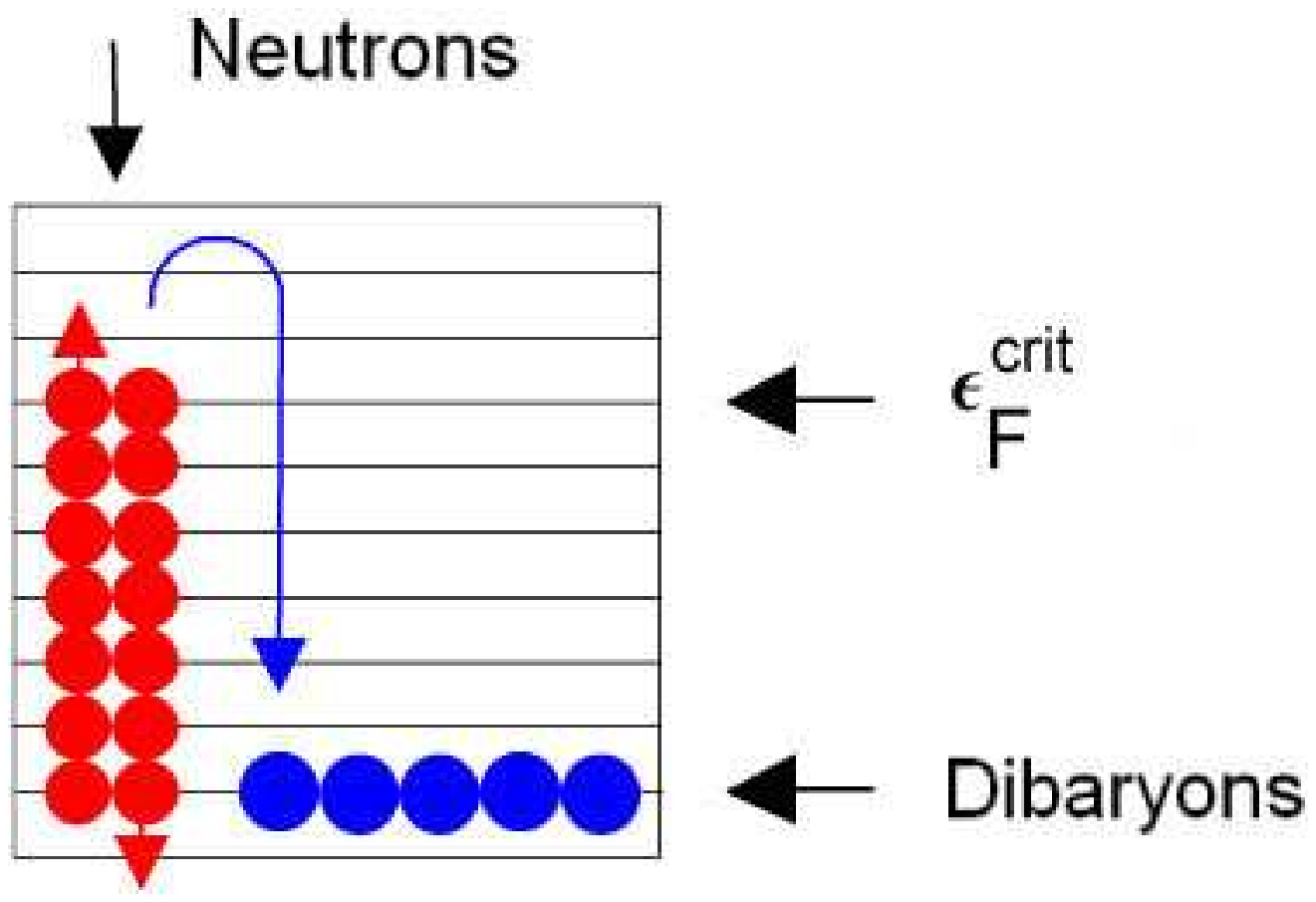}
\caption{Illustration of transition to the heterophase state of nuclear matter with an admixture of dibaryons.}
\label{boxx}
\end{figure}

Dibaryons are Bose particles and form a Bose condensate. Dibaryons in the condensate
have zero velocity, so they do not collide with the boundary of the box
and do not contribute to the pressure.

On the other hand, the Fermi energy of neutrons is frozen, so the pressure does
not increase with increasing density (see Fig. \ref{press}). As a result,
a binary mixture of neutrons and dibaryons loses its elasticity. Nuclear matter
with such properties can not protect neutron stars against gravitational
compression and the subsequent collapse.

\begin{figure} [htb!]
\includegraphics[width = 0.382\textwidth]{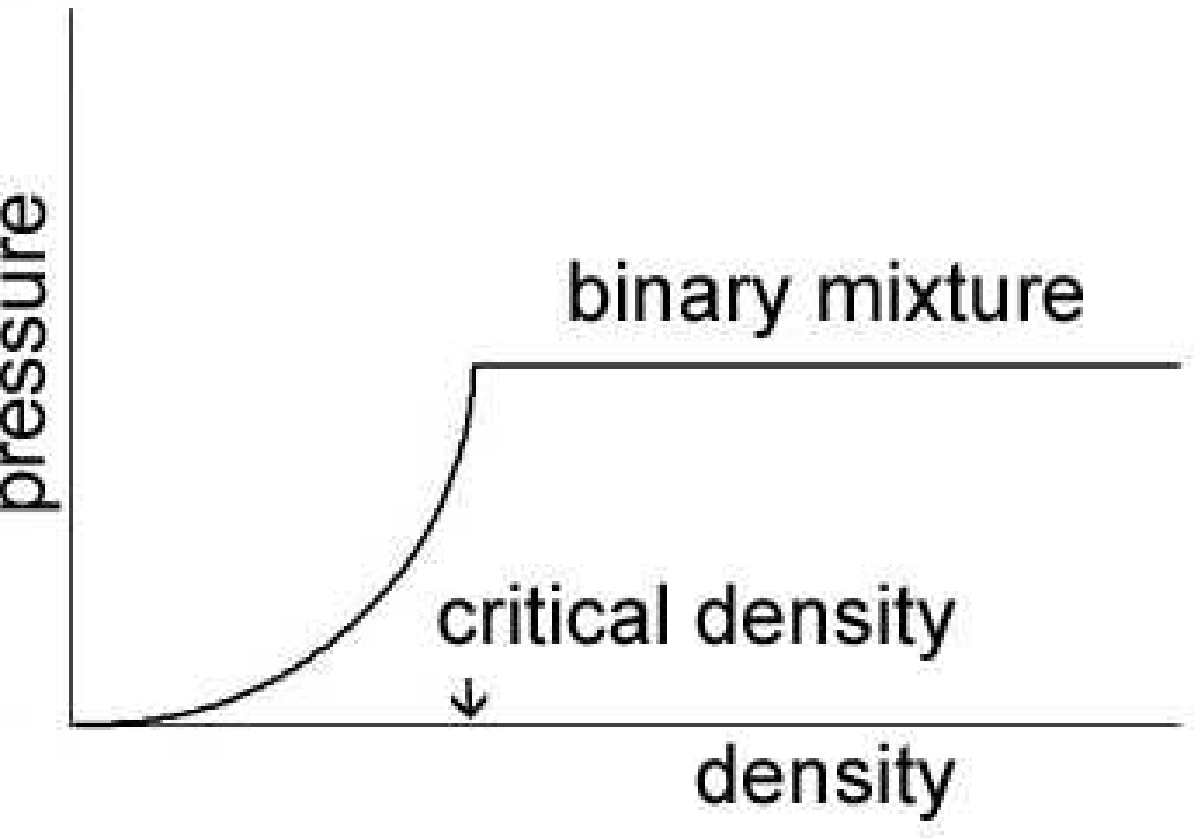}
\caption{Schematic view of EoS of the heterophase nuclear matter
with an admixture of dibaryons in the absence of interactions.}
\label{press}
\end{figure}

Incompressibility of nuclear matter at saturation density is well known experimentally,
it is certainly not equal to zero, so the dibaryon Bose condensate, apparently, does not exist
in normal nuclei (at least in the ideal gas model
and some simple modifications of this model, see below). Because of the charge
symmetry of nuclei, $\mu _n=\mu _p$. Assuming further that the shell potential of dibaryons
is two times deeper than for nucleons, we can conclude that the masses of primitives
(which turn in nuclear medium to dibaryons, i.e., resonances) associated with the $NN$-channel
must be greater than
\begin{equation}
\label{NC}m_D> 2 \varepsilon _F = 1.96\;{\rm GeV},
\end{equation}
where $\varepsilon _F = m­_N + 40$ MeV is the Fermi energy of
nucleons in nuclei. The primitive masses we derived earlier from
the $NN$ scattering, $m_D =$ 2047 and 2006 MeV, satisfy this constraint.

In the nonrelativistic approximation, the evolution of the local velocity  ${\bf v}$ of matter
inside the newly born protoneutron star is described by Euler's equation
$$
\rho \frac{\partial {\bf v}}{\partial t}+ \rho ({\bf v\cdot \nabla }){\bf v}%
=-{\bf \nabla }p-\rho {\bf \nabla }\Phi,
$$
where $ \Phi $ is gravitational potential, $\rho $ is density, and $p$ is pressure.
In the presence of the dibaryon Bose condensate, pressure remains constant, as indicated
by the horizontal line in Fig.~\ref{press}, and hence ${\bf \nabla }p=0$.
Gauss' law implies that $\int d{\bf S\cdot \nabla }\Phi =4\pi
GM(r)$, where $M(r)$ is the mass inside a sphere of radius $r$, so ${\bf \nabla }\Phi \neq 0$.
Thus, we conclude that ${\bf v} \neq 0$. In the approximation of an ideal gas,
the static solutions disappear because of formation of the dibaryon Bose condensate.
These arguments are valid in the general relativity. Formation of the dibaryon condensate
can be considered as a possible mechanism for the phase transition to quark matter.

Dibaryon-nucleon and dibaryon-dibaryon interactions contribute to the pressure. EoS of the binary mixture becomes
stiffer, ensuring the stability of neutron stars in
a certain range of densities. If the formation of the dibaryon Bose condensate is a first-order phase transition
and the density jump $ \Delta \rho $ at the phase transition is large enough,
the neutron stars become unstable when the central density hits the critical value
\cite{Lighthill 1950, Blinnikov 1975,Bisnovatyi-Kogan 1989}.

In a Bose gas of interacting bosons, a fraction of bosons is outside the condensate. These particles
move, collide with a boundary,
and contribute to the pressure. The stronger the repulsive interaction, the more stable can
be a Bose gas of interacting bosons. With increasing density, the interaction energy is growing faster
than the kinetic energy, so the difference between fermions and bosons becomes asymptotically less significant.
Coupling constants and masses of dibaryons can be constrained from the existence and stability of massive
neutron stars ~\cite{Faessler:1997jg}.

In the RMF model of Ref. \cite{Faessler:1997jg}
dibaryons with masses close to the ones we derived from the phase
shifts analysis ($m_D =$ 2047 and 2006 MeV in the $^3
S_1$ and $^1 S_0$ channels, respectively) with sufficiently weak $\sigma$-meson couplings
(weak attraction force) appear at high
densities without destabilizing neutron stars. Dibaryons with
strong $\sigma$-meson couplings appear at densities below $0.3$ fm$^{-3}$.
Such dibaryons generate instability (square of sound velocity gets negative),
providing phase transition to more dense phase of nuclear matter, e.g., quark matter.

There is therefore the interesting connection between parameters of
the lightest dibaryons and structure and stability of neutron
stars.

\section{Nuclear phase transition in collapsed stellar cores}
\setcounter{equation}{0}

The behavior of stars under a phase transition instability is determined by
the jump of density $\lambda$. If $\lambda$ exceeds the critical value
$\WID{\lambda}{c}=\frac{3}{2}$, the star becomes unstable at the moment the phase
transition starts in the center. This result is remarkably independent on the specific
properties of EoS. The magnitude of the density jump is the only relevant parameter \cite{Lighthill 1950}.

In general, the problem of stellar stability with a phase transition is more complicated as discussed,
for instance, in Refs. \cite{Blinnikov 1975,Bisnovatyi-Kogan 1989}. A star can lose stability even
when $\lambda < \frac{3}{2}$ (subcritical regime) on conditions that the interface between two phases
is not exactly at the stellar center and a new-phase inner core of finite size has already been formed.
The critical size of the new-phase core depends on the stiffness of EoS which is determined by the adiabatic
index averaged over the star $\overline{\gamma}$. The lower the stability factor $(\overline{\gamma}-\frac{4}{3})$
the smaller density jump is required to destabilize the star. The hydrostatic equilibrium of a star at
the boundary of stability ($\overline{\gamma}=\frac{4}{3}$) can be disturbed by phase transition with
an arbitrary small jump of density.

\subsection{Equation of state}

Since our goal is to investigate only the main features of the hydrodynamic behavior
of the instability we use a simplified approximation for EoS as described below.

The free energy is approximated in the following way
\begin{equation}
F_\mathrm{L,H}=\WID{F}{id}(T,\xi)+\triangle F_\mathrm{L,H}(n), \quad
n=\WID{n}{id}(T,\xi),
\label{EoS-General}
\end{equation}
where $F$ is the specific (per particle) free energy, $T$ is
temperature, and $n$ is the particle number density. The indices
$L$ and $H$ correspond to low and high density phases,
respectively. For simplicity, the matter is assumed to consist of
neutrons only. The functions $\WID{F}{id}$ and $\WID{n}{id}$ are
defined by the Fermi-Dirac statistics for ideal (noninteracting)
nonrelativistic particles and depend on the temperature and
parameter $\xi$ that failing the interaction is
simply the chemical potential.

The additive terms $\triangle F(n)$ in (\ref{EoS-General}) describe
the interaction and depend only on $n$:

\begin{equation}\label{DeltaF}
\triangle F_\mathrm{L,H}(n)=C_\mathrm{L,H}\,
n^{{\textstyle\gamma}_{\mathrm{L,H}} -1}-\WID{C}{id}n^{2/3}\, .
\end{equation}

Thus, Eqs.$\,$(\ref{EoS-General},\ref{DeltaF}) define $F$
as a function of $n$ and $T$ parametrically (parame\-ter $\xi$).
All the necessary for hydrodynamical calculations of thermodynamical quantities
such as the pressure, internal energy, entropy, chemical potential etc.
can be easily found from $F$ by standard way. For example, the pressure
$P$ and chemical potential $\mu$ are
\begin{equation}\label{Pmu}
 P=n^2\left(\frac{\partial F}{\partial n}\right)_T ,\quad\mu= F+\frac{P}{n}\, .
\end{equation}

The region of transition between the two phases can be found from
the equations of phase co-existence
\begin{equation}
\WID{P}{L}(T,\WID{n}{L})=\WID{P}{H}(T,\WID{n}{H}),\quad
\WID{\mu}{L}(T,\WID{n}{L})=\WID{\mu}{H}(T,\WID{n}{H}),
\label{PTregion}
\end{equation}
which require the continuity of pressure and chemical potentials
at the phase interface. These equations determine the particle
density of the phase transition beginning $\WID{n}{L}$ and that of its ending
$\WID{n}{H}$, and so the magnitude of density jump
$\lambda=\WID{n}{H}/\WID{n}{L}$, as functions of the temperature.

The constant $\WID{C}{id}$ corresponds to ideal degenerate nonrelativistic
neutron gas at zero temperature. Therefore, at $T=0$ the second term in
the right hand side of Eq.$\,$(\ref{DeltaF}) cancels the term
$\WID{F}{id}$ in Eq.$\,$(\ref{EoS-General}).
Thus, at zero temperature we have a polytropic EoS
(for both the phases) whereas for $T>0$ it has more general
form defined by Eq.$\,$(\ref{EoS-General}). What why we use
the term ``quasi-polytropic star''.

We start hydrodynamic calculations from a hydrostatic configuration
at $T=0$. According to Eqs.$\,$(\ref{EoS-General},\ref{DeltaF},\ref{Pmu})
at zero temperature we have
\begin{eqnarray}
 & & F_\mathrm{L,H}(n)=C_\mathrm{L,H}\;
n^{{\textstyle\gamma}_{\mathrm{L,H}} -1},\quad
P_\mathrm{L,H}=({\textstyle\gamma}_{\mathrm{L,H}} -1)\, C_\mathrm{L,H}\;
n^{{\textstyle\gamma}_{\mathrm{L,H}}},\label{EoST0FP}\\
& & \mu_\mathrm{L,H} =\frac{{\textstyle\gamma}_{\mathrm{L,H}}}
{{\textstyle\gamma}_{\mathrm{L,H}} -1}\frac{P_\mathrm{L,H}}{n}\, .
\label{EoST0mu}
\end{eqnarray}

 From Eqs.$\,$(\ref{PTregion}) it follows that
for a polytropic EoS, such as given by Eqs.$\,$(\ref{EoST0FP},\ref{EoST0mu}),
the phase transition density jump $\lambda$ is
unambiguously determined by adiabatic indices of L and H phases:
\begin{equation}
\lambda \equiv\frac{\WID{n}{H}}{\WID{n}{L}} =
\frac{\WID{\gamma}{H}(\WID{\gamma}{L}-1)}{\WID{\gamma}{L}(\WID{\gamma}{H}-1)}\, .
\label{Pol_jump}
\end{equation}

\subsection{Initial model}

 Our calculations of the phase transition hydrodynamics were performed in dimensionless
 variables. Everywhere below (in the text and figures) the dimensionless
 values will be implied if not specified explicitly.

 The following units for the temperature $T$, density $\rho$,
 velocity $V$, time $t$, pressure $P$, and energy $E$ were used
\begin{equation}
\begin{split}
&T:\left[\frac{\WID{m}{u}}{\WID{k}{b}}\frac{G\WID{M}{s}}{\WID{R}{s}}\right],\quad
\rho: \left[\frac{\WID{M}{s}}{4\pi\WIT{R}{s}{3}}\right],\quad V:
\left[\sqrt{\frac{G\WID{M}{s}}{\WID{R}{s}}}\right],\\
& t:\left[\sqrt{\frac{\WIT{R}{s}{3}}{G\WID{M}{s}}}\right], \quad
P:\left[\frac{G\WIT{M}{s}{2}}{4\pi\WIT{R}{s}{4}}\right],\quad
E:\left[\frac{G\WIT{M}{s}{2}}{\WID{R}{s}}\right], \label{Units}
\end{split}
\end{equation}
where $\WID{k}{b}$ is the Boltzmann constant,  $G$ --- gravitational
constant, $\WID{M}{s}$ and $\WID{R}{s}$ ---
the mass and initial radius of the star, respectively.

As an example, $\WID{M}{s}$ and $\WID{R}{s}$ for a hot protoneutron
star formed in the collapse of the SN core in several tens of
ms after bounce can be estimated as
$\WID{M}{s}= 1.4 M_{\odot}$ и $\WID{R}{s}\approx 39~\mbox{km}$.
Substituting these values for $\WID{M}{s}$ and $\WID{R}{s}$ in
Eqs.$\,$(\ref{Units}) we obtain
\begin{equation}
\begin{split}
[\ T\ ]\approx 49\,\mbox{MeV}\, ,\ \  [\ \rho\ ]\approx\xmn{3.7}{12}\,\mbox{g/cm}^3,
\ \  [\ V\ ]\approx 69\ 000~\mbox{km/s}\, ,\\
[\ t\ ]\approx 0.57~\mbox{ms}\, ,\ \ [\ P\ ]\approx\xmn{1.7}{32}\,\mbox{erg/cm}^3,
\ \ [\ E\ ]\approx\xmn{1.3}{53}\,\mbox{erg}\, . \label{Unitsproto}
\end{split}
\end{equation}

Instead of the particle number density $n$, we will use hereafter
the mass density $\rho$ that is connected with
$n$ by the relation $\rho =\WID{m}{u}n$ where
$\WID{m}{u}$ is the atomic mass unit.

The initial configuration in our calculations was
represented by a hydrosta\-tically equilibrium star consisting
of L-phase matter at $T=0$ with the adiabatic index
$\WID{\gamma}{L}=5/3$.  The dimensionless form of
Eq.$\,$(\ref{EoST0FP}) for pressure reads as
\begin{equation}\label{PotRho}
P=A\, \rho^{5/3}\, , \quad
A = \frac{2}{3}\,\frac{\WID{C}{L}\,
\WID{m}{u}^{5/3}}{G (4\pi)^{2/3}M^{1/3}_s R_s}\, .
\end{equation}
So the initial model is a polytrope of index $k=1.5$ with $A=0.078484$
and dimensionless central density $\rho_{\mathrm c}=17.97$.
The initial central density at $t=0$ was assumed to be just
at the beginning of the phase transition.

In order to simulate the fact that in a core-collapse SN the stellar core
is actually rather hot and the pressure can exceed that for ideal
Fermi gas at zero temperature, the factor $\WID{C}{L}$ was chosen
to be equal to $2.83\times\WID{C}{id}$. If the density jump $\lambda$
and $\WID{\gamma}{L}$ are specified at $T=0$ then one can find
$\WID{\gamma}{H}$ from Eq.$\,$(\ref{Pol_jump}).
Next, the factor  $\WID{C}{H}$ can be estimated from the relation
\begin{equation}\label{PLPH0}
\WID{P}{L}(0,\WID{\rho}{c})=\WID{P}{H}(0,\lambda\WID{\rho}{c})\, .
\end{equation}
For the main value $\lambda=1.525$ (at $T=0$) in our calculations
and $\WID{\gamma}{L}=5/3$ one can derive from Eq.$\,$(\ref{Pol_jump})
$\WID{\gamma}{H}\approx 1.355$.

\begin{figure}[!htb]
\begin{center}
\includegraphics[width=0.618\textwidth]{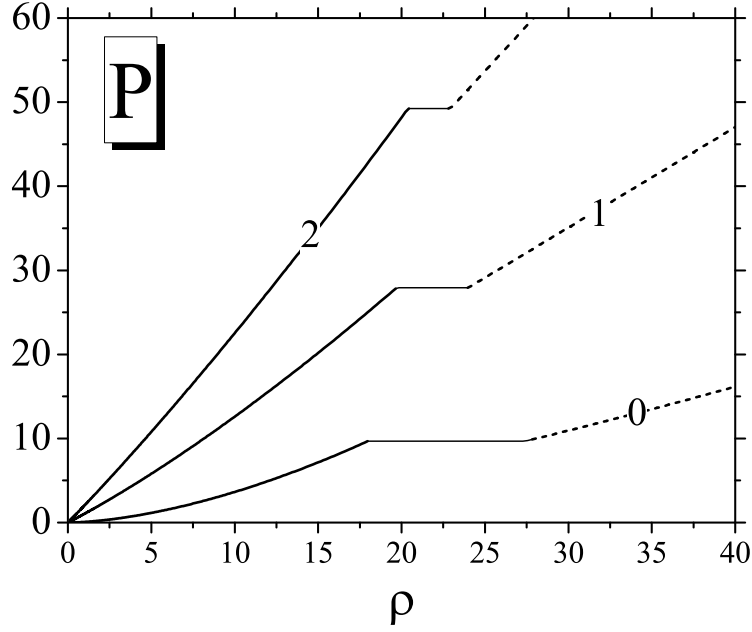}
\end{center}
\caption{Pressure as a function of density at different temperatures $T=0, 1, 2$}
\label{Pic-Pressure}
\end{figure}
\vspace{5 mm}
\begin{figure}[!hbt]
\begin{center}
\includegraphics[width=0.618\textwidth]{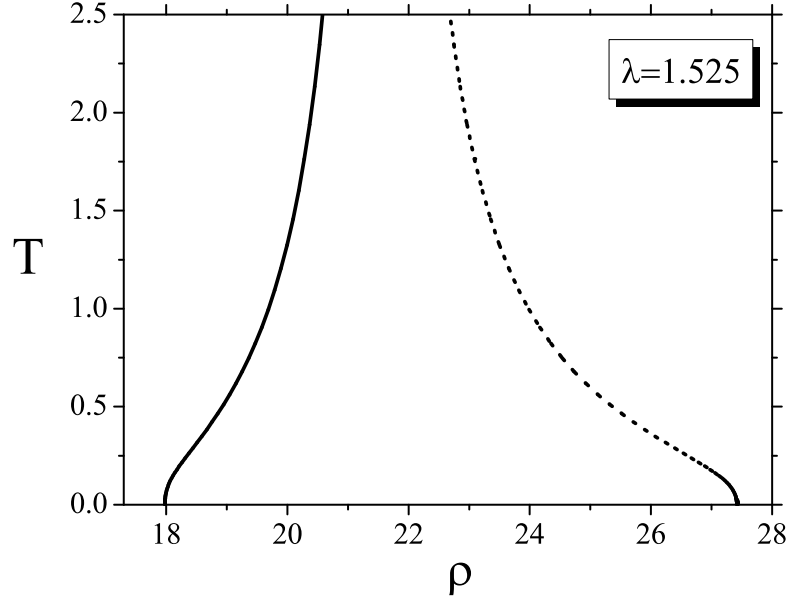}
\end{center}
\caption{The phase diagram for $\lambda=1.525$ at $T=0$}\label{Pic-PhaseDiagram}
\end{figure}

 In subsequent hydrodynamic calculations the quantities
 $\WID{C}{L,H}$ and $\WID{\gamma}{L,H}$ at any temperatures
 remain fixed and equal to their values estimated at $T=0$
 as is described above.

 The only parameter that we change in different versions
 of our calculations is the value of $\lambda$ at $T=0$.

Figure$\,$\ref{Pic-Pressure} shows typical dependence of
pressure on density at different dimensionless temperatures
$T=0$ (initial model), 1 and 2
(for version $\lambda=1.525$ at $T=0$).

An example of the phase diagram in our calculations is shown in
Fig.$\,$\ref{Pic-PhaseDiagram} for $\lambda=1.525$ at $T=0$.
The domains at the left of the solid curve and from the right
of the dashed one cover the low-density and high-density
single-phase states, respectively.
The mixed-phase states occur everywhere over the region between
those curves. Here the pressure does not depend on density
at given temperature.

One can see that the density jump $\lambda$ decreases with increasing
temperature.

\subsection{Results of hydrodynamical calculations}

The hydrodynamic calculations were run with fully implicit
Lagrangian code, the star being divided onto 1000 mass zones.

In order to activate the phase transition the hydrostatic
equilibrium was perturbed by imposing a small inward velocity
$V_0$ to stellar matter $(V_0 = -\alpha r)$.
The total kinetic energy injected in the star
by such a perturbation was as small as $\xmn{5}{-8}$ of the star
gravitational energy $\WID{E}{g}$ for $\lambda=1.525$.
In this way several models were calculated
for different values of $\lambda$ at $T=0$.
Below we discuss the results for the model with
supercritical density jump $\lambda=1.525$ at $T=0$.

\begin{figure}[!htb]
\begin{center}
\includegraphics[width=0.618\textwidth]{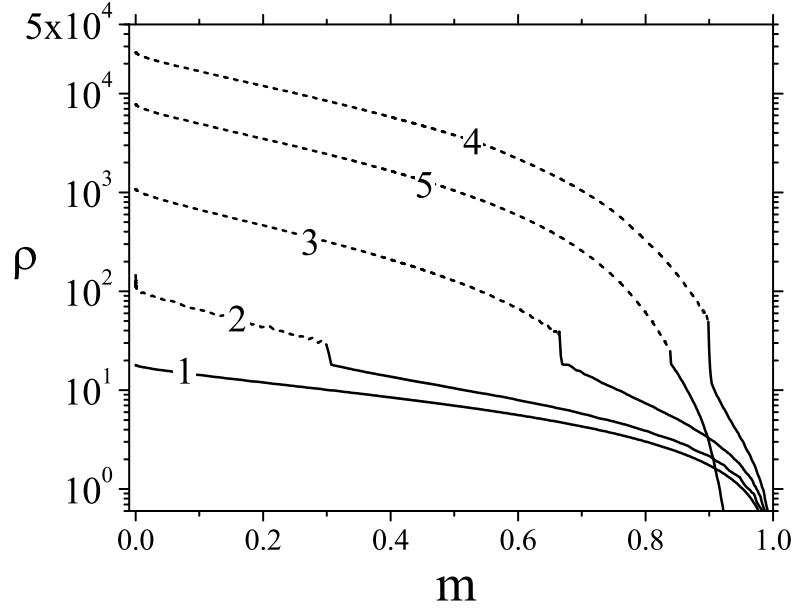}
\end{center}
\caption{Temporal evolution of density}
\label{Pic-Density}
\end{figure}

Figures$\,$\ref{Pic-Density}--\ref{Pic-Velocity} show
the temporal evolution of density, temperature and velocity
from the time of the loss of stability $t=0$ (curve labelled 1)
on account of the appearance of new phase to the approach
to final quasi-equilibrium configuration $t=10$ (curve labelled 5).
Intermediate curves correspond to times $t=6.7\, (2)$,
$t=7.4\, (3)$, and $t=7.7\, (4)$.
Solid sections of the curves belong to domains contain
low-density phase whereas dashed ones correspond to
high-density phase.
All the quantities are shown as the functions of dimensionless
mass coordinate $0\leq m\leq1$.

\begin{figure}[!htb]
\begin{center}
\includegraphics[width=0.618\textwidth]{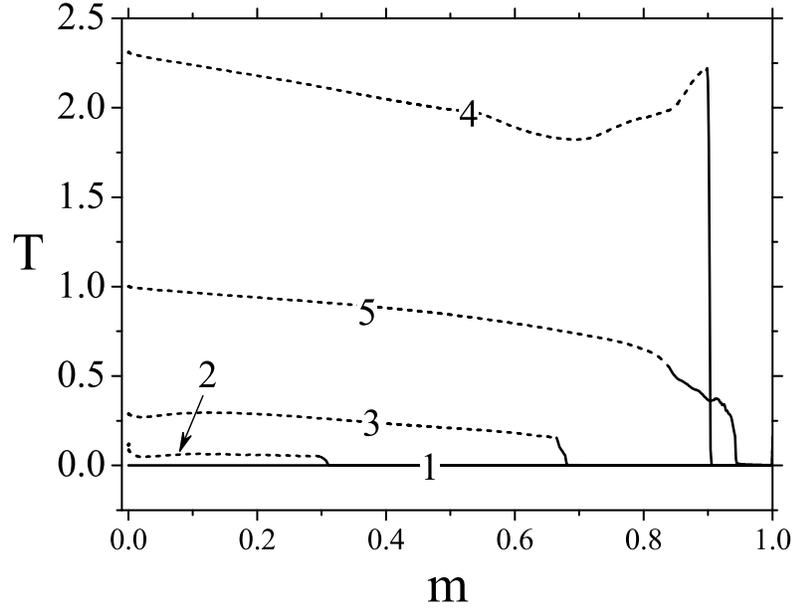}
\end{center}
\caption{Temporal evolution of temperature}
\label{Pic-Temperature}
\end{figure}
\vspace{5mm}
\begin{figure}[!htb]
\begin{center}
\includegraphics[width=0.618\textwidth]{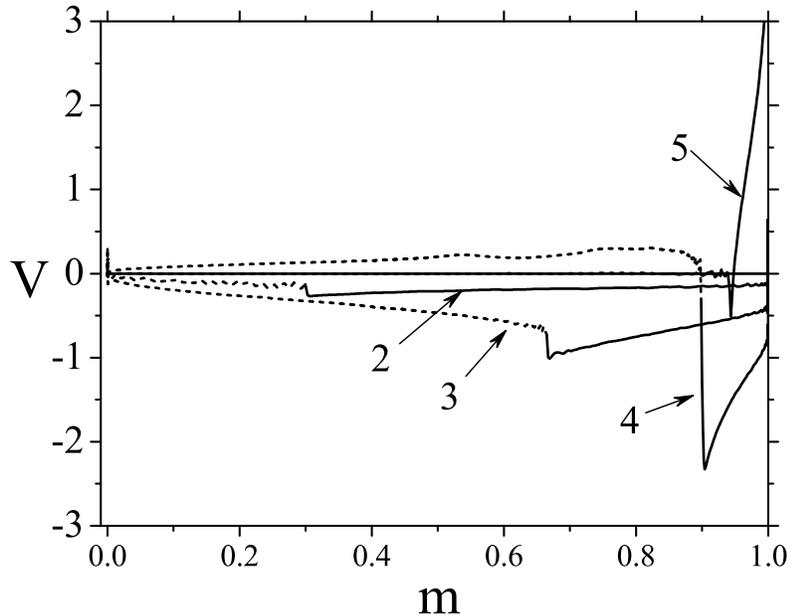}
\end{center}
\caption{Temporal evolution of velocity}
\label{Pic-Velocity}
\end{figure}

 One can clearly observe that the conversion of low-density
 matter into high-density one has a nature of mini collapse
 and finally results in hydrodynamic transition of the star
 into new equilibrium.
 The onset of a new phase at stellar center causes pulsational
 instability. The interface between two phases generates
 progressive waves that excite oscillation of the new-phase core.
 According to our numerical experiments, the heat-transport processes
 turn out to be of significant importance here for to convert the kinetic
 energy of pulsations in heat and to redistribute it in matter.
 To include convection in the mixing length approximation
 (see e.g. \cite{CoxGuili 2004}) proves to be sufficient in our case.

As the core of new phase increases the collapse gathers strength
and outgoing shock front appears at the phase interface.
Both the density and temperature are growing quickly behind the front.
This process becomes similar to standard bounce during the collapse
of stellar cores without phase transition. Ultimately the core stops to compress
and rebounds under the action of ram pressure creating outgoing
shock wave that reaches outermost low density layers, cumulates
and ejects the envelope of a mass $\WID{M}{ejc}\lesssim 0.05
\WID{M}{s}$.

We have calculated a number of models with different values
of the initial density jump $\lambda$ at $T=0$.

\begin{figure}[!htb]
\begin{center}
\includegraphics[width=0.618\textwidth]{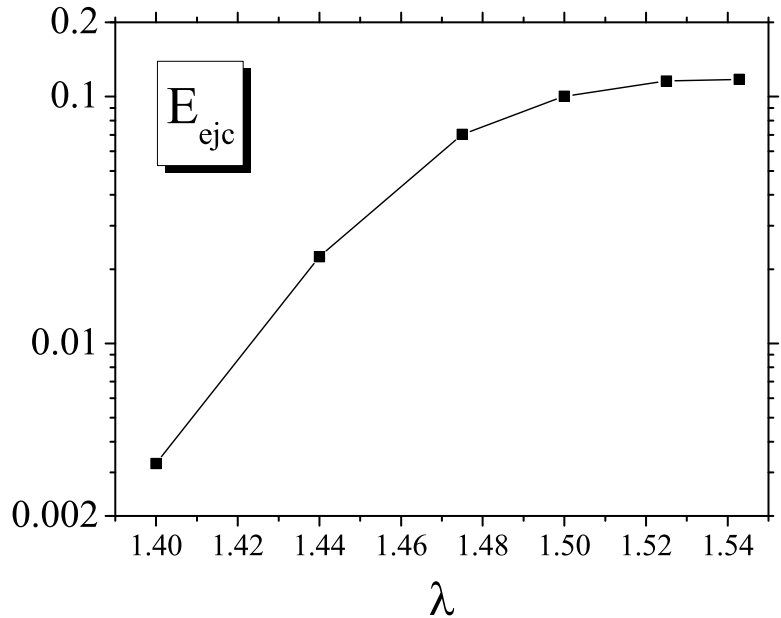}
\end{center}
\caption{Kinetic energy of ejected matter versus $\lambda$ at $T=0$}
\label{Pic-Erelease}
\end{figure}

Figure$\,$\ref{Pic-Erelease} shows kinetic energy of ejected
matter $\WID{E}{ejc}$ in terms of the initial gravitation energy
of the star as a function of $\lambda$. For
$\lambda\gtrsim\frac{3}{2}$,  $\WID{E}{ejc}$ has virtually
constant value of about 12\% of initial gravitational energy since
the initial configuration is a polytrope of index 1.5 with the
dimensionless gravitational energy  $\WID{E}{g}=-6/7$. When
$\lambda$ decreases below critical value 3/2, $\WID{E}{ejc}$ falls
down abruptly --- at subcritical $\lambda=1.4$ it only amounts to
0.4\%. It is significant that in the subcritical regime one has to
increase the initial inward velocity in order to excite undamped
(converting finally in collapse) oscillations of the new-phase
core. For $\lambda=1.4$ the kinetic energy of the initial
perturbation of about $\xmn{2}{-5}|\WID{E}{g}|$ proved to be
sufficient.

The lower $\lambda$, the larger
the new-phase core should be. This is in full
agreement with the work of Blinnikov \cite{Blinnikov 1975}.
If the mass of a subcritical core is not large enough
the core remains stable.
Moreover, for $\lambda < 3/2$ the ejection of stellar envelope
proceeds owing to numerous pulsations each entailing
small leakages of matter rather than a single outflow
forced by a shock wave.\\

\subsection{Discussion}

 Our calculations provide an example of hydrodynamical development
 of the phase transition-induced instability in collapsing stellar cores.
 The onset of the new-phase inner central core gives rise to
 a shock wave that ejects though low mass  but of high
 velocity envelope. For our model the kinetic energy of the ejection
 can reach about 12\% of the initial gravitational energy of
 the core.
 Taking into account the unit of energy
 from Eq.\,(\ref{Unitsproto}) we obtain
 $\WID{E}{ejc}\approx\xmn{1.3}{52}$erg ($\lambda=1.525$).
 This is even more than necessary to conform
 to astronomical observations.

 In reality the final ejection of SN envelope
 will occur after the following interaction of the phase transition induced
 shock wave with an outer standing accretion shock that
 decomposes infalling iron-peak nuclei into free nucleons.
 Thus, behind the accreting shock there exists a layer
 enriched with free neutrons and protons.
 When crossing this layer the phase transition shock has only to
 slightly accelerate it to cause its expansion
 accompanied  by cooling and recombination of
 neutrons and proton back into heavy nuclei.
 The energy liberation due to such a recombination
 (about $7-8\,$MeV per recombined nucleon) could
 facilitate the final expulsion of SN envelope
 \cite{Nadyozhin1978}. So, the phase transition shock can act at least as
 a trigger of the SN explosion.

  The gravitational energy released in the hydrodynamical transition
  of the star into a new-phase state converts mostly to heat while
  the density increases by a factor of $10^3$
  (Figs.\,\ref{Pic-Density},\,\ref{Pic-Temperature}).
  Therefore, the processes of heat transport becomes of primary
  importance. The major factors here are expected to be convection
  and neutrino transport in quark-gluon plasma.
  The latter was neglected in our calculations whereas
  convection was included. Hence, our explosion energy $\WID{E}{ejc}$
  seems to be underestimated because additional energy supply
  provided by neutrino heat transport to outer layers can only
  contribute to the ejection energy.

\section{Conclusions and outlook}

We briefly reviewed in this paper vast material related to the structure of QCD vacuum and
the properties of neutron stars and presented new results on the physics of the density-induced deconfinement
(Sect. III), relativistic QCB model (Sects.V-VII), EoS of nuclear matter with an admixture
of MQS and possible phase transitions (Sects. IX-X), and finally and most importantly, we have
considered in Section XI the evolution of newly born protoneutron stars under the influence
of phase transition in nuclear matter.

We have shown that the phase transition curve in $ (\mu, T) $ plane of states of the hadron-quark-gluon matter
can be understood and predicted on the basis of the vacuum energy dominance in good agreement with available lattice data,
however, the resulting critical density for $ T \la 20 $ MeV is very high, $ \rho_c \sim 30 \rho_0 $,
which leaves room for possible intermediate phase transitions in normal nuclear matter below the deconfined quark phase.

To describe this region, one needs a more adequate description of the nucleon-nucleon forces,
which includes quark degrees of freedom. We envisaged old QCB model, which was reformulated in
modern relativistic form and well describes the experimental data. Having at hands new $NN$
(and, in principle, $3N, 4N $, ..) forces, we have formulated new system of equations for the EoS of
nuclear matter with an admixture of 6QS, and here the main question is whether MQS play
a role of "primitives'', i.e., the instantaneous objects generating hard $NN$ repulsive cores,
or they exist further in the environment on the mass shell and reveal themselves as a new
physical component of high-density nuclear matter. We have shown that this new phase
may appear when the density increases up to $ (2 \div 3) \rho_0 $, while more quantitative analysis is currently in progress.
From this side, one can apply the resulting EoS to describe the interiors of neutron stars,
the cumulative effect in reactions on nuclei, and study the role of MQS in phase transitions in nuclear matter.

 One should emphasize a twofold role of
the phase transition of the considered type in physics of supernovae. On the
one hand, the phase transition could be a missing ingredient necessary to
resolve a 30-years puzzle of supernova explosion in spherical symmetry. On the
other hand, as demonstrated by recent calculations in Ref. \cite{Sagert 2009a},
supernova outburst induced by the phase transition is accompanied by a second
peak on its neutrino light curve that occurs in about 250 ms after
core-collapse bounce. The temporal and neutrino-flavor properties of such a
peak provide unique information about the EoS of superdense matter. Thus,
nuclear and particle physicists have a good chance to experimentally check
their findings in case of detection the neutrino signal from stellar core
collapse in our Galaxy. As for astrophysicists, further detailed and systematic
modeling of the phase transition effects in core-collapse supernovae is of
great importance.

\begin{acknowledgments}

The authors wish to thank Natali Igumnova for help in preparation of the manuscript.
D.K.N. and A.V.Yu. are supported by the SNSF grant (SCOPES project
No.~IZ73Z0--128180/1) and Russian Federal Agency for Science and
Innovations (Contract No.~02.740.11.0250). D.K.N. is supported by the RFBR grant No.~09-02-12168-ophi\_m.
M.I.K. is supported by the RFBR grant No. 09-02-91341 and DFG grant No. 436 RUS 113/721/0-3,
Yu.A.S. and M.A.T. are supported by the RFBR grant No. 09-02-00629.

\end{acknowledgments}

\newpage

\section{Appendix A: The $6q$ description of $^1S_0$ and $^3S_1$ states}
\setcounter{equation}{0}

We  follow here results of Ref.~\cite{SIMO81} and the last reference in~\cite{NARO94}.
For $~^1 S_0$ and $~^3 S_1$ we neglect $V_{hh}$ and keep only two
levels of the $6q MS$. The following form of $V_{hqh}$ is used
\begin{eqnarray}
V_{hqh}  (r,r') &=& \frac{[(E_1-E)\eta_1(r) -c_1 \delta(r-b)][(E_1-E)\eta_1(r') -c_1 \delta(r'-b)]}{E-E_1}
+ \frac{c^2_2 \delta(r-b) \delta(r'-b)}{E-E_2}.\label{A1.1}
\end{eqnarray}
\end{widetext}
The $P$-matrix for this potential has the form
\be P(b) =k + \frac{2\mu
c_1}{\eta_1 (b)} + \frac{2\mu c^2}{E-E_1} + \frac{2\mu
c^2_2}{E-E_2}.
\label{A1.2}
\ee

The quantity $k$ is defined due to orthogonality of $\eta_\nu (r)$
to be \cite{NARO94}
\be \left.\frac{1}{\eta_\nu(r)}\frac{d}{dr}
\eta_\nu (r) \right|_{r=b} = k.
\label{A1.3}
\ee

 From the Schr\"odinger equation for the total wave function
$\psi(r)$ one obtains $\beta_1$ \be \eta_1 (b) c_1 =-
\frac{\beta^2_1}{2\mu} + E_1.\nonumber\ee

Then the values of all parameters reproducing the phases with good
accuracy in the interval $0\leq T_{Kin} \leq 515 $ MeV are as follows
\begin{eqnarray*} ~^1S_0: ~~ b &=& 7.42{\rm ~GeV}^{-1},~~ \beta_1=0.27{\rm ~GeV},     \\
                           c_1 &=& 0.34{\rm~GeV}^{1/2},~~ c_2 = 0.371{\rm ~GeV}^{1/2},\\
                           ~E_1&=& 0.23{\rm ~GeV},~~ E_2 =1{\rm ~GeV},                \\
                  ~^3S_1: ~~ b &=& 7.025{\rm ~GeV}^{-1},~~ \beta_1=0.268{\rm ~GeV},   \\
                        ~~ c_1 &=&0.343{\rm~GeV}^{1/2},~~c_2=0.445{\rm ~GeV}^{1/2},   \\
                            E_1&=&0.243{\rm ~GeV},~~ E_2 =1{\rm ~GeV}.
\end{eqnarray*}

\section{Appendix B: Transformation properties of dibaryon currents}
\setcounter{equation}{0}

In this section, transformation properties of the nucleon wave functions and the operators $O$ entering two-nucleon currents
under the charge conjugation are given.
The nucleon wave functions $\Psi =\psi \otimes \chi $ are products of bispinors $\psi$ and isospinors $\chi$.

\subsection{Isospin $C$-parity}

The isospin $C$-parity operator and its transformation properties under
the main algebraic operations are given by
\begin{eqnarray*}
C_{I} &=&i\tau ^{2} = \left(
\begin{array}{ll}
 0 & 1 \\
-1 & 0
\end{array}
\right), \\
C_{I}^{T} &=&C_{I}^{-1}=C_{I}^{+}=-C_{I}, \\
C_{I}^{*} &=&C_{I}.
\end{eqnarray*}

The $C$-conjugation of isospinors is defined by
\begin{eqnarray*}
\chi _{c} &=&C_{I}\chi ^{+T}, \\
\chi _{c}^{+} &\equiv &(\chi _{c})^{+}=-\chi ^{T}C_{I}.
\end{eqnarray*}

The isospin hermitian matrices are expanded over the set $\Xi = (1,\tau^{\alpha})$,
the components of the set transform as follows
\begin{eqnarray}
C_{I}^{T}\left(
\begin{array}{l}
1 \\
\tau ^{\alpha }
\end{array}
\right) ^{T}C_{I}=\left(
\begin{array}{l}
1 \\
-\tau ^{\alpha }
\end{array}
\right).
\label{AAPI}
\end{eqnarray}
The equivalent form reads $C_{I}^{T}\Xi ^{T}C_{I}=(-)^{I}\Xi$,
where $I$ is isospin of the element.
If $\chi_1$ and $\chi_2$ are $q$-numbers,
\[
\chi _{1c}^{+}\Xi \chi _{2} = \chi _{2c}^{+}C_{I}^T\Xi ^{T}C_{I}\chi
_{1}=(-)^{I}\chi _{2c}^{+}\Xi \chi _{1}.
\]

\subsection{Lorentz $C$-parity:}

In the standard representation \cite{BJDR}, the $C$-conjugation matrix
and its transformation properties under the main algebraic operations
are given by
\begin{eqnarray*}
C_{L} &=&-i\gamma ^{0}\gamma ^{2}=-i\alpha ^{2} = -\left(
\begin{array}{ll}
0 & i\sigma ^{2} \\
i\sigma ^{2} & 0
\end{array}
\right), \\
C_{L}^{T} &=&C_{L}^{-1}=C_{L}^{+}=-C_{L}, \\
C_{L}^{*} &=&\bar{C}_{L}=C_{L}.
\end{eqnarray*}

The $C$-conjugated bispinors are defined as follows
\begin{eqnarray*}
\psi _{c} &=&i\gamma ^{2}\psi ^{*} = C_{L}\bar{\psi}^{T}, \\
\bar{\psi}_{c} &\equiv &\overline{(\psi _{c})}=\psi ^{T}C_{L}.
\end{eqnarray*}

Matrices acting on the bispinors can be expanded over the
set $\Gamma = (1,\gamma_{5},\gamma ^{\mu },\ldots )$, the components of the set obey
\begin{eqnarray}
C_{L}^{T}\left(
\begin{array}{l}
1 \\
i\gamma _{5} \\
\gamma ^{\mu } \\
\gamma _{5}\gamma ^{\mu } \\
\sigma _{\mu \nu } \\
i\gamma _{5}\sigma _{\mu \nu }
\end{array}
\right) ^{T}C_{L} &=&\left(
\begin{array}{l}
1 \\
i\gamma _{5} \\
-\gamma ^{\mu } \\
\gamma _{5}\gamma ^{\mu } \\
-\sigma _{\mu \nu } \\
-i\gamma _{5}\sigma _{\mu \nu }
\end{array}
\right).
\label{AAPP}
\end{eqnarray}
The bilinear forms are transformed as follows
\[
\bar{\psi}_{1c}\Gamma \psi _{2}=\bar{\psi}_{2c}C_{L}^{T}\Gamma ^{T}C_{L}\psi
_{1}=\pm \bar{\psi}_{2c}\Gamma \psi _{1},
\]
the sign can be read off from Eq.~(\ref{AAPP}).

\subsection{Combined $C$-parity}

The $C$-conjugation operator acting on the nucleon wave functions
$\Psi =\psi \otimes \chi $ has the following properties:
\begin{eqnarray*}
C &=&C_{L}C_{I}, \\
C^{T} &=&C^{-1}=C^{+}=C^{*}=C, \\
\bar{C} &=&-C.
\end{eqnarray*}

The $C$-conjugated nucleon wave functions are given by
\begin{eqnarray*}
\Psi _{c} &=&C\bar{\Psi}^{T}, \\
\bar{\Psi}_{c} &\equiv &\overline{\left( \Psi _{c}\right) }=-\Psi ^{T}C.
\end{eqnarray*}

The matrices $O$ entering two-nucleon currents can be expanded over
products of the matrices of $\Xi$ and $\Gamma$. They transform
according to the rule
\begin{eqnarray*}
C^{T}\Xi ^{T}\Gamma ^{T}C &=&\pm \Xi \Gamma.
\end{eqnarray*}
The bilinear forms are transformed under the permutations as follows
\begin{eqnarray*}
\bar{\Psi}_{1c}\Xi \Gamma \Psi _{2} &=& \bar{\Psi}_{2c} C^{T}\Xi ^{T}\Gamma ^{T}C \Psi _{1} = \pm \bar{\Psi}_{2c}\Xi \Gamma \Psi_{1}.
\end{eqnarray*}
The signs are fixed by Eqs.~(\ref{AAPI})~and~(\ref{AAPP}).

\subsection{Even and odd bilinear forms}

Using the above rules, one finds even bilinear forms under permutations
of the nucleon fields. These forms are constructed using matrices
\begin{eqnarray*}
\left( 1, ~i\gamma_5, ~\gamma_5 \gamma_{\mu}\right),\\
\tau^{\alpha} \left(\gamma_{\mu}, ~\sigma_{\mu \nu},~i\gamma_5 \sigma_{\mu \nu}\right).
\end{eqnarray*}
The odd bilinear forms are constructed using matrices
\begin{eqnarray*}
\tau^{\alpha} \left( 1, ~i\gamma_5, ~\gamma_5 \gamma_{\mu}\right),\\
\left(\gamma_{\mu}, ~\sigma_{\mu \nu},~i\gamma_5 \sigma_{\mu \nu}\right).
\end{eqnarray*}
The odd structures
including the first-order differential operator, that
enter two-nucleon currents, are listed in Table \ref{tab:2}.

\end{document}